\documentclass{jfm}

\usepackage{subfiles}
\usepackage{graphicx}
\usepackage{epstopdf,epsfig,subfloat,subfig,floatrow}
\usepackage{newtxtext}
\usepackage{newtxmath}
\usepackage{natbib}
\usepackage{amsfonts}
\usepackage{amsmath}
\usepackage{amssymb,color,mathtools}
\usepackage{bm}
\usepackage{hyperref}
\usepackage{multirow,makecell,array,booktabs}

\usepackage{ulem}

\hypersetup{
    colorlinks = true,
    urlcolor   = blue,
    citecolor  = blue,
}

\newcommand{\RomanNumeralCaps}[1]
\linenumbers
\allowdisplaybreaks[4]

\usepackage{soul,cancel}

\title{Kinetic modelling of rarefied gas flows with radiation}

\author{Qi Li,
		Jianan Zeng
        \and Lei Wu
        \corresp{\email{wul@sustech.edu.cn}}
        }

\affiliation{
  Department of Mechanics and Aerospace Engineering, Southern University of Science and Technology, Shenzhen 518055, China
  }

\begin{document}
\maketitle

\begin{abstract}

Two kinetic models are proposed for high-temperature rarefied (or non-equilibrium) gas flows with radiation. One of the models uses the Boltzmann collision operator to model the translational motion of gas molecules, which has the ability to capture the influence of intermolecular potentials, while the other adopts the relaxation time approximations, which has higher computational efficiency. In the kinetic modelling, not only the transport coefficients such as the shear/bulk viscosity and thermal conductivity but also their fundamental relaxation processes are recovered. Also, the non-equilibrium dynamics of gas flow and radiation are coupled in a self-consistent manner. The two proposed kinetic models are first validated by the direct simulation Monte Carlo method in several non-radiative rarefied gas flows, including the normal shock wave, Fourier flow, Couette flow, and the creep flow driven by the Maxwell demon. Then, the rarefied gas flows with strong radiation are investigated, not only in the above one-dimensional gas flows, but also in the two-dimensional radiative hypersonic flow passing cylinder. In addition to the Knudsen number of gas flow, the influence of the photon Knudsen number and relative radiation strength is scrutinised. It is found that the radiation can make a profound contribution to the total heat transfer on obstacle surface.

\end{abstract}


\section{Introduction}

The non-equilibrium dynamics of molecular (diatomic/polyatomic) gas is commonly encountered in aerospace engineering. For example, at a high Mach number, the air surrounding an aircraft decelerates and heats up rapidly after compression by shock waves, which causes strong conversion from the translational energy into the internal energy. The temperature may reach thousands of degrees Kelvin and leads to significant changes in the physical and chemical properties of the gas \citep{Anderson2019, Ivano1998ARFM}. Meanwhile, the thermal radiation, which is induced by transitions between excited states of gas molecules, makes a significant contribution to the overall heat load on hypersonic aircraft. For instance, in the reentry into Mars' atmosphere, the shock layer radiation can constitute a larger portion than convective heating in the stagnation region of spacecraft~\citep{Edquist2014JSR}, and an order of magnitude analysis suggests that the absolute radiative power at the
stagnation point scales as eighth-power of entry velocity \citep{daSilva2011AIAA}. Therefore, accurate predictions of molecular gas flow and radiative energy transfer are essential for developing thermal protection systems for aircraft entering planetary atmospheres. Additionally, radiative gas dynamics also plays important role in interpretation of spectrometer measurements \citep{Horvath2010AIAA} and technologies of gas dynamics lasers.

Under the assumption of thermodynamic equilibrium, the traditional Navier-Stokes-Fourier equations are used to predict the thermal environment and aerodynamic characteristics of the aircraft. The influence of internal degrees of freedom (DoF) is taken into account by the variations of heat capacity and transport properties of molecular gas \citep{Malik1991}. On the other hand, when the thermodynamic non-equilibrium occurs, gases with different temperatures associated with various relaxation processes needs to be considered. Several sets of Navier-Stokes-type equations have been developed with multi-temperatures of different types of kinetic modes \citep{Taylor1969RMP, Colonna2006JTHT, Bruno2011PoF, Aoki2020PRE}, and applied in the simulations of thermal non-equilibrium gas dynamics \citep{Kustova2011SW, Armenise2016JTHT}. Meanwhile, studies of radiative gas flow are largely oriented towards hypersonic reentry vehicles \citep{Vincent1971ARFM}. The radiation field and the state of gas flow must be determined self-consistently, and the coupling of gas dynamics and radiative heat transfer is achieved by integrating the radiative heat transfer as a heat source into the total energy conservation equation.

Since the macroscopic models are obtained at small Knudsen number, they are only applicable in the near-continuum flow regime, where the mean free path of gas molecules is much smaller than the characteristic flow length. However, the gas could be in highly thermal non-equilibrium in many realistic situations, such as the reentry of aircraft into the atmosphere, where the gas flow changes from the continuum to the free-molecular regimes. Therefore, the treatment based on gas kinetic theory from mesoscopic perspective is inevitable, as the molecular dynamics simulation at microscopic level is limited to small spatial and temporal domains. The fundamental equation in gas kinetic theory is the Boltzmann equation, but it is only rigorously established for monatomic gas. For the molecular gas, its internal DoF pose difficulties in the modelling of rarefied gas dynamics. The heuristic way to describe the molecular gas dynamics in all flow regimes is the \cite{WangCS} (WCU) equation, which treats the internal DoF quantum mechanically and assigns each internal energy level an individual velocity distribution function. A general framework for the kinetic modelling of molecular gases is proposed recently based on a set of allowed internal states endowed with a suitable measure \citep{Borsoni2022CMP}. However, the complexity and excessive computational burden prevent their engineering applications.


The direct simulation Monte Carlo (DSMC) method \citep{Bird1994} is prevailing in simulating the rarefied gas dynamics. Although it is proven that DSMC is equivalent to the Boltzmann equation for monatomic gas~\citep{wagner_consist}, there are some drawbacks when applied to radiative molecular gas flows. First, the bulk viscosity and the thermal conductivities cannot be recovered simultaneously. The reason lies in its phenomenological collision model of \cite{Borgnakke1975}, which realizes the correct exchange rate between the translational and internal energies to exactly recover the bulk viscosity \citep{Haas1994, Lavin2002PoF}; however, it cannot guarantee that the thermal conductivity, or its translational and internal components, is recovered at the same time \citep{Wu2020JFM, Li2021JFM}. Second, the DSMC method has been coupled with photon Monte Carlo method to simulate  radiative gas flow, by calculating the rates of radiative heating/cooling at each time step~\citep{Sohn2012JTHT}. However, because of the high sensitivity of radiation rates on temperature, the fluctuation of temperature sampled in simulation cells may lead to significant instabilities \citep{Prem2019Icarus}. Third, DSMC is not well suited to the simulation of low speed or low Knudsen number flows due to its intrinsic stochastic nature. For instance, it has been found that the computational cost increases as $\text{Ma}^{-2}$ ($\text{Ma}$ is the Mach number) when the flow speed is approaching zero \citep{Hadjiconstantinou2003}. Therefore, the multiscale feature in non-equilibrium hypersonic flow passing spacecrafts makes DSMC method time-consuming and even intractable in some cases.

Alternatively, kinetic models are proposed to imitate as closely as possible the behaviour of the WCU equation, and multiscale deterministic methods are developed to solve those kinetic models. The Bhatnagar-Gross-Krook (BGK) type kinetic models~\citep{Bhatnagar1954,Holway1966,Shakhov1968}, which replace the Boltzmann collision operator with a single relaxation approximation, has achieved notable success in the modelling of monatomic rarefied gas flows. These kinetic models have been extended to polyatomic rarefied gas by introducing additional internal energy variables in the distribution function \citep{Morse1964, Rykov, andries2000gaussian, Tantos2016IJHMT, Rahimi2016JFM, Wang2017JCP, Bernard2019JSC}, as well as the gas mixture of polyatomic molecules \citep{Pirner2018JSP}.  
Besides, the Fokker–Planck models have been proposed \citep{Gorji2013, Mathiaud2020}, which take advantage of the continuous distribution functions in terms of stochastic velocity processes to speed up the stochastic particle methods. 

However, these models do not reduce to the Boltzmann equation for monatomic gases when the translational-internal energy exchange is absent. Therefore, these models cannot distinguish the influence of different intermolecular potentials. For example, the uncertainties caused by different intermolecular potentials has been demonstrated in calculation of the thermal creep slip on diffuse walls \citep{loyalka1990slip}, the thermal creep and Poiseuille flows \citep{Sharipov2009,Takata2011,wuPoF2015}, the viscous slip of the Couette flow \citep{Su2019JCP2}, and the slip coefficient and mass flow rate fo thermal transpiration \citep{Wang2020PoF}. 
On the other hand, all these kinetic model equations  concern only the transport coefficients, such as the thermal conductivity and bulk viscosity, while their fundamental relaxation processes are not captured, which are found to be important in rarefied molecular gas dynamics. For example, the relaxation rates of heat flux can significantly affect the creep flow driven by molecular velocity-dependent external force~\citep{Li2021JFM}. Thus, it is necessary to tackle the two difficulties when building a gas kinetic model for molecules with rational and vibrational DoF.

Additionally, kinetic model of molecular gas flow with the radiative heat transfer is still far from being well developed. \cite{Groppi1999JMC} proposed a kinetic model with radiation transitions between energy levels by absorption and emission of photons, as an extension of WCU equation with radiative field. However, it is unlikely to be solved practically due to its even much higher complexity than WCU equation. Therefore, the tractable kinetic model equations incorporating radiative heat transfer is highly desirable. 




Hence the present work is dedicated to developing general kinetic models of molecular gas in radiative rarefied flow, which include translational, rotational and discrete vibrational modes with correct rates of relaxation processes. The rest of the paper is organized as follows. In \S\ref{sec:kinetic_model}, the kinetic models with both relaxation time approximation (RTA) and Boltzmann collision operator (BCO) are proposed, and the transport coefficients and their intrinsic relation to relaxation rates are discussed. In \S\ref{sec:validation}, the kinetic models without radiation field are validated by DSMC in typical rarefied gas flows. Then, in \S\ref{sec:radiation}, the influence of radiative heat transfer is examined by solving the kinetic models. Furthermore, the kinetic models are applied to solve hypersonic flow passing cylinder where the radiative heat transfer becomes essential in \S\ref{sec:2D}. Finally, conclusions are presented in \S\ref{sec:conclusion}.

\section{Kinetic model}\label{sec:kinetic_model}

In gas kinetic theory, the distribution function is used to describe the status of dilute gaseous system at the mesoscopic level. The evolution of the distribution function of molecular gas is governed by the WCU equation, which is too complicated to be applied in realistic problems. Therefore, kinetic models are urgently needed to simplify its collision operator. A fundamental requirement in constructing a kinetic model is that all the transport coefficients are consistent with those obtained from the Boltzmann equation for monatomic gas or the WCU equation for molecular gas. Due to the excitation of internal DoF in molecular gas, additional relaxation processes occur between different type of energies, which lead to exclusively  transport coefficients in molecular gas such as the bulk viscosity and internal thermal conductivity. The recovery of these transport coefficients in kinetic model is crucial to accurately describe rarefied gas dynamics. For instances, the modelling of the shock wave requires correct bulk viscosity due to its high compressibility, while the modelling of thermal transpiration requires the recovery of translational thermal conductivity, rather than the total thermal conductivity~\citep{Mason1963JCP,porodnov1978thermal,Loyalka1979Polyatomic,Wu2020JFM,Li2021JFM}.

Well-known kinetic models are the stochastic~\cite{Borgnakke1975} model in DSMC and the deterministic \cite{Rykov} and ellipsoidal-statistical BGK models~\citep{Holway1966,andries2000gaussian}, with the emphasis to recover the transport coefficients, rather than the rates of the essential relaxation process. To be specific, although the total thermal conductivity can be recovered in the ellipsoidal-statistical BGK model, this model cannot give correct translational and internal thermal conductivities respectively; the Rykov model can recover each component of the thermal conductivity, and therefore has flexibility in the simulation of thermal transpiration, but in the rarefied flow driven by the Maxwell demon the flow velocity and heat flux are incorrect~\citep{Li2021JFM,Zeng2022AAS}.  

Meanwhile, transitions between different energy states of a gas molecule lead to absorption and emission of photons, and therefore the interactions between gas and photons of the self-consistent radiation have to be taken into account. The kinetic equations of gas and photon are coupled to each other through gas-photon collision terms, which are related to the radiation intensity and photon absorptivity that depends on gas distribution function. Thus, simplifications of kinetic equations of both gas and photon are required to reduce the complexity.

We start from a generalized kinetic model based on each discrete vibrational energy level, which includes not only the intermolecular collisions but also the interactions between vibrational transition and radiation. By introducing the reduced velocity distribution function and total radiative intensity, the evolution of the system can be governed by four coupled equations, three for distribution function of gas and one for photon intensity. Meanwhile, we modify the Rykov model to recovery the correct thermal relaxation rates and then give a kinetic model based on RTA. Furthermore, by adopting the BCO for elastic intermolecular collisions, the second kinetic model, which can accurately distinguish the intermolecular potentials, is proposed. 

\subsection{Distribution functions and macroscopic quantities}

When the temperature is higher than tens of degrees Kelvin, the translational and rotational DoFs are fully activated. Therefore, it is a common choice to use constant values of DoF for these modes. Since the gap between two subsequent discrete levels is much lower for rotational energy than for vibrational energy, the rotational energy of gas molecules takes continuous values approximately. However, the vibrational DoF is temperature dependent, due to large energy gaps between discrete   vibrational energy levels~\citep{Anderson2019}. Suppose there are $N$ energy levels allowed in vibrational mode, then $N$ distribution functions $f_i(t,\bm{x},\bm{v},I_r)$ are needed to identify the states of molecular gas, where $t$ is the time, $\bm{x}$ is the spatial coordinates, $\bm{v}$ is the translational molecular velocity, $I_r$ is the rotational energy, and $i=1,2,...,N$ indicates the vibrational level with energy $\varepsilon_i$. Macroscopic variables, such as the number density $n$, flow velocity $\bm{u}$, pressure tensor $\bm{P}$, temperatures $T_t,T_r,T_v$, and heat fluxes $\bm{q}_t,\bm{q}_r,\bm{q}_v$, are obtained by taking the moments of distribution function and summation over $N$ vibrational states:
\begin{equation}\label{eq:macroscopic_variables_f}
	\begin{aligned}[b]
		\left(n, n\bm{u},\bm{P}\right)=\sum_{i}^{N}\int_{0}^{\infty}\int_{-\infty}^{\infty}\left(1,\bm{v},m\bm{c}\bm{c}\right){f_i}\mathrm{d}\bm{v}\mathrm{d}I_r, \\
		\left(\frac{3}{2}k_BT_t,\frac{d_r}{2}k_BT_r,\frac{{d_v(T_v)}}{2}k_BT_v\right)=\frac{1}{n}\sum_{i}^{N}\int_{0}^{\infty}\int_{-\infty}^{\infty}{\left(\frac{1}{2}mc^2,I_r,\varepsilon_i\right)f_i}\mathrm{d}\bm{v}\mathrm{d}I_r, \\
		\left(\bm{q}_t,\bm{q}_r,\bm{q}_v\right)=\sum_{i}^{N}\int_{0}^{\infty}\int_{-\infty}^{\infty}\bm{c}\left(\frac{1}{2}mc^2,I_r,\varepsilon_i\right){f_i}\mathrm{d}\bm{v}\mathrm{d}I_r,
	\end{aligned}
\end{equation}
where the subscripts $t,~r,~v$ denote the translational, rotational and vibrational components, respectively; $\bm{c}=\bm{v}-\bm{u}$ is the peculiar velocity, $m$ is the molecular mass, and $k_B$ is the Boltzmann constant; $d_r$ and $d_v(T_v)$ are the rotational and vibrational number of DoF, respectively. For a simple harmonic oscillator, 
\begin{equation}\label{eq:harmonic_oscillator_dv}
	\begin{aligned}[b]
		d_v(T_v)=\frac{2T_{\text{ref}}/T_v}{\exp({T_{\text{ref}}/T_v})-1},
	\end{aligned}
\end{equation}
where $T_{\text{ref}}$ is the characteristic temperature of the active vibrational mode.
 
We also define the temperature $T_{tr}$ to be the equilibrium temperature between the translational and rotational modes, $T_{tv}$ the equilibrium temperature between the translational and vibrational modes, and $T$ the equilibrium temperature over all DoF:
\begin{equation}\label{eq:Ttr_Ttv}
	\begin{aligned}[b]
		T_{tr}=\frac{3T_t+d_rT_r}{3+d_r}, \quad T_{tv}=\frac{3T_t+{{d_v(T_v)}}T_v}{3+{{d_v(T_{tv})}}}, \quad T=\frac{3T_t+d_rT_r+{{d_v(T_v)}}T_v}{3+d_r+{{d_v(T)}}},
	\end{aligned}
\end{equation}
and the corresponding pressures are $(p_t, p_r, p_v, p, p_{tr}, p_{tv}) = nk_B(T_t, T_r, T_v, T, T_{tr}, T_{tv})$.

\subsection{Generalized kinetic model with vibrational radiation transition}

The mechanisms changing the distribution function $f_i(t,\bm{x},\bm{v},I_r)$ include the gas-gas interactions $J_{gas}$, the gas-photon interactions $J_{photon}$ exchanging energy between vibrational mode and radiation field, and the streaming $\mathcal{D}$. Ignoring the momentum exchange between gas and photon, (which is an acceptable approximation in most of the practical situations, since the momentum exchange rate is proportional to the ratio of radiative heat flux to the speed of light), the evolution of distribution function under external body acceleration $\bm{a}$ is governed by
\begin{equation}\label{eq:gas_equation}
	\begin{aligned}
		\underbrace{\frac{\partial{f_{i}}}{\partial{t}}+\bm{v} \cdot \frac{\partial{f_{i}}}{\partial{\bm{x}}}+ \frac{\partial{(\bm{a}f_{i})}}{\partial{\bm{v}}}}_{\mathcal{D}f_i}=J_{gas,i} + J_{photon,i}.
	\end{aligned}
\end{equation}

The kinetic model of binary gas-gas collisions has been well established by the WCU equation. When the rotational mode is treated by classical mechanics, it can be simplified as:
\begin{equation}\label{eq:J_gas}
	\begin{aligned}
		J_{gas,i} = \sum_{i'j'}\sum_{j}\int_{-\infty}^{\infty}\int_{4\pi}{\left(\frac{g_{i}g_{j}}{g_{i'}g_{j'}}f_{i'}f_{j'}-f_{i}f_{j}\right)|\bm{v}-\bm{v}_*|\sigma_{ij}^{i'j'}\mathrm{d}\Omega\mathrm{d}\bm{v}_*},
	\end{aligned}
\end{equation}
where $\bm{v}$ and $\bm{v}_*$ are the velocities of the two molecules with vibrational states $i$ and $j$, respectively, before collision, superscript $'$ indicates the state after collision, $g_i$ is the degeneracy of the vibrational states $i$, $\sigma_{ij}^{i'j'}$ is the scattering cross-section, and $\Omega$ is the solid angle. Since the total energy is conserved during collision, such a transition occurs only when 
\begin{equation}
v'^{2}_r=|\bm{v}-\bm{v}_*|^2+\frac{4}{m}(e_{i}+e_{j}-e_{i'}-e_{j'})>0.
\end{equation}
When the collision is admissible, the molecular velocity after collision is
\begin{equation}
\begin{aligned}[b]
\bm{v}'=\frac{\bm{v}+\bm{v}_\ast}{2}+\frac{v'_r}{2}\bm{\Omega},
\quad
\bm{v}'_\ast=\frac{\bm{v}+\bm{v}_\ast}{2}-\frac{v'_r}{2}\bm{\Omega}.
\end{aligned}
\end{equation}
It is clear that, the computational cost will be $N^4$ times higher than that of the Boltzmann collision operator, posing urgently needs to develop simplified gas kinetic models. Also, it should be noted that, the collision is called elastic when $v_r'=|\bm{v}-\bm{v}_*|$ since the kinetic energy is conserved. Otherwise, it is inelastic.

Molecular bound-bound radiation occurs as a result of radiative transitions between the quantized energy levels of gas molecules. These radiative events are determined by the molecular structure itself, but not associated directly with the intermolecular collisions. For gas-photon interactions, the following three types of vibrational energy transition that is related to the radiation are considered in the present work:
\begin{equation}\label{eq:vib_photon}
	\begin{aligned}[b]
		\text{spontaneous emission:}&~ M_j \rightarrow M_i+h\nu_{ij}, \\
		\text{stimulated emission:}&~ M_j+h\nu_{ij} \rightarrow M_i+2h\nu_{ij}, \\
		\text{absorption:}&~ M_i+h\nu_{ij} \rightarrow M_j, \\
	\end{aligned}
\end{equation}
where $\varepsilon_j>\varepsilon_i$ are the corresponding energy of the vibrational energy levels $i$ and $j$ of the molecule $M$. Thus, the photon absorbed or emitted during these transitions has a frequency $\nu_{ij}=(\varepsilon_j-\varepsilon_i)/h$, where $h$ is the Planck number. Since the photon frequencies have $N(N-1)/2$ discretized values, the radiation field is described by the same number of radiation intensity functions $I_{\nu_{ij}}^{R}(t,\bm{x},\bm{\Omega})$, which measure the energy fluxes per unit solid angle of photon propagating along the direction $\bm{\Omega}$ with the frequency $\nu_{ij}$.

The transition changes the population number of the energy levels $i$ and $j$. The rates of change of distribution function $f_i$ during processes \eqref{eq:vib_photon} can be calculated by the Einstein coefficients, $A_{ji},B_{ji},B_{ij}$ for the spontaneous emission, stimulated emission and absorption, respectively:
\begin{equation}\label{eq:dfi}
	\begin{aligned}[b]
		&\left(\frac{\mathrm{d}{f_i}}{\mathrm{d}{t}}\right)^{sp}_j = \int_{4\pi}{A_{ji}f_j}\mathrm{d}\Omega, \\
		&\left(\frac{\mathrm{d}{f_i}}{\mathrm{d}{t}}\right)^{st}_j = \int_{4\pi}{B_{ji}I_{\nu_{ij}}^{R}\mathrm{d}\Omega}f_j, \\
		&\left(\frac{\mathrm{d}{f_i}}{\mathrm{d}{t}}\right)^{ab}_j = -\int_{4\pi}{B_{ij}I_{\nu_{ij}}^{R}\mathrm{d}\Omega}f_i. 
	\end{aligned}
\end{equation}
Therefore, when the Doppler and aberration effect are ignored, the rates of change of the distribution function $f_i$ due to all possible gas-photon interactions yield \citep{Groppi1999JMC},
\begin{equation}\label{eq:J_photon}
	\begin{aligned}[b]
		J_{photon,i} &= \sum_{j>i}\int_{4\pi}{\left[A_{ji}f_j+\left(B_{ji}f_j-B_{ij}f_i\right)I_{\nu_{ij}}^{R}\right]\mathrm{d}\Omega} \\
		&-\sum_{j<i}\int_{4\pi}{\left[A_{ij}f_i+\left(B_{ij}f_i-B_{ji}f_j\right)I_{\nu_{ji}}^{R}\right]\mathrm{d}\Omega}.
	\end{aligned}
\end{equation}

Next, we consider the evolution of radiation intensity due to the interaction with gas molecules $J_{\nu_{ij}}^{R}$. Ignoring the photon scattering (which is an accepted assumption when no particles or droplets are considered in the gas flow, and the radiation occupies a spectral range such that wave lengths are much larger compared with molecular size), the evolution of intensity  $I_{\nu_{ij}}^{R}$ is:
\begin{equation}\label{}
	\begin{aligned}
		\underbrace{\frac{1}{c_l}\frac{\partial{I_{\nu_{ij}}^{R}}}{\partial{t}}+\bm{n} \cdot \frac{\partial{I_{\nu_{ij}}^{R}}}{\partial{\bm{x}}}}_{\mathcal{D}I_{\nu_{ij}}^{R}}=J_{\nu_{ij}}^{R},
	\end{aligned}
\end{equation}
where $c_l$ is the speed of light and $\bm{n}$ is unit vector of photon propagation direction. When the radiation passes through the matter over distance $c_l\mathrm{d}{t}$, it is attenuated by a constant fraction $k_{\nu_{ij}}$, and it has a gain part $j_{\nu_{ij}}$ that does not depend on the intensity \citep{Casto2004Radiation}: the rate of change of intensity $I_{\nu_{ij}}^{R}$ due to the gas-photon interaction can be written as: 
\begin{equation}\label{eq:dI}
	\begin{aligned}
		\frac{1}{c_l}\left(\frac{\mathrm{d}{I_{\nu_{ij}}^{R}}}{\mathrm{d}{t}}\right)^{loss} &= -k_{\nu_{ij}}I_{\nu_{ij}}^{R}, \\
		\frac{1}{c_l}\left(\frac{\mathrm{d}{I_{\nu_{ij}}^{R}}}{\mathrm{d}{t}}\right)^{gain} &= j_{\nu_{ij}}.
	\end{aligned}
\end{equation}
It shows from processes \eqref{eq:vib_photon} that the gain part comes from the spontaneous emission alone, and the loss part is the difference between the absorption and stimulated emission, which indicates the decay of radiation intensity caused by the gas-photon interactions during propagation. Based on \eqref{eq:dfi}, the frequency-dependent absorptivity $k_{\nu_{ij}}$ and emissivity $j_{\nu_{ij}}$ are determined by the Einstein coefficients as:
\begin{equation}\label{eq:absorptivity_Einstein_coefficient}
	\begin{aligned}
		k_{\nu_{ij}}=h\nu_{ij}\left(B_{ij}\int{f_i\mathrm{d}\bm{v}}-B_{ji}\int{f_j\mathrm{d}\bm{v}}\right), \quad 
		j_{\nu_{ij}}=h\nu_{ij}A_{ji}\int{f_j\mathrm{d}\bm{v}}.
	\end{aligned}
\end{equation}
Therefore, the rate of change of the photon intensity $I_{\nu_{ij}}^{R}$ due to gas-photon interactions is
\begin{equation}\label{}
	\begin{aligned}
		J_{\nu_{ij}}^{R} = j_{\nu_{ij}}-k_{\nu_{ij}}I_{\nu_{ij}}^{R}.
	\end{aligned}
\end{equation}

According to the Kirchhoff's law, the emissivity is related to the absorptivity as $j_{\nu_{ij}}=B^R_{\nu_{ij}}(T_v)k_{\nu_{ij}}$ with the following Planck function:
\begin{equation}\label{}
	\begin{aligned}
		B^R_{\nu_{ij}}(T)=\frac{2h\nu_{ij}^3}{c_l^2}\frac{1}{\exp\left(h\nu_{ij}/k_BT\right)-1}.
	\end{aligned}
\end{equation}
Then, the evolution of intensity due to the interaction with gas molecules is:
\begin{equation}\label{eq:photon_equation}
	\begin{aligned}
		\frac{1}{c_l}\frac{\partial{I_{\nu_{ij}}^{R}}}{\partial{t}}+\bm{n} \cdot \frac{\partial{I_{\nu_{ij}}^{R}}}{\partial{\bm{x}}}=k_{\nu_{ij}}\left(B^R_{\nu_{ij}}(T_v)-I_{\nu_{ij}}^{R}\right).
	\end{aligned}
\end{equation}

The macroscopic quantities, radiative temperature $T_R$ and heat flux $\bm{q}_R$, are defined as:
\begin{equation}\label{eq:photon_macroscopic_variables}
\left(4\sigma_RT_R^4,\bm{q}_R\right)=\sum_{\nu_{ij}}\int_{4\pi}{I_{\nu_{ij}}^{R}(1,\bm{n})}\mathrm{d}\Omega,
\end{equation}
where $\sigma_R=2\pi^5k_B^4/(15h^3c_l^2)$ is the Stefan-Boltzmann constant. Since the energy exchange between gas and photon results from the radiation transition of vibrational energy, the radiative temperature $T_R$ approaches vibrational temperature $T_v$ at equilibrium state, and hence the equilibrium temperature in Planck function in the photon kinetic equation \eqref{eq:photon_equation} is $T_v$.

Therefore, the governing equations of the rarefied molecular gas flow with radiation  consist of the $N$ equations \eqref{eq:gas_equation} of gas distribution functions with collision terms \eqref{eq:J_gas} and \eqref{eq:J_photon}, and $N(N-1)/2$ equations \eqref{eq:photon_equation} of radiative intensities. Then, simplifications of collision terms will be introduced to develop tractable model equations.

\subsection{Kinetic model with relaxation time approximation}

\subsubsection{Modified Rykov model for gas-gas collisions}

Here we build a kinetic model to simplify the WCU collision operator~\eqref{eq:J_gas} for gas-gas interactions based on the Rykov model, not only due to its more freedom to reflect the relaxation process of heat fluxes, but also due to its much reduced computational complexity. In this model, the elastic and inelastic collisions are considered separately with different relaxation times, which can be adjusted to give a correct bulk viscosity. And the reference distribution functions to which the distribution function relaxes contain the heat fluxes, so that the thermal conductivity can be recovered. Although the Rykov model is initially proposed for diatomic gas without vibrational modes, it has been extended to polyatomic gas~\citep{LeiJFM2015} and gases with vibrational modes~\citep{Titarev2018}. 

However, to recover the correct thermal relaxation rates other than thermal conductivities, the heat fluxes in the reference distribution functions in original Rykov model have to be adjusted. Thus, the modified Rykov model for gas-gas collisions with discretized vibrational states becomes 
\begin{equation}\label{eq:J_gas_Rykov}
	\begin{aligned}
		J_{gas,i} = \frac{g_{t,i}-f_i}{\tau} + \frac{g_{r,i}-g_{t,i}}{Z_r\tau} + \frac{g_{v,i}-g_{t,i}}{Z_{v}\tau} ,
	\end{aligned}
\end{equation} 
where $Z_r$ and $Z_v$ are the rotational and vibrational collision number, respectively. The reference distribution functions $g_{t,i}, g_{r,i}, g_{v,i}$ are expanded about the equilibrium distributions $E_t(T)\cdot E_r(T)\cdot E_v(T)$ in a series of orthogonal polynomials in variables peculiar velocity $\bm{c}$, rotational energy $I_r$, vibrational energy $\varepsilon_i$ and corresponding moments $\bm{q_t}, \bm{q_r}, \bm{q_v}$:
\begin{equation}\label{eq:gt_gr_gv}
	\begin{aligned}
		g_{t,i} =~&nE_t(T_t) \cdot E_r(T_r) \cdot E_{v,i}(T_v) \cdot \left[{1 + \frac{2m\bm{q}_t\cdot{\bm{c}}}{15{k_B}{T_t}{p_t}} \left(\frac{mc^2}{2k_BT_t}-\frac{5}{2}\right)} \right. \\ 
		&\left. {+\frac{2m\bm{q}_r\cdot{\bm{c}}}{d_rk_BT_tp_r}\left(\frac{I_r}{k_BT_r}-\frac{d_r}{2}\right) + \frac{2m\bm{q}_v\cdot{\bm{c}}}{{{d_v(T_v)}}k_BT_tp_v}\left(\frac{\varepsilon_i}{k_BT_v}-\frac{{d_v(T_v)}}{2}\right)}\right], \\
		g_{r,i} =~&nE_t(T_{tr}) \cdot E_r(T_{tr}) \cdot E_{v,i}(T_v) \cdot \left[{1 + \frac{2m\bm{q}_0\cdot{\bm{c}}}{15{k_B}{T_{tr}}{p_{tr}}} \left(\frac{mc^2}{2k_BT_{tr}}-\frac{5}{2}\right)} \right. \\ 
		&\left. {+\frac{2m\bm{q}_1\cdot{\bm{c}}}{d_rk_BT_{tr}p_{tr}}\left(\frac{I_r}{k_BT_{tr}}-\frac{d_r}{2}\right) + \frac{2m\bm{q}_2\cdot{\bm{c}}}{{{d_v(T_v)}}k_BT_{tr}p_v}\left(\frac{\varepsilon_i}{k_BT_v}-\frac{{d_v(T_v)}}{2}\right)}\right], \\
		g_{v,i} =~&nE_t(T_{tv}) \cdot E_r(T_{r}) \cdot E_{v,i}(T_{tv}) \cdot \left[{1 + \frac{2m\bm{q}_0\cdot{\bm{c}}}{15{k_B}{T_{tv}}{p_{tv}}} \left(\frac{mc^2}{2k_BT_{tv}}-\frac{5}{2}\right)} \right. \\ 
		&\left. {+\frac{2m\bm{q}_1\cdot{\bm{c}}}{d_rk_BT_{tv}p_{r}}\left(\frac{I_r}{k_BT_{r}}-\frac{d_r}{2}\right) + \frac{2m\bm{q}_2\cdot{\bm{c}}}{{{d_v(T_{tv})}}k_BT_{tv}p_{tv}}\left(\frac{\varepsilon_i}{k_BT_{tv}}-\frac{{d_v(T_{tv})}}{2}\right)}\right], 
	\end{aligned}
\end{equation}
with the equilibrium distribution functions,
\begin{equation}\label{eq:Et_Er_Ev}
	\begin{aligned}
		E_t(T)&={\left(\frac{m}{2\pi k_BT}\right)}^{3/2}\exp{\left(-\frac{mc^2}{2k_BT}\right)}, \\
		E_r(T)&=\frac{I^{d_r/2-1}_{r}}{\Gamma(d_r/2)(k_BT)^{d_r/2}}\exp{\left(-\frac{I_r}{k_BT}\right)}, \\
		E_{v,i}(T)&=\frac{g_i}{\sum_i{g_i\exp{\left(-\frac{\varepsilon_i}{k_BT}\right)}}}\exp{\left(-\frac{\varepsilon_i}{k_BT}\right)},
	\end{aligned}
\end{equation}
where $\Gamma$ is the gamma function, ${\bm{q}_{0}}$, ${\bm{q}_{1}}$, and ${\bm{q}_{2}}$ are linear combinations of translational, rotational and vibrational heat fluxes, which will be designed to recover the exact thermal relaxation rates. 

\subsubsection{Reduced distribution functions and radiative intensity}

In practical numerical simulations, it is almost impossible to solve $N(N+1)/2$ equations even with RTA, especially when the number of vibrational energy levels $N$ considered is large. However, the complexity arising from the discrete vibrational energy can be eliminated with the reduced distribution technique \citep{Chu1965pof, Mathiaud2020}. Therefore, we eliminate the rotational energy variable ${I_r}$ and vibrational energy level index $i$ by introducing the following reduced velocity distribution functions ${f_0,~f_1,~f_2}$:
\begin{equation}\label{eq:f0_f1_f2}
	\begin{aligned}
		\left(f_0,f_1,f_2\right)=&\sum_{i}^{N}\int_{0}^{\infty}\left(1,I_r,\varepsilon_i\right)f_i\left(t,\bm{x},\bm{v},I_r\right)\mathrm{d}{I_r}.
	\end{aligned}
\end{equation}
Similarly, the total radiative intensity $I^{R}$ will be solved instead of the frequency dependent ones by using 
\begin{equation}\label{eq:IR}
	\begin{aligned}
		I^{R}=&\sum_{\nu_{ij}}I^{R}_{\nu_{ij}}, \\ B^R(T)=&\sum_{\nu_{ij}}B^R_{\nu_{ij}}(T)=\frac{1}{\pi}\sigma_RT^4,
	\end{aligned}
\end{equation}

Giving the fact that the radiation transitions are not correlated with the translational motion of the gas molecules, then, the governing equations~\eqref{eq:gas_equation} and \eqref{eq:photon_equation} with modified Rykov model \eqref{eq:J_gas_Rykov} can be transferred to four coupled equations:
\begin{equation}\label{eq:governing_equation_Rykov}
	\begin{aligned}[b]
	&\mathcal{D}f_0 =~ \frac{g_{0t}-f_0}{\tau} + \frac{g_{0r}-g_{0t}}{Z_r\tau} + \frac{g_{0v}-g_{0t}}{Z_v\tau}, \\
	&	\mathcal{D}f_1 =~ \frac{g_{1t}-f_1}{\tau} + \frac{g_{1r}-g_{1t}}{Z_r\tau} + \frac{g_{1v}-g_{1t}}{Z_v\tau}, \\
	&	\mathcal{D}f_2 =~ \frac{g_{2t}-f_2}{\tau} + \frac{g_{2r}-g_{2t}}{Z_r\tau} + \frac{g_{2v}-g_{2t}}{Z_v\tau} 
		-\frac{f_0}{n}\int_{4\pi}{\left(k^{e}B^R(T_v)-k^{ne}I^{R}\right)\mathrm{d}\Omega},  \\
	&	\frac{1}{c_l}\frac{\partial{I^{R}}}{\partial{t}}+\bm{n} \cdot \frac{\partial{I^{R}}}{\partial{\bm{x}}} =~ k^{e}B^R(T_v)-k^{ne}I^{R},
	\end{aligned}
\end{equation}
with the reduced reference velocity distribution functions
\begin{align}\label{eq:glt_glr_glv}
		g_{0t}&=nE_t(T_t)\left[1+\frac{2m\bm{q}_t\cdot{\bm{c}}}{15{k_B}{T_t}{p_t}} \left(\frac{mc^2}{2k_BT_t}-\frac{5}{2}\right)\right], \notag \\
		g_{0r}&=nE_t(T_{tr})\left[1+\frac{2m\bm{q}_0\cdot{\bm{c}}}{15{k_B}{T_{tr}}{p_{tr}}} \left(\frac{mc^2}{2k_BT_{tr}}-\frac{5}{2}\right)\right], \notag \\
		g_{0v}&=nE_t(T_{tv})\left[1+\frac{2m\bm{q}_0\cdot{\bm{c}}}{15{k_B}{T_{tv}}{p_{tv}}} \left(\frac{mc^2}{2k_BT_{tv}}-\frac{5}{2}\right)\right], \notag \\
		g_{1t}&=\frac{d_r}{2}k_BT_rg_{0t}+\frac{m\bm{q}_r\cdot{\bm{c}}}{k_BT_t}E_t(T_t), \notag \\
		g_{1r}&=\frac{d_r}{2}k_BT_{tr}g_{0r}+\frac{m\bm{q}_1\cdot{\bm{c}}}{k_BT_{tr}}E_t(T_{tr}), \notag \\
		g_{1v}&=\frac{d_r}{2}k_BT_{r}g_{0v}+\frac{m\bm{q}_1\cdot{\bm{c}}}{k_BT_{tv}}E_t(T_{tv}), \notag \\
		g_{2t}&=\frac{{d_v(T_v)}}{2}k_BT_vg_{0t}+\frac{m\bm{q}_v\cdot{\bm{c}}}{k_BT_t}E_t(T_t), \notag \\
		g_{2r}&=\frac{{d_v(T_v)}}{2}k_BT_{v}g_{0r}+\frac{m\bm{q}_2\cdot{\bm{c}}}{k_BT_{tr}}E_t(T_{tr}), \notag \\
		g_{2v}&=\frac{{d_v(T_{tv})}}{2}k_BT_{tv}g_{0v}+\frac{m\bm{q}_2\cdot{\bm{c}}}{k_BT_{tv}}E_t(T_{tv}),
\end{align}
and the effective absorptivities
\begin{equation}\label{eq:ke_kne}
	\begin{aligned}[b]
		k^e=\frac{\sum_{\nu_{ij}}k_{\nu_{ij}}B^R_{\nu_{ij}}(T_v)}{\frac{1}{\pi}\sigma_RT_v^4}, \\
		k^{ne}=\frac{\sum_{\nu_{ij}}k_{\nu_{ij}}B^R_{\nu_{ij}}(T_R)}{\frac{1}{\pi}\sigma_RT_R^4}.
	\end{aligned}
\end{equation}
From equation \eqref{eq:absorptivity_Einstein_coefficient}, it can be seen that the absorptivity $k_{\nu_{ij}}$ is proportional to the population of molecules at vibrational energy levels $i$ and $j$, which is determined by the number density as well as the vibrational DoF of the gas. When the gray model of photon is adopted, the absorptivity $k_{\nu_{ij}}=k_{\text{gray}}$ is  frequency-independent, thus both $k^e$ and $k^{ne}$ reduce to $k_{\text{gray}}$.

The macroscopic quantities of gas flow in \eqref{eq:macroscopic_variables_f} and radiative field in \eqref{eq:photon_macroscopic_variables} can be calculated based on the reduced velocity distribution functions and intensity:
\begin{equation}\label{eq:macroscopic_variables_f0_f1_f2}
	\begin{aligned}[b]
		\left(n, n\bm{u},p_{ij}\right)=\int_{-\infty}^{\infty}\left(1,\bm{v},mc_ic_j\right){f_0}\mathrm{d}\bm{v}, \\
		\left(\frac{3}{2}k_BT_t,\frac{d_r}{2}k_BT_r,\frac{{d_v(T_v)}}{2}k_BT_v\right)=\frac{1}{n}\int_{-\infty}^{\infty}{\left(\frac{1}{2}mc^2f_0,f_1,f_2\right)}\mathrm{d}\bm{v}, \\
		\left(\bm{q}_t,\bm{q}_r,\bm{q}_v\right)=\int_{-\infty}^{\infty}{\bm{c}\left(\frac{1}{2}mc^2f_0,f_1,f_2\right)}\mathrm{d}\bm{v}, \\
		\left(4\sigma_RT_R^4,\bm{q}_R\right)=\int_{4\pi}{I^{R}(1,\bm{n})}\mathrm{d}\Omega.
	\end{aligned}
\end{equation}

\subsection{Kinetic model with Boltzmann collision operator}\label{FurtherModel}

Obviously, all molecules relax with the same speed in the RTA~\eqref{eq:J_gas_Rykov}, which is not very physical, since in general molecules with larger peculiar velocity has larger collision probability and hence smaller relaxation time; in fact, when \eqref{eq:J_gas_Rykov} is used, the temperature of normal shock wave will be overpredicted in the upstream~\citep{LeiJFM2015}.  To circumvent this problem, by observing that the elastic collision term in the first equation of \eqref{eq:governing_equation_Rykov} is just the Shakhov-type approximation of the BCO for monatomic gas~\citep{Shakhov1968,Shakhov_S}, we replace the elastic collision term $(g_{0t}-f_0)/\tau$ back with the BCO in monatomic gas:
\begin{equation}\label{eq:Boltzmann_collision_operator}
	Q_B(f_0) = \int_{-\infty}^{\infty}\int_{4\pi}{B(\cos{\theta},{\left| \bm{v}-\bm{v}_* \right|})[f_0(\bm{v}'_*)f_0(\bm{v}')-f_0(\bm{v}_*)f_0(\bm{v})]\mathrm{d}{\Omega}}\mathrm{d}{\bm{v}_*},
\end{equation} 
so that the relaxation time depends on the molecular velocity through the collision kernel $B(\cos{\theta},{\left| \bm{v}-\bm{v}_* \right|})$ that is determined by the intermolecular potential. Note that in~\eqref{eq:Boltzmann_collision_operator}, $\theta$ is the deflection angle of collision,  $\bm{v}$ and $\bm{v}_*$ are the velocities of the two molecules before collision, while $\bm{v}'$ and $\bm{v}'_*$ are the velocities of the two molecules after collision.  When the inverse power-law potential is considered, the collision kernel is modelled as~\citep{Lei2013,lei_Jfm}:
\begin{equation}
	\begin{aligned}[b]
		B&=\frac{5\sqrt{\pi{m}k_BT_0}(4k_BT_0/m)^{(2\omega-1)/2}}{64\pi\mu(T_0)\Gamma^2(9/4-\omega/2)}\sin^{(1-2\omega)/2}\left(\frac{\theta}{2}\right)
		\cos^{(1-2\omega)/2}\left(\frac{\theta}{2}\right)
		\left| \bm{v}-\bm{v}_* \right|^{2(1-\omega)},
	\end{aligned}
\end{equation}
where $\omega$ is the viscosity index, that is,
\begin{equation}\label{eq:viscosity_temperature}
	\mu(T)=\mu(T_0)\left(\frac{T}{T_0}\right)^\omega.
\end{equation}

Meanwhile, $g_{1t}$ and $g_{2t}$ are modified correspondingly~\citep{LeiJFM2015}, resulting in the following kinetic model for molecular gas:
\begin{equation}\label{eq:governing_equation_Boltzmann}
	\begin{aligned}[b]
 &	\mathcal{D}f_0 =~ Q_B(f_0) + \frac{g_{0r}-g_{0t}}{Z_r\tau} + \frac{g_{0v}-g_{0t}}{Z_v\tau}, \\
	&	\mathcal{D}f_1 =~ \frac{g_{1t}'-f_1}{\tau} + \frac{g_{1r}-g_{1t}}{Z_r\tau} + \frac{g_{1v}-g_{1t}}{Z_v\tau}, \\
	&		\mathcal{D}f_2 =~ \frac{g_{2t}'-f_2}{\tau} + \frac{g_{2r}-g_{2t}}{Z_r\tau} + \frac{g_{2v}-g_{2t}}{Z_v\tau} 
		-\frac{f_0}{n}\int_{4\pi}{\left(k^{e}B^R(T_v)-k^{ne}I^{R}\right)\mathrm{d}\Omega},  \\
	&	\frac{1}{c_l}\frac{\partial{I^{R}}}{\partial{t}}+\bm{n} \cdot \frac{\partial{I^{R}}}{\partial{\bm{x}}} =~ k^{e}B^R(T_v)-k^{ne}I^{R},
	\end{aligned}
\end{equation}
with
\begin{equation}\label{eq:glt_v}
	\begin{aligned}[b]
		g_{1t}'&=\frac{d_r}{2}k_BT_r[\tau Q(f_0)+f_0]+\frac{m\bm{q}_r\cdot{\bm{c}}}{k_BT_t}E_t(T_t), \\
		g_{2t}'&=\frac{{d_v(T_v)}}{2}k_BT_v[\tau Q(f_0)+f_0]+\frac{m\bm{q}_v\cdot{\bm{c}}}{k_BT_t}E_t(T_t). 
	\end{aligned}
\end{equation}

The kinetic model with the BCO is able to distinguish the role of intermolecular potentials~\citep{Sharipov2009,Takata2011,lei_Jfm,wuPoF2015}, while the models based on the RTA do not have this capability, but keep the advantage of computational efficiency. To be specific, the collision terms of RTA can be solved by discrete velocity method with a computational cost of $O(N_v^3)$, and the BCO can be solved by the fast spectral method~\citep{Lei2013} with a cost of $O(M^2N_v^3\ln{N_v})$, where $N_v$ and $M^2$ are the numbers of grid points in one velocity direction and the solid angle, respectively; both the computational costs are much smaller than solving the collision operator~\eqref{eq:J_gas} in the WCU equation.

\subsection{Determination of parameters from relaxation properties}

So far, both the kinetic model equations with the RTA \eqref{eq:governing_equation_Rykov} and the BCO \eqref{eq:governing_equation_Boltzmann} have been established, which contain the free parameters $\bm{q}_0$, $\bm{q}_1$, $\bm{q}_2$, $Z_r$, $Z_v$, and $\tau$. In addition to the shear viscosity and translational heat conductivity in monatomic gas, the molecular gas possesses the bulk viscosity and internal thermal conductivities. The essence of these new transport coefficients are the relaxation of internal temperature and heat fluxes. Therefore, the free parameters will be determined by the recovery of relaxation rates of shear stress, energy, and heat fluxes, which corresponding to the recover of shear viscosity, bulk viscosity, and translational/internal thermal conductivities, respectively.

Since the Shakhov model and the Boltzmann equation for monatomic gas have the same shear viscosity and translational thermal conductivity, it can be shown that the second model~\eqref{eq:governing_equation_Boltzmann} has the same transport coefficients with the first model~\eqref{eq:governing_equation_Rykov}. Therefore, for the sake of simplicity, we use the first model with RTA to discuss the determination of free parameters.

\subsubsection{Relaxation of shear stress}


Consider a spatial-homogeneous system without external acceleration. Multiplying the first equation in \eqref{eq:governing_equation_Rykov} by $mc^2$ and $mc_ic_j$, and integrating them with respect to $\bm{v}$, yields,
\begin{equation}\label{eq:dpt}
	\begin{aligned}[b]
		\frac{\partial p_t}{\partial t} = &~\frac{p_{tr}-p_t}{Z_r\tau} + \frac{p_{tv}-p_t}{Z_v\tau}, \\
		\frac{\partial p_{ij}}{\partial t} = &~\frac{p_{t}\delta_{ij}-p_{ij}}{\tau} + \frac{p_{tr}\delta_{ij}-p_t\delta_{ij}}{Z_r\tau} + \frac{p_{tv}\delta_{ij}-p_t\delta_{ij}}{Z_v\tau}.
	\end{aligned}
\end{equation}
These two equations lead to the relaxation equation for the components of the non-equilibrium stress tensor,
\begin{equation}\label{eq:dpij}
	\frac{\partial p_{ij}}{\partial t} = -\frac{1}{\tau}{p_{ij}},
\end{equation}
which indicates the relaxation time of shear stress is the mean collision time due to molecular translational motion. Therefore, the shear viscosity can be recovered by adjusting the relaxation time $\tau$ as (see Appendix \ref{app:A}):
\begin{equation}
\mu=p_t\tau.
\end{equation}
The dependence of shear viscosity on mean molecular collision time in molecular gas is the same as that in monatomic gas. 

\subsubsection{Relaxation of energy}

During the contraction or expansion of gas, the work done by pressure is converted to the translational energy immediately. However, in molecular gas, the molecules exhibit internal relaxation that exchanges the translational and internal energies in a finite time, which gives rise to the resistance that opposes the volume change. This is the origin of bulk viscosity.

The energy relaxation can be obtained by multiplying the first equation in \eqref{eq:governing_equation_Rykov} by $\frac{1}{2}mc^2$ and integrating the first three equations with respect to $\bm{v}$,
\begin{equation}\label{eq:dTt_dTr_dTv}
	\begin{aligned}[b]
		\frac{\partial T_t}{\partial t} =& ~\frac{T_{tr}-T_{t}}{Z_r\tau} + \frac{T_{tv}-T_{t}}{Z_v\tau}, \\
		\frac{\partial T_r}{\partial t} =& ~\frac{T_{tr}-T_{r}}{Z_r\tau}, \\
		\frac{\partial ({{d_v(T_v)}}T_v)}{\partial t} =& ~\frac{{{d_v(T_{tv})}}T_{tv}-{{d_v(T_v)}}T_{v}}{Z_v\tau}-\frac{2}{nk_B}\int_{4\pi}{\left(k^{e}B^R(T_v)-k^{ne}I^{R}\right)\mathrm{d}\Omega}. 
	\end{aligned}
\end{equation}
When the radiation is absent, equations \eqref{eq:dTt_dTr_dTv} reduce to the Jeans-Landau-Teller equations for the rotational and vibrational relaxation at macroscopic level,
\begin{equation}\label{eq:Jeans_Landau_Teller}
	\frac{\mathrm{d}{T_r}}{\mathrm{d}{t}} = \frac{T_t-T_r}{\tau_r}, \quad \frac{\mathrm{d}{T_v}}{\mathrm{d}t} = \frac{T_t-T_v}{\tau_v}.
\end{equation}
Define the rotational and vibrational collision numbers relate to the corresponding relaxation time $\tau_r$ and $\tau_v$ as:
\begin{equation}\label{eq:Zr_Zv}
	\begin{aligned}[b]
		Z_r = \frac{3\tau_r}{(3+d_r)\tau}, \quad Z_v = \frac{3\tau_v}{(3+d_v)\tau},
	\end{aligned}
\end{equation}
the bulk viscosity can be derived based on the Chapman-Enskog expansion in the Appendix \ref{app:A}: 
\begin{equation}
	\begin{aligned}[b]
		\mu_b(T_t)&=2p_t\tau\frac{(3+d_r)d_rZ_r+(3+d_v)d_vZ_v}{3\left(3+d_r+d_v\right)^2}.
	\end{aligned}
\end{equation}

It is shown that the ratio $\mu_b/\mu$ depends only on the numbers of internal DoF and the corresponding collision numbers. Larger $Z_r$ or $Z_v$ makes the energy exchange between translational and internal motions more difficult, thus lead to higher bulk viscosity. Meanwhile, since the radiative energy does not exchange with translational energy of gas directly, the presence of radiation field will not affect the bulk viscosity.

On the other hand, during the radiation transitions of the vibrational mode, the energy exchange between gas and photon alters the vibrational energy immediately, while the corresponding changes in translational and rotational modes is delayed due to the finite time of internal energy relaxation. Therefore, a new type of bulk viscosity arises, denoted as $\mu_b^{R}$, which gives the resistance of the translational-internal energy changes due to the vibrational-radiation transitions. Consider the energy conservation in the radiative gas,
\begin{equation}\label{eq:energy_change_radiation}
	\begin{aligned}[b]
		\frac{nk_B}{2}\left(3\frac{\partial T_t}{\partial t}+d_r\frac{\partial T_r}{\partial t}+\frac{\partial ({{d_v(T_v)}}T_v)}{\partial t}\right) = -\int_{4\pi}{\left(k^{e}B^R(T_v)-k^{ne}I^{R}\right)\mathrm{d}\Omega}. 
	\end{aligned}
\end{equation}
When the deviation between equilibrium temperature $T$ and $T_t,~T_r,~T_v$ are small, the higher order terms of $T-T_t$ can be ignored, thus we have
\begin{equation}\label{eq:pressure_change_radiation}
	\begin{aligned}[b]
		p_t-p=2p_v\frac{3\tau_v+d_r(\tau_v-\tau_r)}{\left(3+d_r+d_v\right)^2}\frac{1}{T_v}\int_{4\pi}{\left(k^{e}B^R(T_v)-k^{ne}I^{R}\right)\mathrm{d}\Omega}.
	\end{aligned}
\end{equation}
Then the bulk viscosity resisting the vibrational radiation transitions is obtained,
\begin{equation}
	\begin{aligned}[b]
		\mu_b^{R}(T_v)&=2p_v\tau\frac{(3+d_r)(3+d_v)Z_v-(3+d_r)d_rZ_r}{3\left(3+d_r+d_v\right)^2}.
	\end{aligned}
\end{equation}
Clearly, both $\mu_b$ and $\mu_b^{R}$ are determined by the internal collision numbers $Z_r$ and $Z_v$, which describe how rapidly the equipartition of kinetic energy among different modes can be reached. Meanwhile, the collision numbers $Z_r$ and $Z_v$ can be obtained simultaneously, when both $\mu_b$ and $\mu_b^{R}$ of a system are measured.

\subsubsection{Relaxation of heat flux}

The rotational and vibrational modes in molecular gas carry the thermal energy and contribute also to the heat flux, while the conductance can be quite different from that of the translational one. In the continuum flow limit, the total thermal conductivity determines the gas dynamics in addition to the viscosity and diffusivity. However, the thermal conductivity of a single type  mode may be important and even dominant when the gas is in a non-equilibrium state. For example, the mass flow rate in thermal transpiration is found to depend on the translational thermal conductivity of gas rather than the total thermal conductivity \citep{Mason1963JCP}.

In the original Rykov model, the relaxation of translational heat flux is independent of the rotational one, and vice versa. However, due to the energy exchange between different modes, it is necessary to consider the fact that the relaxations of heat fluxes are coupled within all the DoF. In general, the relaxation of translational, rotational and vibrational heat fluxes satisfies the following relation in spatially-homogeneous system \citep{Mason1962},
\begin{equation}\label{eq:heat_flux_relaxation}
	\left[ 
      \begin{array}{ccc} 
        \partial{\bm{q}_{t}}/{\partial{t}} \\ \partial{\bm{q}_{r}}/{\partial{t}} \\ \partial{\bm{q}_{v}}/{\partial{t}}
      \end{array}
    \right]
    = -\frac{p_t}{\mu}
    \left[ 
      \begin{array}{ccc} 
        A_{tt} & A_{tr} & A_{tv} \\ A_{rt} & A_{rr} & A_{rv} \\ A_{vt} & A_{vr} & A_{vv}
      \end{array}
    \right]
    \left[ 
      \begin{array}{ccc} 
        \bm{q}_{t} \\ \bm{q}_{r} \\ \bm{q}_{v}
      \end{array}
    \right],
\end{equation}
where the dimensionless relaxation rates $\bm{A}$ is a $3\times3$ matrix encapsulating the dimensionless thermal relaxation rates. Accordingly, $\bm{q}_0$, $\bm{q}_1$, $\bm{q}_2$ in reference distributions \eqref{eq:gt_gr_gv} can be determined in terms of $\bm{q}_t$, $\bm{q}_r$, $\bm{q}_v$ and the thermal relaxation rates $\bm{A}$. To be specific, the first three equations \eqref{eq:governing_equation_Rykov} are multiplied by $\frac{1}{2}mc^2\bm{c}$, $\bm{c}$ and $\bm{c}$, respectively, and then are integrated with respect to $\bm{v}$, yielding
\begin{equation}\label{eq:q0_q1_q2}
    \begin{bmatrix} 
      \bm{q}_{0} \\ \bm{q}_{1} \\ \bm{q}_{2}
	\end{bmatrix}
	= 
    \begin{bmatrix}		(2-3A_{tt})Z_{int}+1 & -3A_{tr}Z_{int} & -3A_{tv}Z_{int} \\		-A_{rt}Z_{int} & -A_{rr}Z_{int}+1 & -A_{rv}Z_{int} \\ 		-A_{vt}Z_{int} & -A_{vr}Z_{int} & -A_{vv}Z_{int}+1
    \end{bmatrix}
    \begin{bmatrix} 
      \bm{q}_{t} \\ \bm{q}_{r} \\ \bm{q}_{v}
    \end{bmatrix},
\end{equation}
where $Z_{int}=\left({1}/{Z_r}+{1}/{Z_v}\right)^{-1}$.

When the gas is stationary,  the thermal relaxation rates are related to the translational, rotational and vibrational thermal conductivities, $\kappa_t$, $\kappa_r$ and $\kappa_v$, respectively (see Appendix~\ref{app:A})
\begin{equation}\label{eq:kappa_A}
	\left[ 
      \begin{array}{ccc} 
        \kappa_t \\ \kappa_r \\ \kappa_v
      \end{array}
    \right]
	= \frac{k_B\mu}{2m}
	\left[ 
      \begin{array}{ccc} 
        A_{tt} & A_{tr} & A_{tv} \\ A_{rt} & A_{rr} & A_{rv} \\ A_{vt} & A_{vr} & A_{vv}
      \end{array}
    \right]^{-1}
    \left[ 
      \begin{array}{ccc} 
        5 \\ d_r \\ d_v(T_v)
      \end{array}
    \right].
\end{equation}
It has shown that the presence of radiation transition does not affect the thermal conductivities.

It will be convenient to use the following dimensionless \cite{Eucken1913} factor $f_{eu}$:
\begin{equation}\label{eq:feu}
	\begin{aligned}[b]
		c_vf_{eu}\equiv\frac{\kappa}{\mu}=\frac{\kappa_t+\kappa_r+\kappa_v}{\mu},
	\end{aligned}
\end{equation}
where $\kappa$ is the total thermal conductivity,  and $c_v$ is the specific heat capacity at constant volume. Similarly, $f_t$, $f_r$ and $f_v$ represent the Eucken factors of the translational, rotational and vibrational modes, respectively,
\begin{equation}\label{eq:ft_fr_fv}
	\begin{aligned}[b]
		f_t = \frac{2}{3}\frac{m\kappa_t}{k_B\mu}, \quad f_r = \frac{2}{d_r}\frac{m\kappa_r}{k_B\mu}, \quad f_v = \frac{2}{d_v}\frac{m\kappa_v}{k_B\mu}.
	\end{aligned}
\end{equation}
Therefore, the Eucken factors are determined by the thermal relaxation rates as
\begin{equation}\label{eq:EuckenFactor_A}
	\left[ 
      \begin{array}{ccc} 
        f_t \\ f_r \\ f_v
      \end{array}
    \right]
	= 
	\left[ 
      \begin{array}{ccc} 
        3A_{tt} & d_rA_{tr} & d_v(T_v)A_{tv} \\ 3A_{rt} & d_rA_{rr} & d_v(T_v)A_{rv} \\ 3A_{vt} & d_rA_{vr} & d_v(T_v)A_{vv}
      \end{array}
    \right]^{-1}
    \left[ 
      \begin{array}{ccc} 
        5 \\ d_r \\ d_v(T_v)
      \end{array}
    \right].
\end{equation}

Clearly, the elements in matrix $\bm{A}$ cannot be fully determined even though all the Eucken factors $f_t, f_r, f_v$ (thermal conductivities $\kappa_t, \kappa_r, \kappa_v$ equivalently) are fixed. In other words, in molecular gas, having all the transport coefficients is not enough to exactly describe the relaxation of heat flux. Therefore, it is necessary to recovery the thermal relaxation rates in the kinetic model correctly. Nevertheless, since the rate of translation-vibration energy exchange is usually much slower than that of translation-rotational energy exchange, the values of $A_{tv},~A_{vt},~A_{rv},~A_{vr}$ in matrix $\bm{A}$ are much smaller than the others and thus can be approximated to be zero practically~\citep{Mason1962}. While the off-diagonal elements $A_{tr},~A_{rt}$ still have to be correctly recovered, which have been found to make uncertainty in predicting macroscopic gas dynamics~\citep{Li2021JFM}.

\subsection{Dimensionless expressions}

Let $L_0,~T_0,~n_0$ be the reference length, temperature and number density, respectively, then the most probable speed is $v_m=\sqrt{2k_BT_0/m}$ and reference pressure is $p_0=n_0k_BT_0$. The dimensionless variables are introduced as,
\begin{equation}\label{eq:dimensionless_variables}
	\begin{aligned}[b]
		&\tilde{\bm{x}}=\bm{x}/L_0, \quad\quad \tilde{T}=T/T_0, \quad\quad \tilde{n}=n/n_0, \quad\quad \tilde{t}=v_mt/L_0, \\
		&\tilde{\bm{v}}=\bm{v}/v_m, \quad\quad \tilde{\bm{c}}=\bm{c}/v_m, \quad\quad \tilde{p}=p/p_0, \quad\quad \tilde{\bm{q}}=\bm{q}/(p_0v_m), \\
		&\tilde{f}_0=v_m^{3}f_0/n_0, \quad\quad \tilde{f}_1=v_m^{3}f_1/p_0, \quad\quad \tilde{f}_2=v_m^{3}f_2/p_0, \quad\quad
		\tilde{I}^R=I^R/(p_0v_m).
	\end{aligned}
\end{equation}
When the gray model is used for photon transport, the Knudsen numbers for gas flow and photon transport are defined, respectively, as,
\begin{equation}\label{eq:Kn}
	\text{Kn}_{\text{gas}}=\frac{\mu(T_0)}{n_0L_0}\sqrt{\frac{\pi}{2mk_BT_0}}, \quad\quad 
	\text{Kn}_{\text{photon}}=\frac{1}{k_{\text{gray}}L_0}.
\end{equation}
The relative strength of the radiative to the convective heat transfer is given by dimensionless parameter,
\begin{equation}\label{eq:sigma_R}
	\tilde{\sigma}_R=\frac{\sigma_RT_0^3}{n_0k_Bv_m},
\end{equation}
which is equivalent to the reciprocal of Boltzmann number commonly used in the radiation hydrodynamics \citep{Casto2004Radiation}. Therefore, the model equations are nondimensionlized,
\begin{equation}\label{eq:governing_equation_dinmensionless}
	\begin{aligned}[b]
	&\tilde{\mathcal{D}}\tilde{f_0}
	 = \tilde{Q}(\tilde{f_0}) + \frac{\sqrt{\pi}\tilde{n}\tilde{T}_t^{1-\omega}}{2\text{Kn}_{\text{gas}}}\left[\frac{\tilde{g}_{0r}-\tilde{g}_{0t}}{Z_r} + \frac{\tilde{g}_{0v}-\tilde{g}_{0t}}{Z_v}\right], \\
&	\tilde{\mathcal{D}}\tilde{f_1} = \frac{\sqrt{\pi}\tilde{n}\tilde{T}_t^{1-\omega}}{2\text{Kn}_{\text{gas}}}\left[\left(\tilde{g}_{1t}-\tilde{f}_1\right)+\frac{\tilde{g}_{1r}-\tilde{g}_{1t}}{Z_r} + \frac{\tilde{g}_{1v}-\tilde{g}_{1t}}{Z_v}\right], \\
&	\tilde{\mathcal{D}}\tilde{f_2} = \frac{\sqrt{\pi}\tilde{n}\tilde{T}_t^{1-\omega}}{2\text{Kn}_{\text{gas}}}\left[\left(\tilde{g}_{2t}-\tilde{f}_2\right)+\frac{\tilde{g}_{2r}-\tilde{g}_{2t}}{Z_r} + \frac{\tilde{g}_{2v}-\tilde{g}_{2t}}{Z_v}\right]  -\frac{\tilde{f}_0}{\tilde{n}\text{Kn}_{\text{photon}}}{\left(4\tilde{\sigma}_R\tilde{T}_v^4-\int{\tilde{I}^{R}\mathrm{d}\Omega}\right)},  \\
	&	\frac{1}{\tilde{c}_l}\frac{\partial{\tilde{I}^{R}}}{\partial{\tilde{t}}}+\bm{n} \cdot \frac{\partial{\tilde{I}^{R}}}{\partial{\tilde{\bm{x}}}} = \frac{1}{\text{Kn}_{\text{photon}}}\left(\frac{1}{\pi}\tilde{\sigma}_R\tilde{T}_v^4-\tilde{I}^{R}\right),
	\end{aligned}
\end{equation}
where $	\tilde{\mathcal{D}}\tilde{f_i}=	\frac{\partial{\tilde{f_i}}}{\partial{\tilde{t}}}+\bm{\tilde{v}} \cdot \frac{\partial{\tilde{f_i}}}{\partial{\tilde{\bm{x}}}}+  \frac{\partial{(\tilde{\bm{a}}\tilde{f_i})}}{\partial{\tilde{\bm{v}}}}$, $\tilde{c}_l=c_l/v_m$ does not affect the results of steady-state problems, and $\tilde{Q}(\tilde{f}_0)$ represents the dinmensionless elastic collision operators: 
\begin{equation}\label{eq:collision_operator_dinmensionless}
	\tilde{Q}(\tilde{f}_0) = \left\{
	\begin{aligned}
		&\frac{\sqrt{\pi}\tilde{n}\tilde{T}_t^{1-\omega}}{2\text{Kn}_{\text{gas}}}\left(\tilde{g}_{0r}-\tilde{g}_{0t}\right), \quad (\text{RTA}) \\
		&\frac{1}{\text{Kn}_{\text{gas}}}\frac{5}{2^{7-\omega}\Gamma^2(9/4-\omega/2)}\iint\sin^{(1-2\omega)/2}\left(\frac{\theta}{2}\right)\cos^{(1-2\omega)/2}\left(\frac{\theta}{2}\right)\left| \tilde{\bm{v}}-\tilde{\bm{v}}_* \right|^{2(1-\omega)}  \\
		&\cdot\left[\tilde{f}_0(\tilde{\bm{v}}'_*)\tilde{f}_0(\tilde{\bm{v}}')-\tilde{f}_0(\tilde{\bm{v}}_*)\tilde{f}_0(\tilde{\bm{v}})\right]\mathrm{d}{\Omega}\mathrm{d}{\tilde{\bm{v}}_*}. \quad (\text{BCO})
	\end{aligned}
	\right.
\end{equation}
The dinmensionless reduced reference velocity distribution functions are
\begin{align}\label{eq:glt_glr_glv_dinmensionless}
	\tilde{g}_{0t}&=\tilde{n}\left(\frac{1}{\pi\tilde{T}_t}\right)^{3/2}\exp{\left(-\frac{\tilde{c}^2}{\tilde{T}_t}\right)}\left[1+\frac{4\tilde{\bm{q}}_t\cdot{\tilde{\bm{c}}}}{15{\tilde{T}_t}{\tilde{p}_t}} \left(\frac{\tilde{c}^2}{\tilde{T}_t}-\frac{5}{2}\right)\right], \notag \\
	\tilde{g}_{0r}&=\tilde{n}\left(\frac{1}{\pi\tilde{T}_{tr}}\right)^{3/2}\exp{\left(-\frac{\tilde{c}^2}{\tilde{T}_{tr}}\right)}\left[1+\frac{4\tilde{\bm{q}}_0\cdot{\tilde{\bm{c}}}}{15{\tilde{T}_{tr}}{\tilde{p}_{tr}}} \left(\frac{\tilde{c}^2}{\tilde{T}_{tr}}-\frac{5}{2}\right)\right], \notag \\
	\tilde{g}_{0v}&=\tilde{n}\left(\frac{1}{\pi\tilde{T}_{tv}}\right)^{3/2}\exp{\left(-\frac{\tilde{c}^2}{\tilde{T}_{tv}}\right)}\left[1+\frac{4\tilde{\bm{q}}_0\cdot{\tilde{\bm{c}}}}{15{\tilde{T}_{tv}}{\tilde{p}_{tv}}} \left(\frac{\tilde{c}^2}{\tilde{T}_{tv}}-\frac{5}{2}\right)\right], \notag \\
	\tilde{g}_{1t}&=
	\left\{
		\begin{aligned}
			&\frac{d_r}{2}\tilde{T}_r\tilde{g}_{0t}+\left(\frac{1}{\pi\tilde{T}_t}\right)^{3/2}\exp{\left(-\frac{\tilde{c}^2}{\tilde{T}_t}\right)}\frac{2\tilde{\bm{q}}_r\cdot{\tilde{\bm{c}}}}{\tilde{T}_t},  \quad (\text{RTA})\\
			&\frac{d_r}{2}\tilde{T}_r\left[\frac{2\text{Kn}_{\text{gas}}}{\sqrt{\pi}\tilde{n}\tilde{T}_t^{1-\omega}} \tilde{Q}_B(\tilde{f}_0)+\tilde{f}_0\right]+\left(\frac{1}{\pi\tilde{T}_t}\right)^{3/2}\exp{\left(-\frac{\tilde{c}^2}{\tilde{T}_t}\right)}\frac{2\tilde{\bm{q}}_r\cdot{\tilde{\bm{c}}}}{\tilde{T}_t},  \quad (\text{BCO}) 
		\end{aligned}
	\right. \notag \\
	\tilde{g}_{1r}&=\frac{d_r}{2}\tilde{T}_{tr}\tilde{g}_{0r}+\left(\frac{1}{\pi\tilde{T}_{tr}}\right)^{3/2}\exp{\left(-\frac{\tilde{c}^2}{\tilde{T}_{tr}}\right)}\frac{2\tilde{\bm{q}}_1\cdot{\tilde{\bm{c}}}}{\tilde{T}_{tr}}, \notag \\
	\tilde{g}_{1v}&=\frac{d_r}{2}\tilde{T}_{r}\tilde{g}_{0v}+\left(\frac{1}{\pi\tilde{T}_{tv}}\right)^{3/2}\exp{\left(-\frac{\tilde{c}^2}{\tilde{T}_{tv}}\right)}\frac{2\tilde{\bm{q}}_1\cdot{\tilde{\bm{c}}}}{\tilde{T}_{tv}}, \notag \\
	\tilde{g}_{2t}&=
	\left\{
		\begin{aligned}
			&\frac{d_v(\tilde{T}_v)}{2}\tilde{T}_v\tilde{g}_{0t}+\left(\frac{1}{\pi\tilde{T}_t}\right)^{3/2}\exp{\left(-\frac{\tilde{c}^2}{\tilde{T}_t}\right)}\frac{2\tilde{\bm{q}}_v\cdot{\tilde{\bm{c}}}}{\tilde{T}_t},  \quad (\text{RTA})\\
			&\frac{d_v(\tilde{T}_v)}{2}\tilde{T}_v\left[\frac{2\text{Kn}_{\text{gas}}}{\sqrt{\pi}\tilde{n}\tilde{T}_t^{1-\omega}} \tilde{Q}_B(\tilde{f}_0)+\tilde{f}_0\right]+\left(\frac{1}{\pi\tilde{T}_t}\right)^{3/2}\exp{\left(-\frac{\tilde{c}^2}{\tilde{T}_t}\right)}\frac{2\tilde{\bm{q}}_v\cdot{\tilde{\bm{c}}}}{\tilde{T}_t},  \quad (\text{BCO}) 
		\end{aligned}
	\right. \notag \\
	\tilde{g}_{2r}&=\frac{{d_v(\tilde{T}_v)}}{2}\tilde{T}_{v}\tilde{g}_{0r}+\left(\frac{1}{\pi\tilde{T}_{tr}}\right)^{3/2}\exp{\left(-\frac{\tilde{c}^2}{\tilde{T}_{tr}}\right)}\frac{2\tilde{\bm{q}}_2\cdot{\tilde{\bm{c}}}}{\tilde{T}_{tr}}, \notag \\
	\tilde{g}_{2v}&=\frac{{d_v(\tilde{T}_{tv})}}{2}\tilde{T}_{tv}\tilde{g}_{0v}+\left(\frac{1}{\pi\tilde{T}_{tv}}\right)^{3/2}\exp{\left(-\frac{\tilde{c}^2}{\tilde{T}_{tv}}\right)}\frac{2\tilde{\bm{q}}_2\cdot{\tilde{\bm{c}}}}{\tilde{T}_{tv}}.
\end{align}

When the gas species and relaxation rates are fixed, the solutions are determined by four dimensionless parameters: $\text{Kn}_{\text{gas}}$, $\text{Kn}_{\text{photon}}$, $\tilde{\sigma}_R$ and $T_0/T_{\text{ref}}$, where $\tilde{\sigma}_R$ gives the relative importance of radiative heat transfer in the system, and $T_0/T_{\text{ref}}$ indicates the degree of vibrational excitation. From the term of energy exchange between vibrational modes and radiative field in governing equations, it can be seen that when $\tilde{\sigma}_R/\text{Kn}_{\text{photon}}\rightarrow 0$, the interaction between gas and photon becomes negligible. 


\section{Validation of the kinetic model}\label{sec:validation}

When the radiation is absent, the accuracy of the proposed kinetic models \eqref{eq:governing_equation_Rykov} and \eqref{eq:governing_equation_Boltzmann} are evaluated by comparing the numerical solutions of one-dimensional Fourier flow, Couette flow, thermal creep flow and normal shock wave in nitrogen with constant vibrational DoF to DSMC solutions. We use kinetic model I and II to represent the model equations with RTA and BCO, respectively. The kinetic model equations are solved by the discretized velocity method with the fast spectral method for the BCO~\citep{Lei2013,lei_Jfm}, while DSMC simulations are conducted using the open source code SPARTA~\citep{SPARTA}.

\subsection{Relaxation rates extracted from DSMC}\label{subsec:A_from_DSMC}

In DSMC simulation for molecular gas flow, generally, not all the transport coefficients of a real gas can be recovered simultaneously in this method. However, the results of DSMC still can be regarded as reference solutions, when we consider a virtual gas with the exact same relaxation rates realized in DSMC simulations. Therefore, we extract the thermal relaxation rates from the DSMC and apply to our kinetic model, to make a fair comparison. With the fixed shear viscosity and self-diffusion coefficient, the collision number $Z_r$ and $Z_v$ are the only parameters that affect the thermal relaxation rates in DSMC. As an example, here we take the collision numbers $Z_r=2.667$ and $Z_v=10Z_r$ for nitrogen.

\begin{figure}[t]
	\centering
	\subfloat[]{\includegraphics[scale=0.22,clip=true]{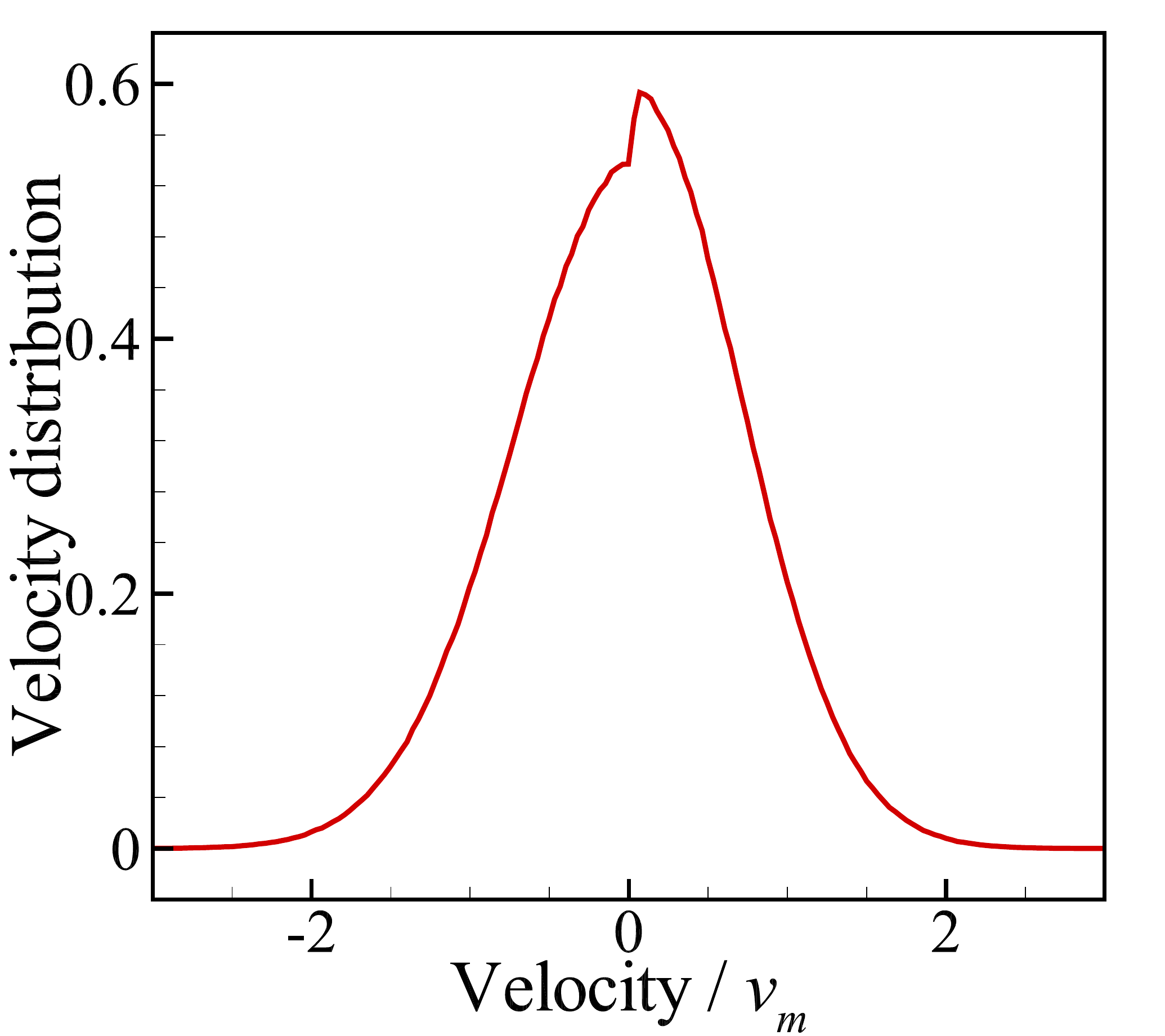}\label{fig:A_DSMC:a}}  \quad
	\subfloat[]{\includegraphics[scale=0.22,clip=true]{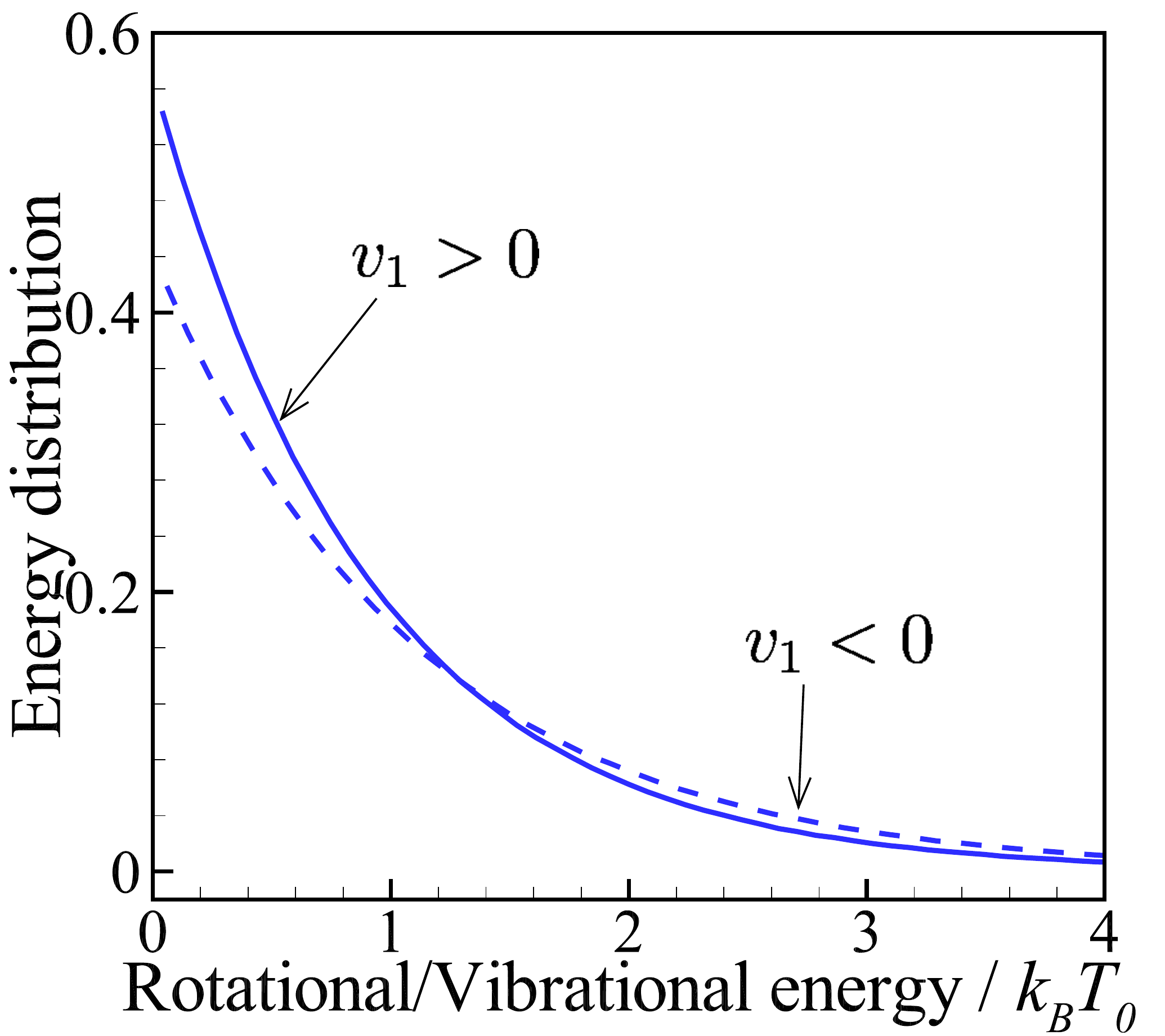}\label{fig:A_DSMC:b}} \\
	\vskip 0.5cm
	\subfloat[]{\includegraphics[scale=0.235,clip=true]{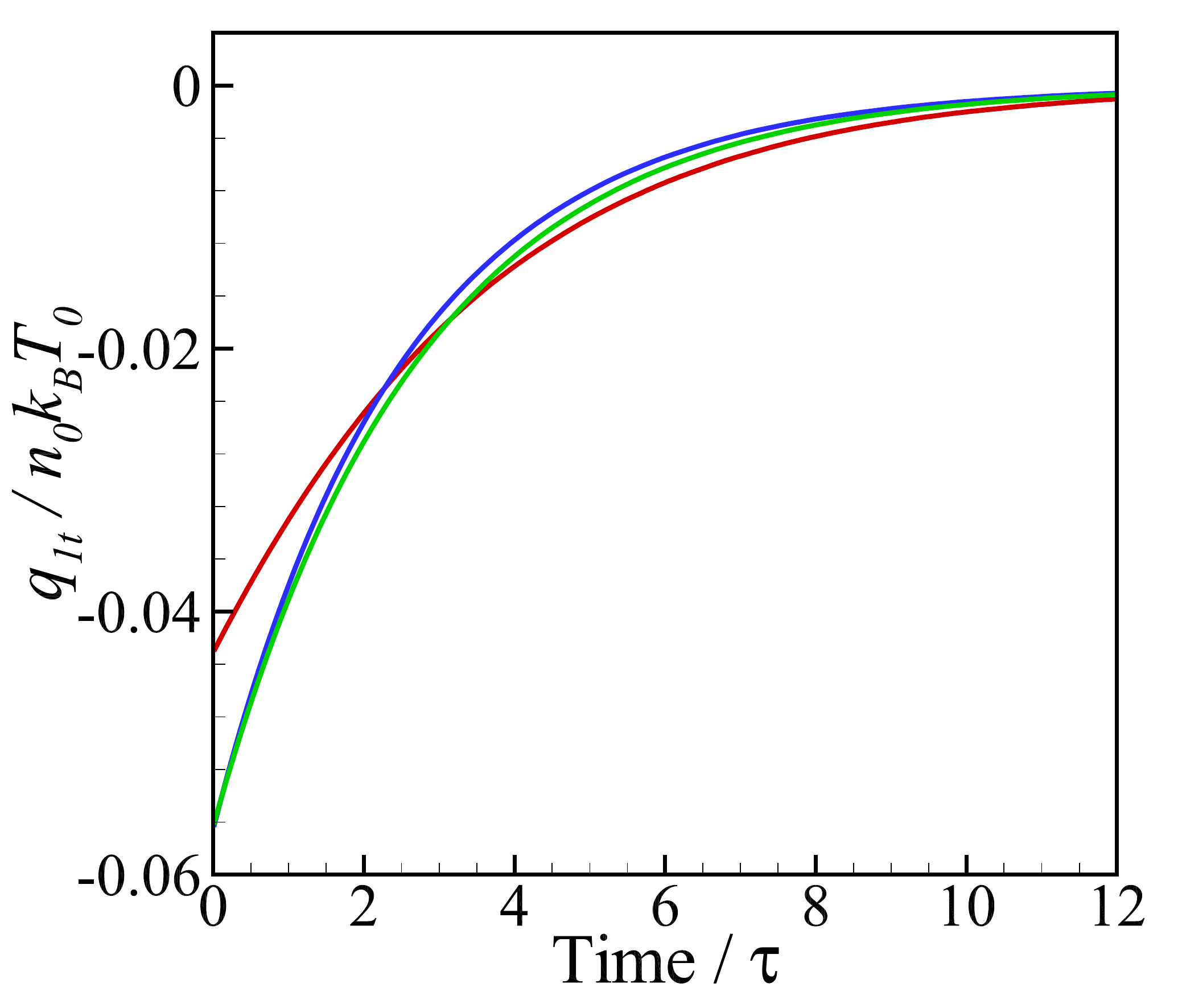}\label{fig:A_DSMC:c}}  \quad
	\subfloat[]{\includegraphics[scale=0.23,clip=true]{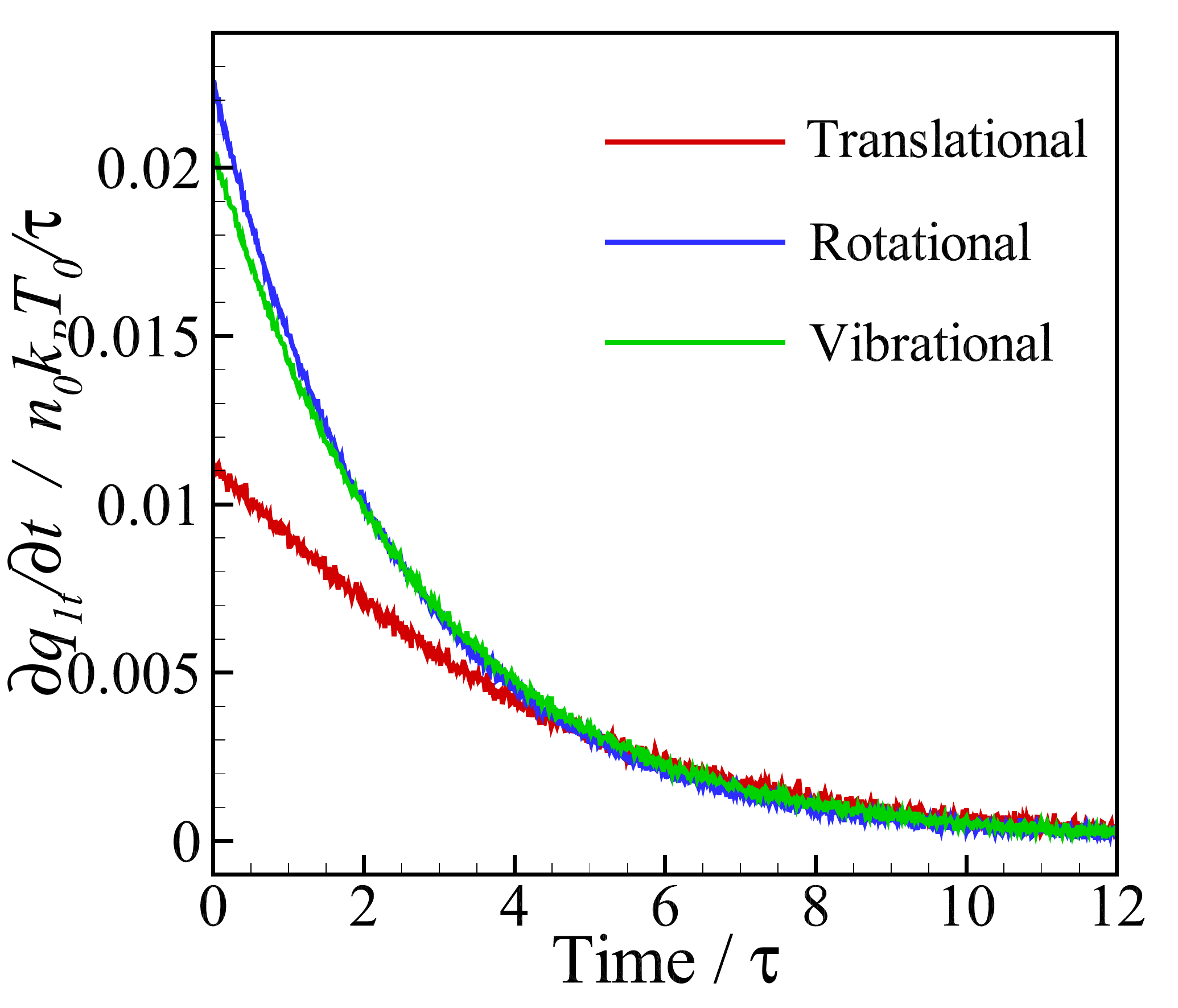}\label{fig:A_DSMC:d}}
	\caption{Extraction of the thermal relaxation rates $\bm{A}$ in~\eqref{eq:heat_flux_relaxation} from the DSMC simulation. Special distributions of (a) the molecular velocity and (b) rotational/vibrational energy (overlap with each other) are designed to generate initial heat flux. (c) The evolution of heat fluxes and (d) their time derivatives are monitored until the system reaches thermal equilibrium.}
	\label{fig:A_DSMC}
\end{figure}

Similar to the procedure of extracting thermal relaxation rates for the translational and rotational DoF from DSMC~\citep{Li2021JFM}, here a homogeneous system of nitrogen is simulated, which consists of $10^6$ simulation particles in a cubic cell of the volume $(10~\text{nm})^3$. The periodic condition is applied at all boundaries. Binary collisions are described by the variable-soft-sphere model, and the system parameters and properties of nitrogen used in the simulations are: $d_r=d_v=2$, $n_0=2.69\times10^{25}~\text{m}^{-3}$,  $T_0=5000~\text{K}$,  $m=4.65\times10^{-26}~\text{kg}$, the molecular diameter is $d=4.11\times10^{-10}~\text{m}$, the viscosity index is $\omega=0.74$,  the angular scattering parameter is $\alpha=1.36$, and the Schmidt number is $Sc=1/1.34$~\citep{Bird1994}. Initially,  simulation particles with positive velocity in the $x_1$ direction follow the equilibrium distribution at $4500$~K, while those moving in the opposite direction follow the equilibrium distribution at $5500$~K, see  figure \ref{fig:A_DSMC:a} and \ref{fig:A_DSMC:b}, so that initial heat fluxes in all DoF are generated. Then the evolution of heat flux is monitored until the entire system reaches thermal equilibrium, see figure \ref{fig:A_DSMC:c}. Ensemble averaged is taken from 3000 independent runs to get the time derivative of heat flux in figure \ref{fig:A_DSMC:d}. Finally, the following relaxation rates are extracted by solving the linear regression problem \eqref{eq:heat_flux_relaxation} with the least squares method:
\begin{equation}\label{eq:relaxation_rates}
    \left[ 
      \begin{array}{ccc} 
        A_{tt} & A_{tr} & A_{tv} \\ A_{rt} & A_{rr} & A_{rv} \\ A_{vt} & A_{vr} & A_{vv}
      \end{array}
    \right]
    =
	\left[ 
      \begin{array}{ccc} 
        ~0.786 & -0.208 & ~0.003 \\ -0.047 & ~0.883 & -0.049 \\ -0.004 & -0.038 & ~0.772
      \end{array}
    \right].
\end{equation}
Hence, according to~\eqref{eq:EuckenFactor_A}, we have $f_t=2.3635$, $f_r=1.3979$, $f_v=1.3825$, and $f_{eu}=1.807$. With these parameters, our kinetic model is uniquely determined.

\subsection{Fourier flow}\label{subsec:validation_FourierFlow}

The heat transfer in the nitrogen gas between two parallel plates located at $x_2=0$ and $L_0$ are considered, where the temperature of the lower and upper plates are $T_l=0.8T_0$ and $T_u=1.2T_0$, respectively. The averaged number density of nitrogen is set to be $n_0$, and the characteristic length $L_0$ is chosen to be the distance between two plates. The Knudsen numbers considered are  $\text{Kn}_{\text{gas}}=0.1$ and 1. The diffuse boundary conditions are adopted, so that the reflected distributions are
\begin{equation}\label{eq:BC_FourierFlow}
	\begin{aligned}[b]
		&x_2=0,~v_2\ge0: \quad f_0=\frac{n_{in}(x_2=0)}{n_0}E_t(T_l), \quad
		f_1=\frac{d_r}{2}k_BT_lf_0, \quad
		f_2=\frac{d_v}{2}k_BT_lf_0, \\
		&x_2=L_0,~v_2\le0: \quad f_0=\frac{n_{in}(x_2=L_0)}{n_0}E_t(T_u), \quad
		f_1=\frac{d_r}{2}k_BT_uf_0, \quad
		f_2=\frac{d_v}{2}k_BT_uf_0,
	\end{aligned}
\end{equation}
where $n_{in}$ is determined by the flux of incident number density of gas at the plates:
\begin{equation}\label{eq:BC_FourierFlow_n_in}
	\begin{aligned}[b]
		n_{in}(x_2=0)&= -\left(\frac{2m\pi}{k_BT_l}\right)^{1/2}\int_{v_2<0}v_2f_0\mathrm{d}\bm{v}, \\
		n_{in}(x_2=L_0)&= \left(\frac{2m\pi}{k_BT_u}\right)^{1/2}\int_{v_2>0}v_2f_0\mathrm{d}\bm{v}.
	\end{aligned}
\end{equation}

\begin{figure}[t]
	\centering
	\subfloat[]{\includegraphics[scale=0.18,clip=true]{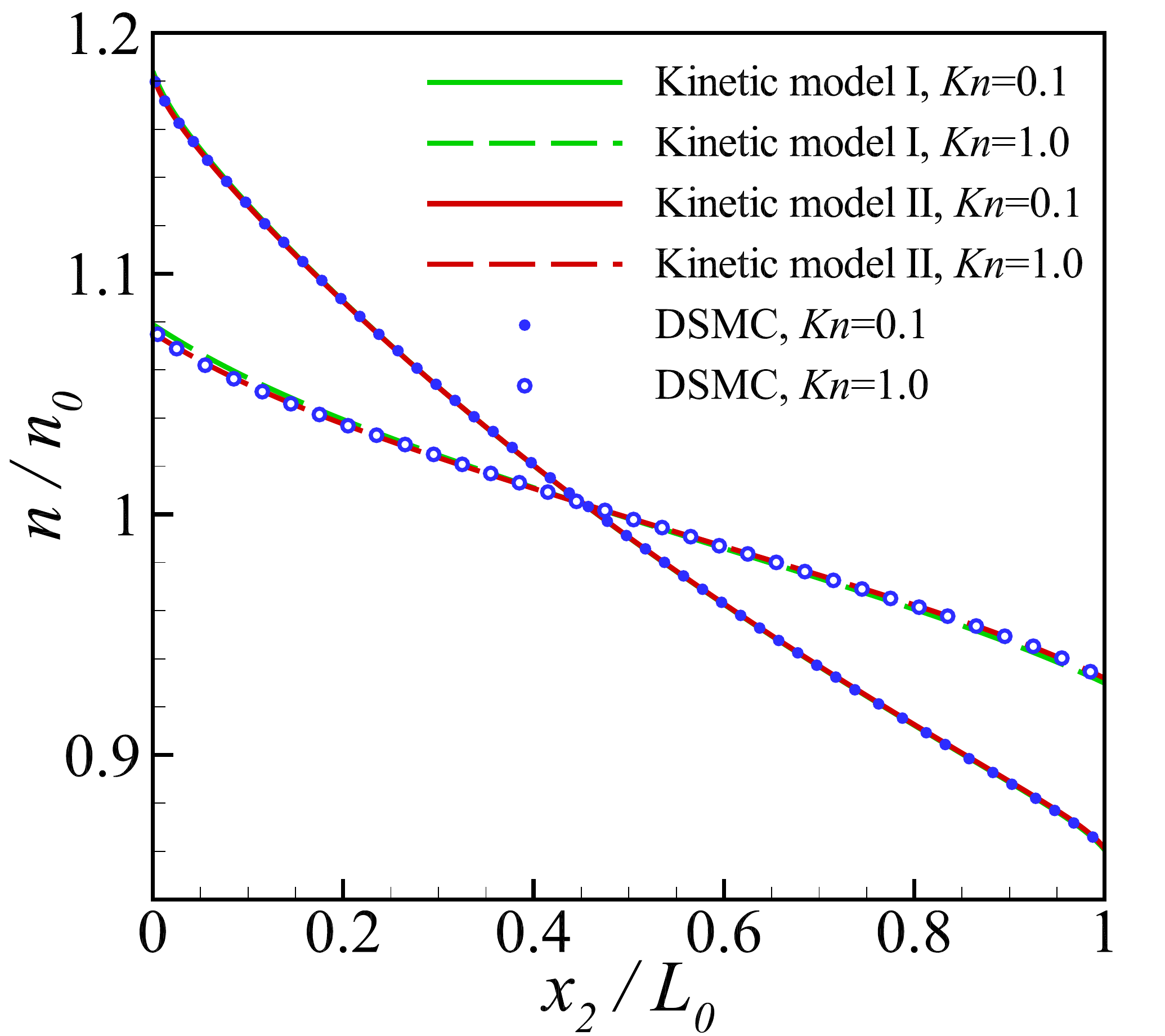}\label{fig:1DFourierFlow:a}}
	\subfloat[]{\includegraphics[scale=0.18,clip=true]{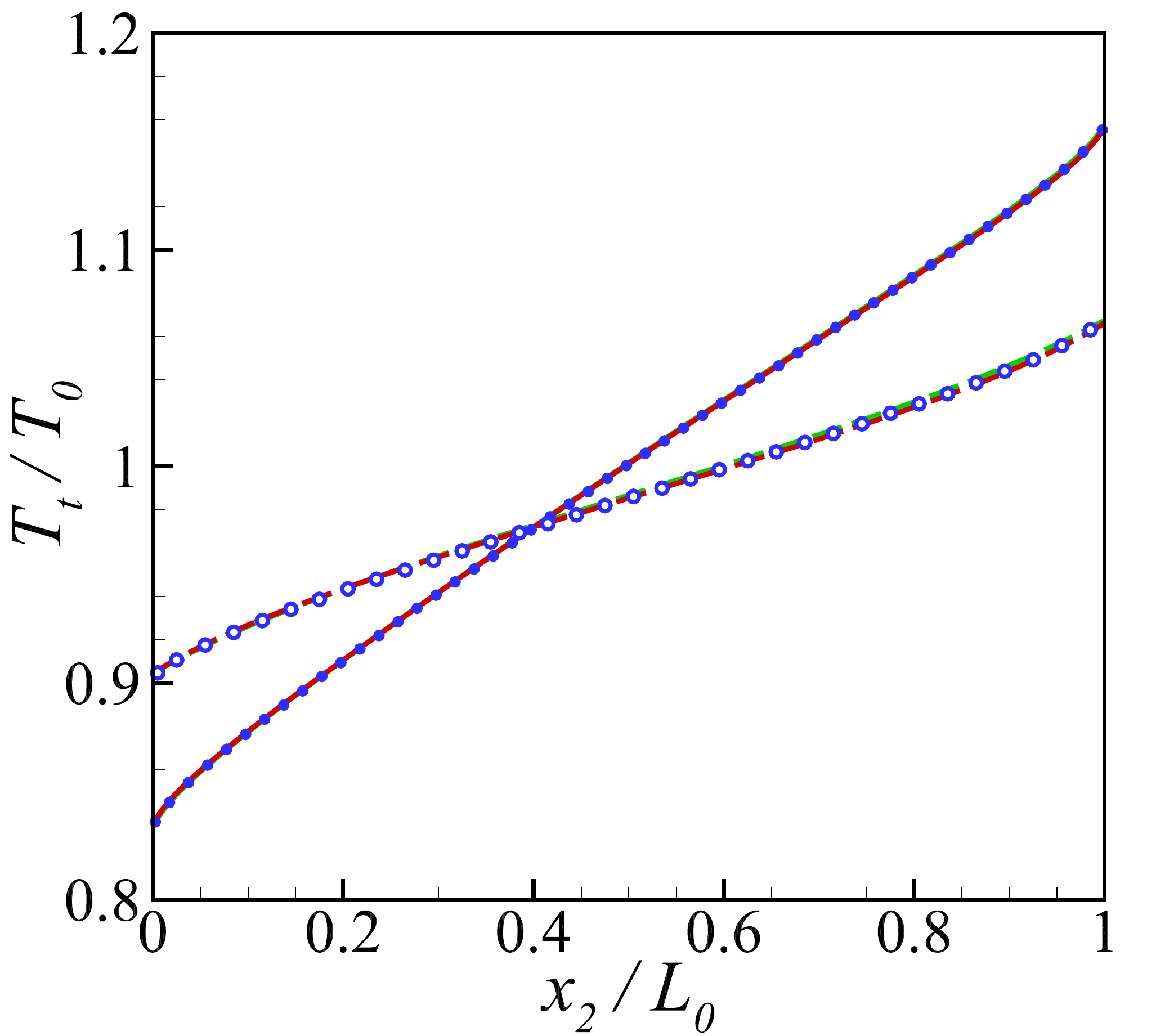}\label{fig:1DFourierFlow:b}}
	\subfloat[]{\includegraphics[scale=0.18,clip=true]{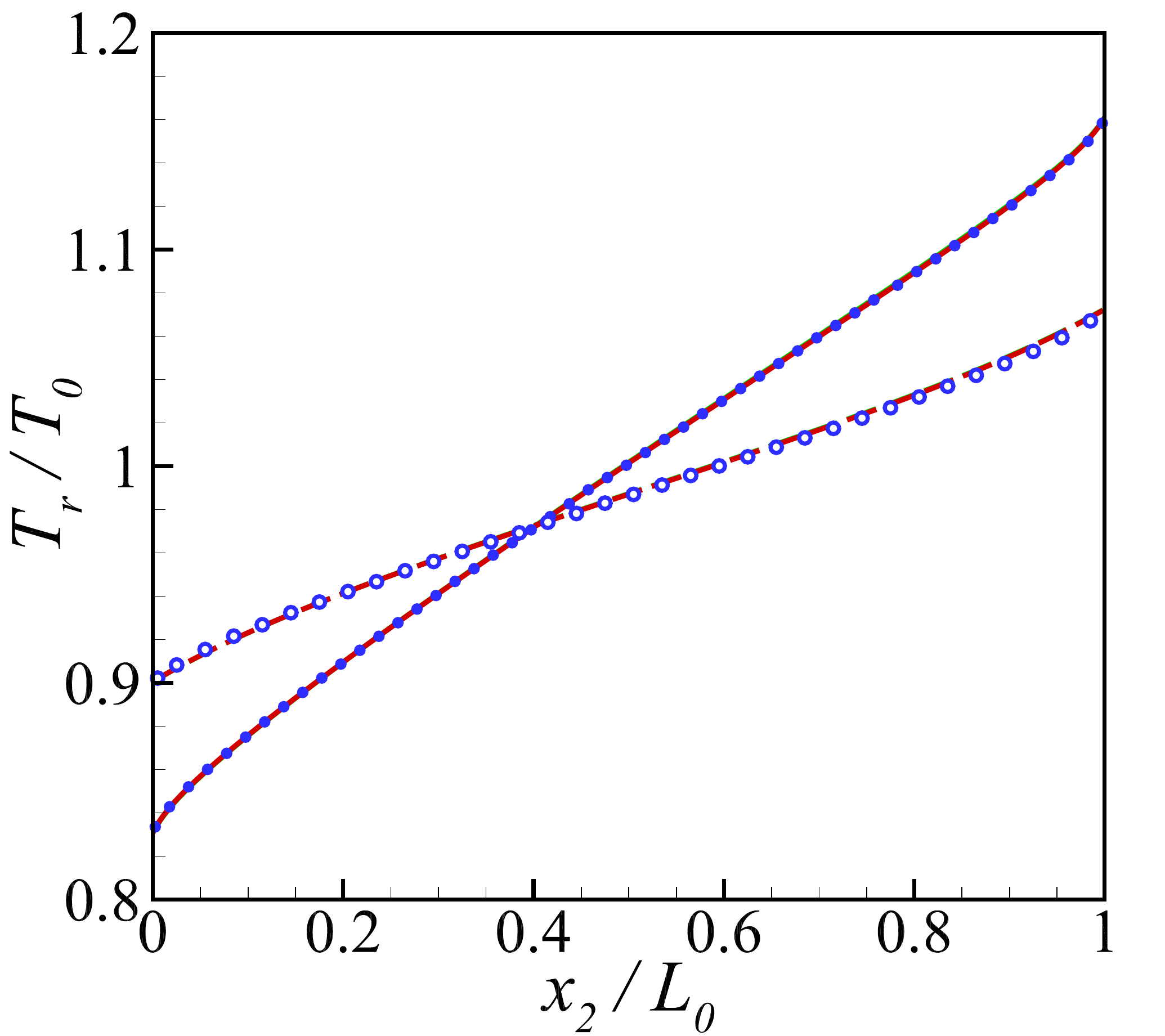}\label{fig:1DFourierFlow:c}} \\
	\subfloat[]{\includegraphics[scale=0.18,clip=true]{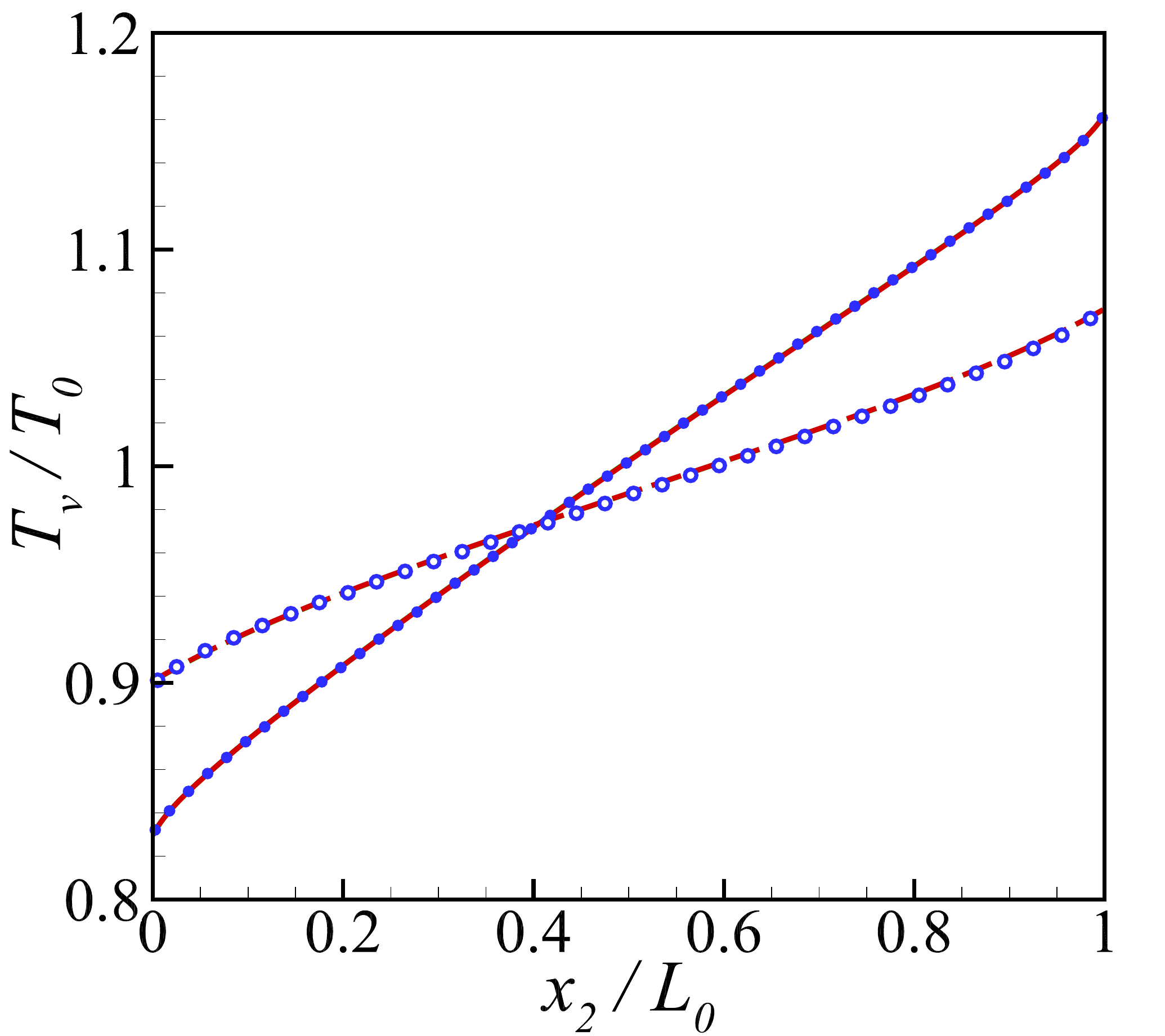}\label{fig:1DFourierFlow:d}}
	\subfloat[]{\includegraphics[scale=0.18,clip=true]{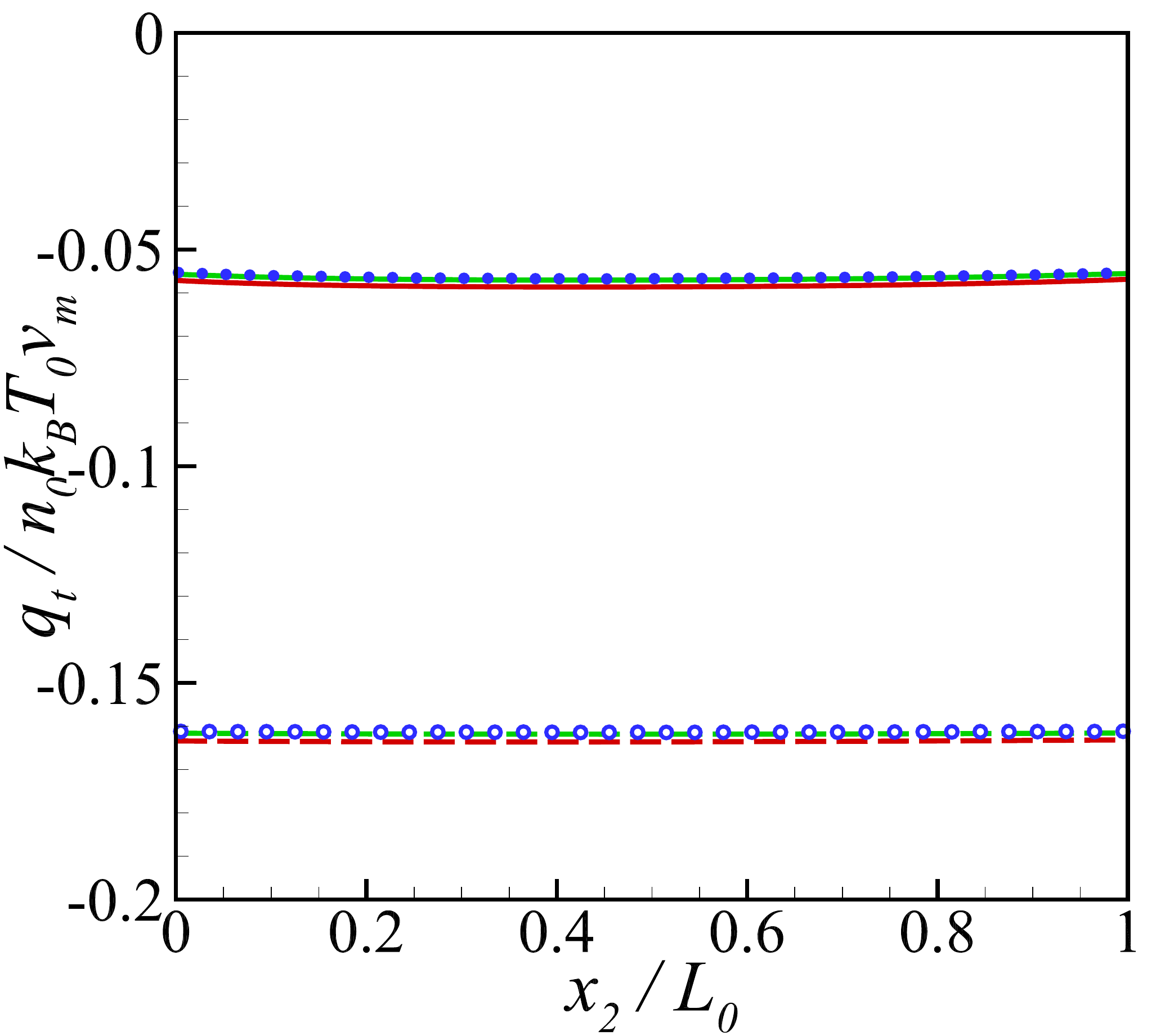}\label{fig:1DFourierFlow:e}}
	\subfloat[]{\includegraphics[scale=0.18,clip=true]{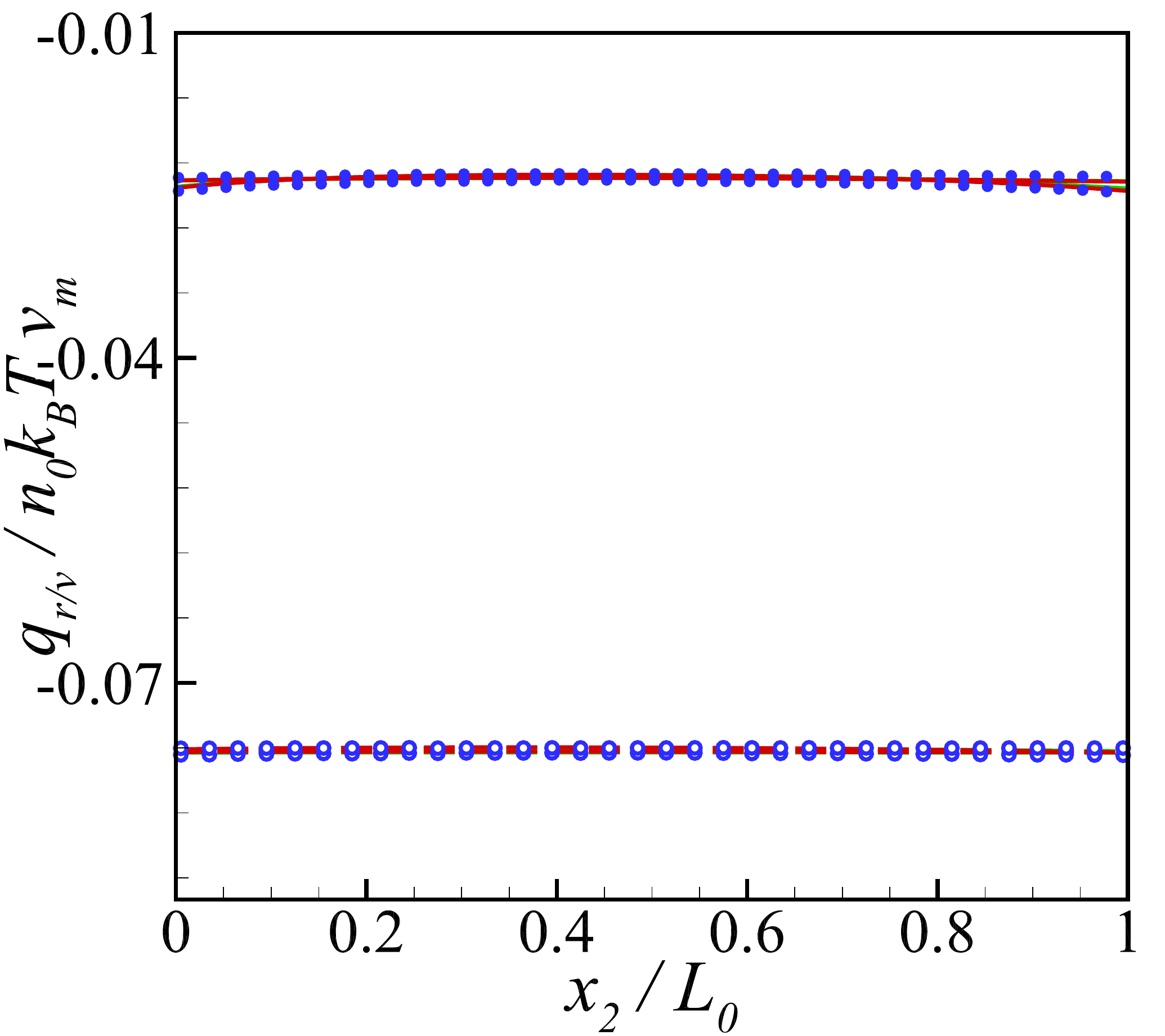}\label{fig:1DFourierFlow:f}}
	\caption{
		Comparisons of the (a) density, (b) translational temperature, (c) rotational temperature, (d) vibrational temperature, (e) translational heat flux and (f) rotational/vibrational heat flux of nitrogen between kinetic model I (green lines), kinetic model II (red lines) and DSMC (blue circles) for the Fourier flows.
	}
	\label{fig:1DFourierFlow}
\end{figure}

Numerical solutions of the kinetic models I and II, as well as the DSMC results are shown in figure \ref{fig:1DFourierFlow}. For both  $\text{Kn}_{\text{gas}}=0.1$ and $\text{Kn}_{\text{gas}}=1$, excellent agreement in the density, temperature and heat flux are observed. Meanwhile, profiles of translational, rotational and vibrational temperatures nearly overlap, although the relaxation times for different modes are different. Additionally, the rotational and vibrational heat flux are almost the same (figure~\ref{fig:1DFourierFlow:f}), due to the close values of the rotational and vibrational thermal conductivities. Thus, it is clearly seen that the values of collision number $Z_r$ and $Z_v$ do not have influence on the distribution of macroscopic quantities for the steady-state planar Fourier flow. 


\begin{figure}[t]
	\centering
	\subfloat[]{\includegraphics[scale=0.24,clip=true]{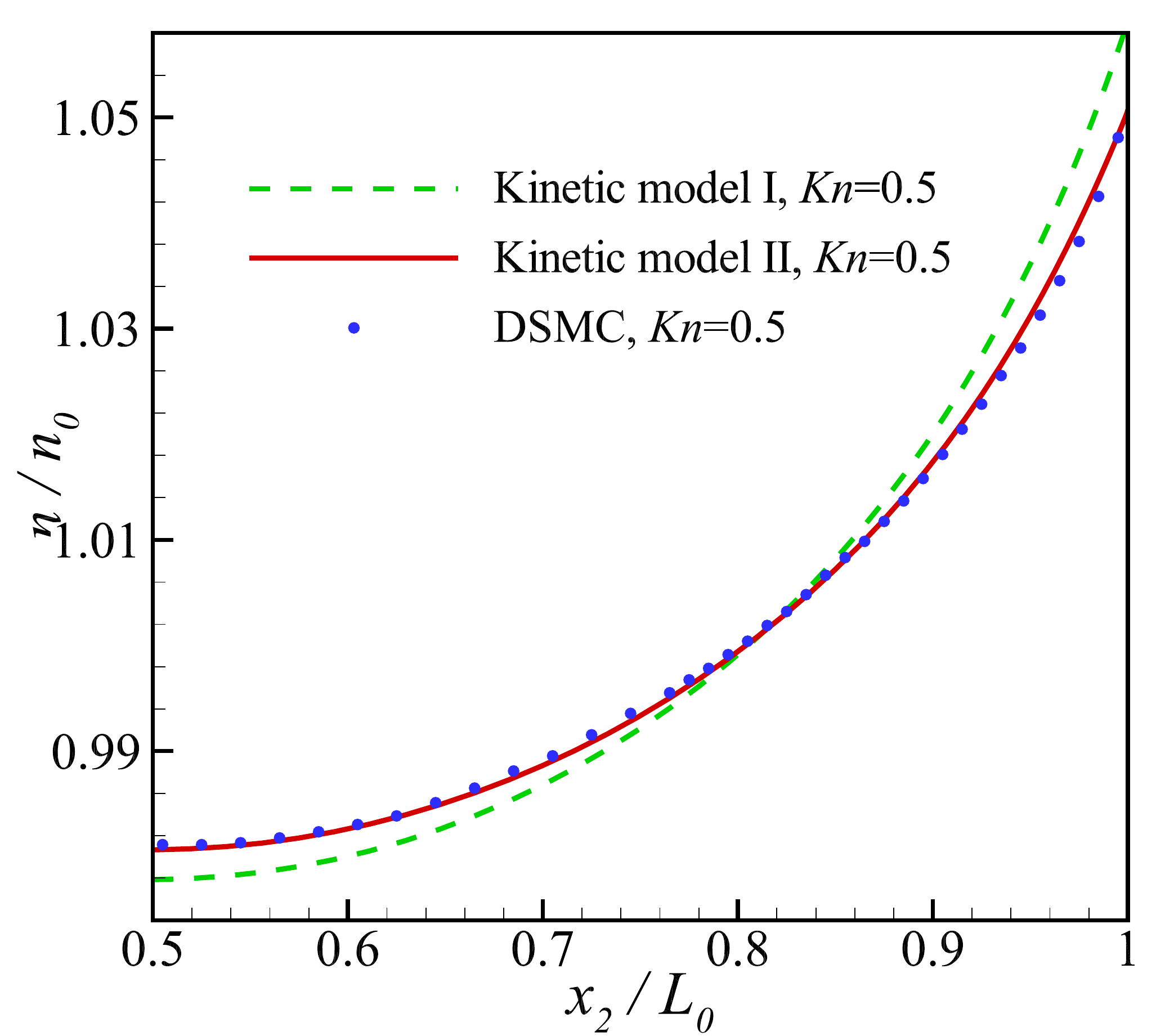}} \quad
	\subfloat[]{\includegraphics[scale=0.24,clip=true]{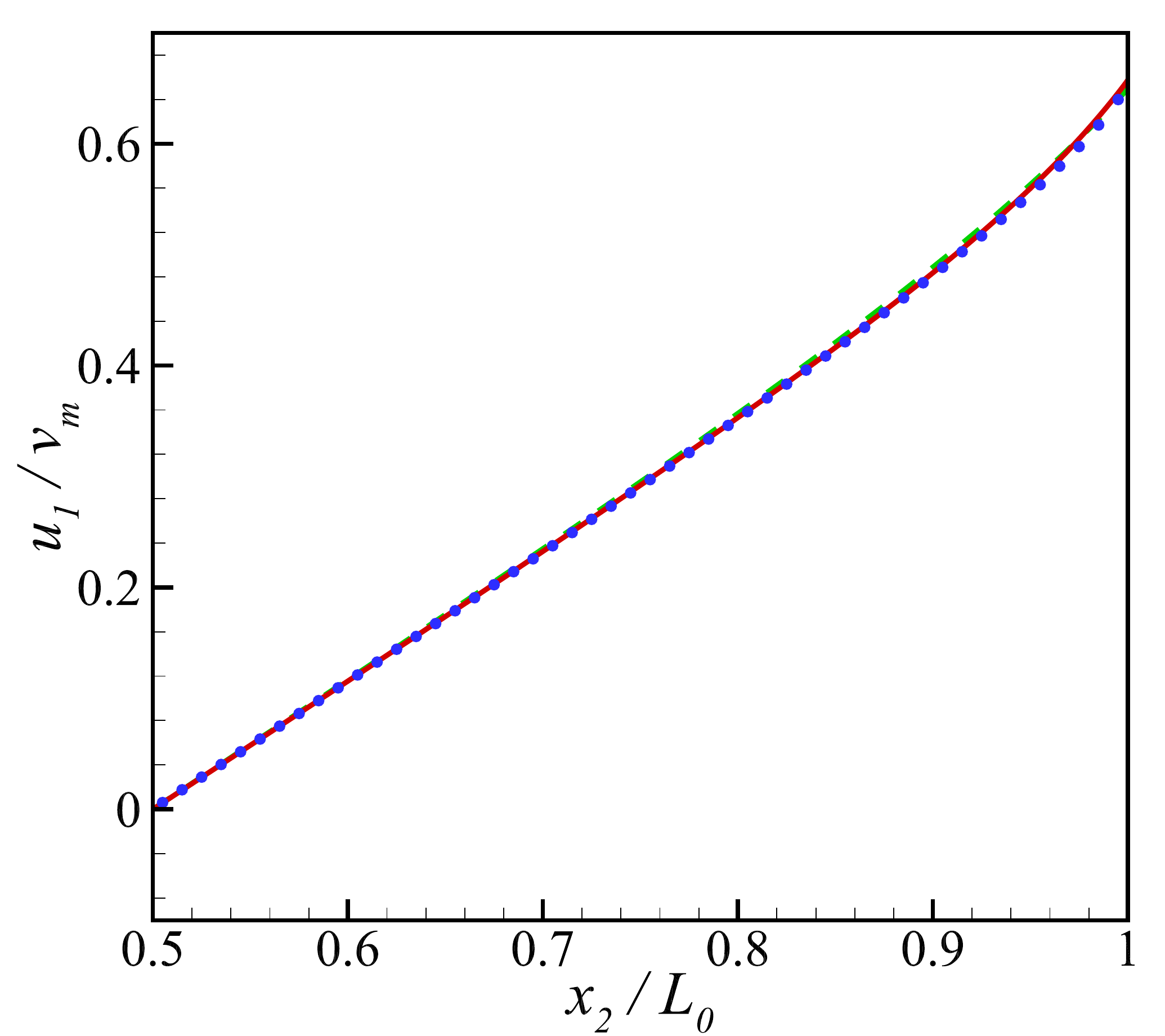}} \\
	\subfloat[]{\includegraphics[scale=0.24,clip=true]{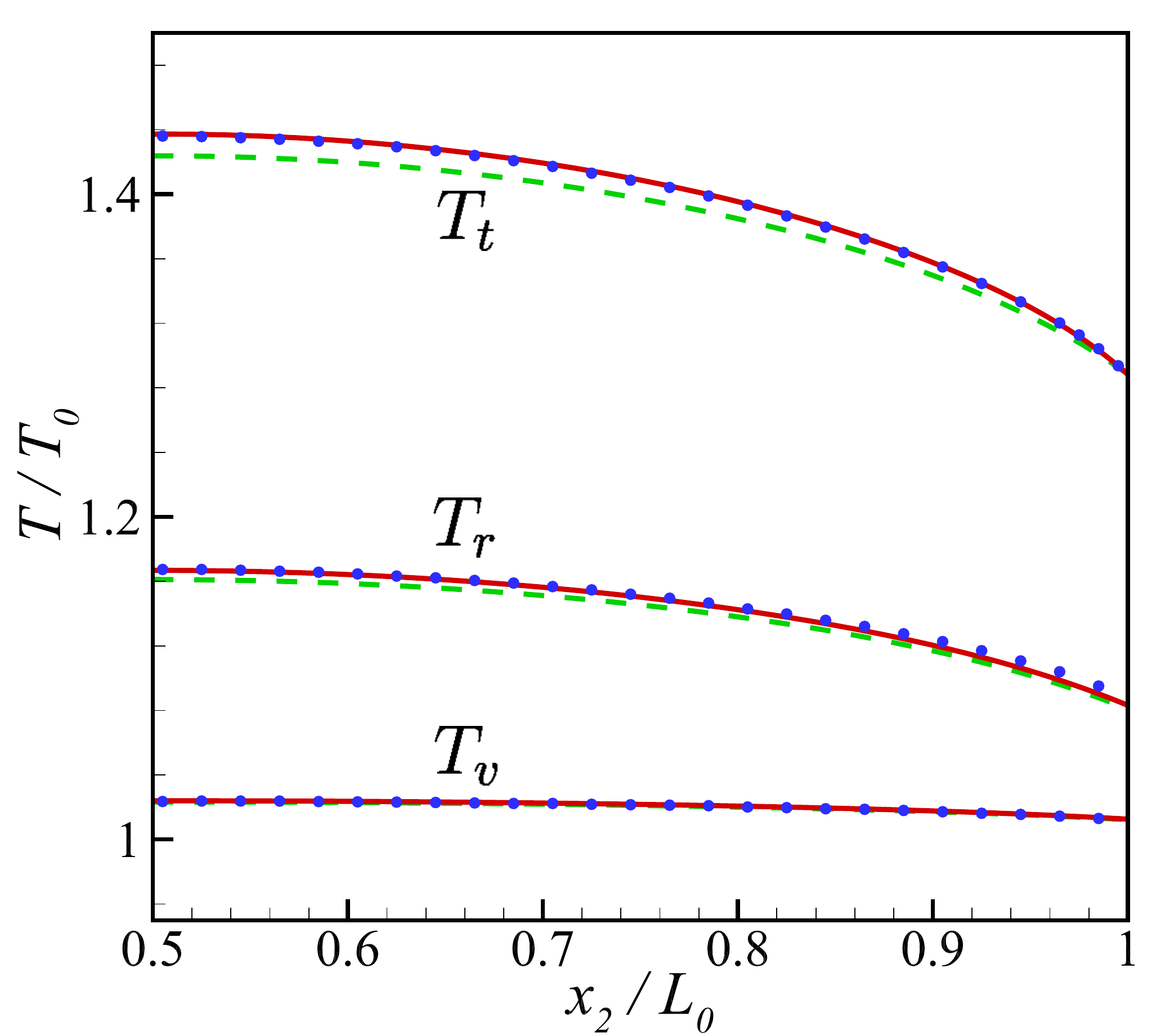}} \quad
	\subfloat[]{\includegraphics[scale=0.24,clip=true]{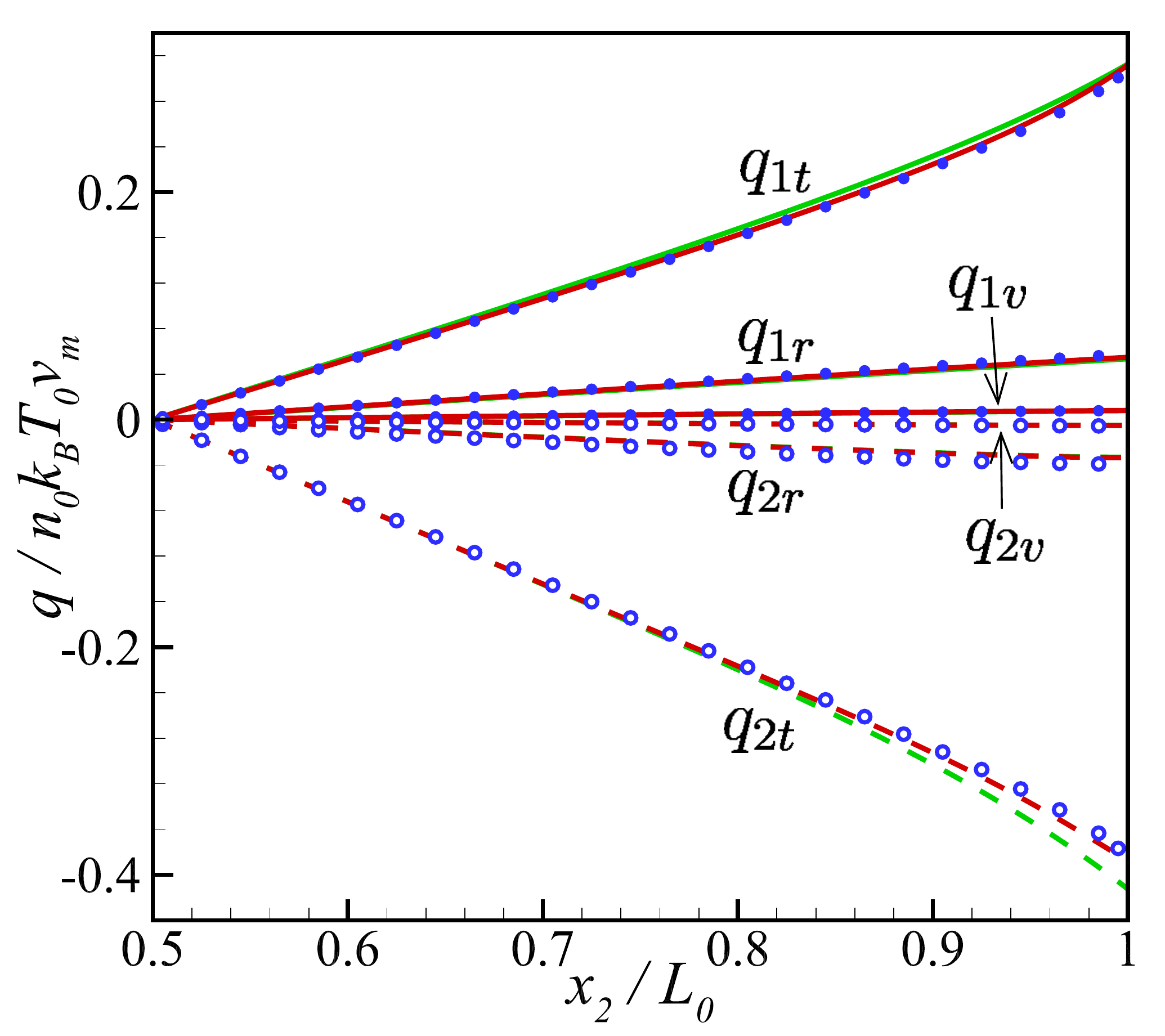}}
	\caption{Comparisons of the (a) density, (b) flow velocity, (c) temperature and (d) heat flux $q_1$ in the flow direction and $q_2$ perpendicular to flow direction of nitrogen, between kinetic model I, kinetic model II and DSMC for the one-dimensional Couette flow at $\text{Kn}_{\text{gas}}=0.5$.}
	\label{fig:1DCouetteFlow}
\end{figure}

\subsection{Couette flow}

The configuration of the Couette flow is the same as the Fourier flow, while the temperature of both plates are kept the same at $T_0$, and the velocity of lower and upper plates are $v_1=-v_m$ and $v_1=v_m$, respectively. Due to the symmetry, only half of the  domain $(L_0/2\le{}x_2\le{}L_0)$ is simulated. The diffuse boundary condition at $x_2=L_0$ yields:
\begin{equation}\label{eq:BC_CouetteFlow}
	v_2\le{}0: \quad 
	f_0=\frac{n_{in}(x_2=L_0)}{n_0}E_t(T_u), \quad
	f_1=\frac{d_r}{2}k_BT_0f_0, \quad
	f_2=\frac{d_v}{2}k_BT_0f_0,
\end{equation}
where $n_{in}(x_2=L_0)$ is determined as the same way as~\eqref{eq:BC_FourierFlow_n_in}, while the symmetrical condition at $x_2=L_0/2$ reads
\begin{equation}
	v_2\ge{}0: \quad 
	f_0=f_0(-v_1,-v_2,v_3), \quad
	f_1=\frac{d_r}{2}k_BTf_0, \quad
	f_2=\frac{d_v}{2}k_BTf_0. 
\end{equation}

The results from our kinetic models I and II, as well as the DSMC simulation at $\text{Kn}_{\text{gas}}=0.5$ are shown in figure~\ref{fig:1DCouetteFlow}. The kinetic model II with BCO shows the better accuracy, while the relative error given by kinetic model I in translational heat flux is around $7\%$. It can be seen that the vibrational temperature is much lower than the rotational one in the Couette flow, since the energy increase in internal DoF only comes from the exchange with translational ones in this problem. Thus, larger collision number leads to less increase in internal temperature at the same distance from the wall (due to the infrequent relaxation with the translational mode), and also contributes less to the heat flux.

\subsection{Creep flow driven by the Maxwell demon}

\begin{figure}[t]
	\centering
	\subfloat[]{\includegraphics[scale=0.24,clip=true]{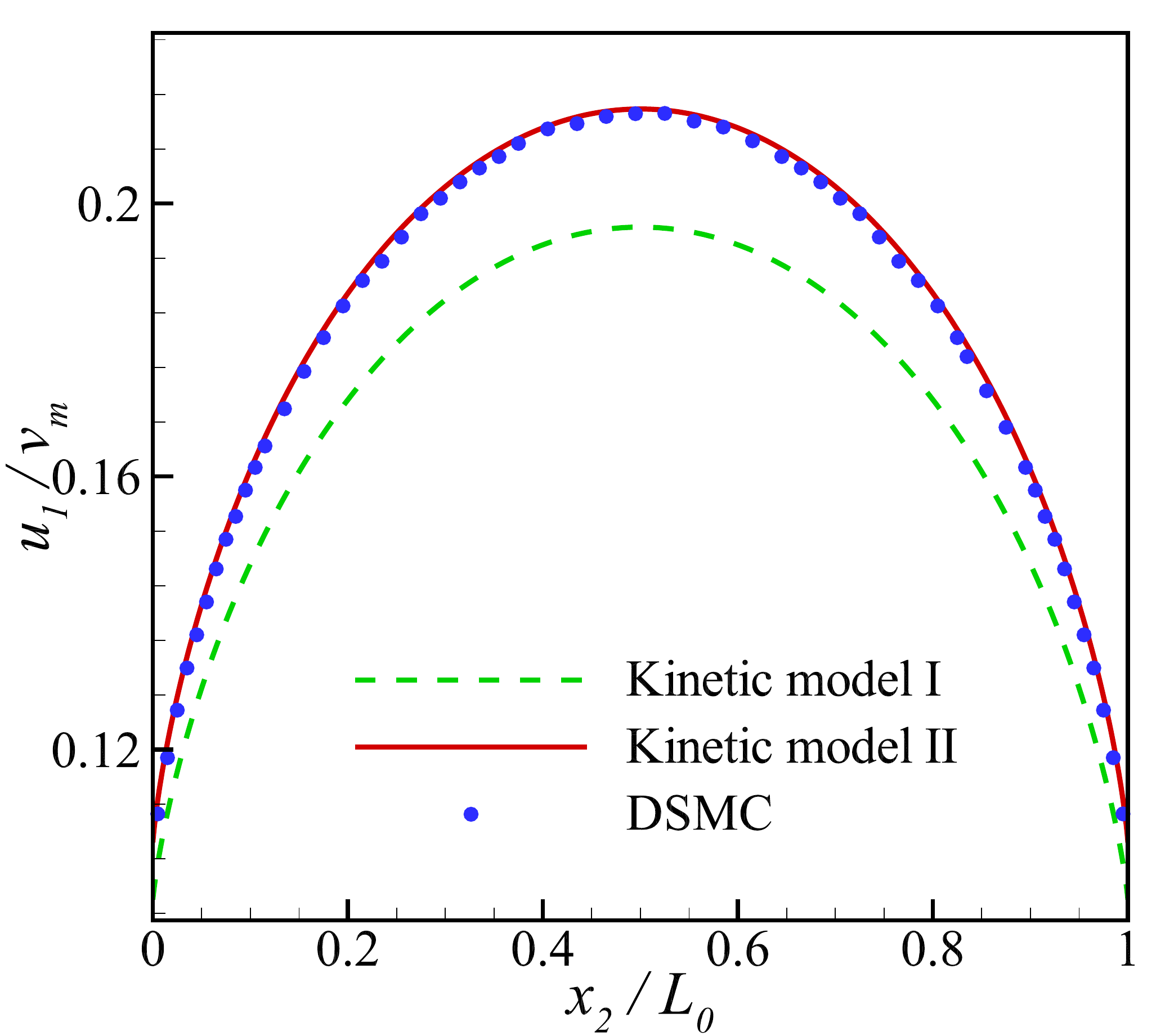}\label{fig:1DCreepFlow:a}} \quad
	\subfloat[]{\includegraphics[scale=0.24,clip=true]{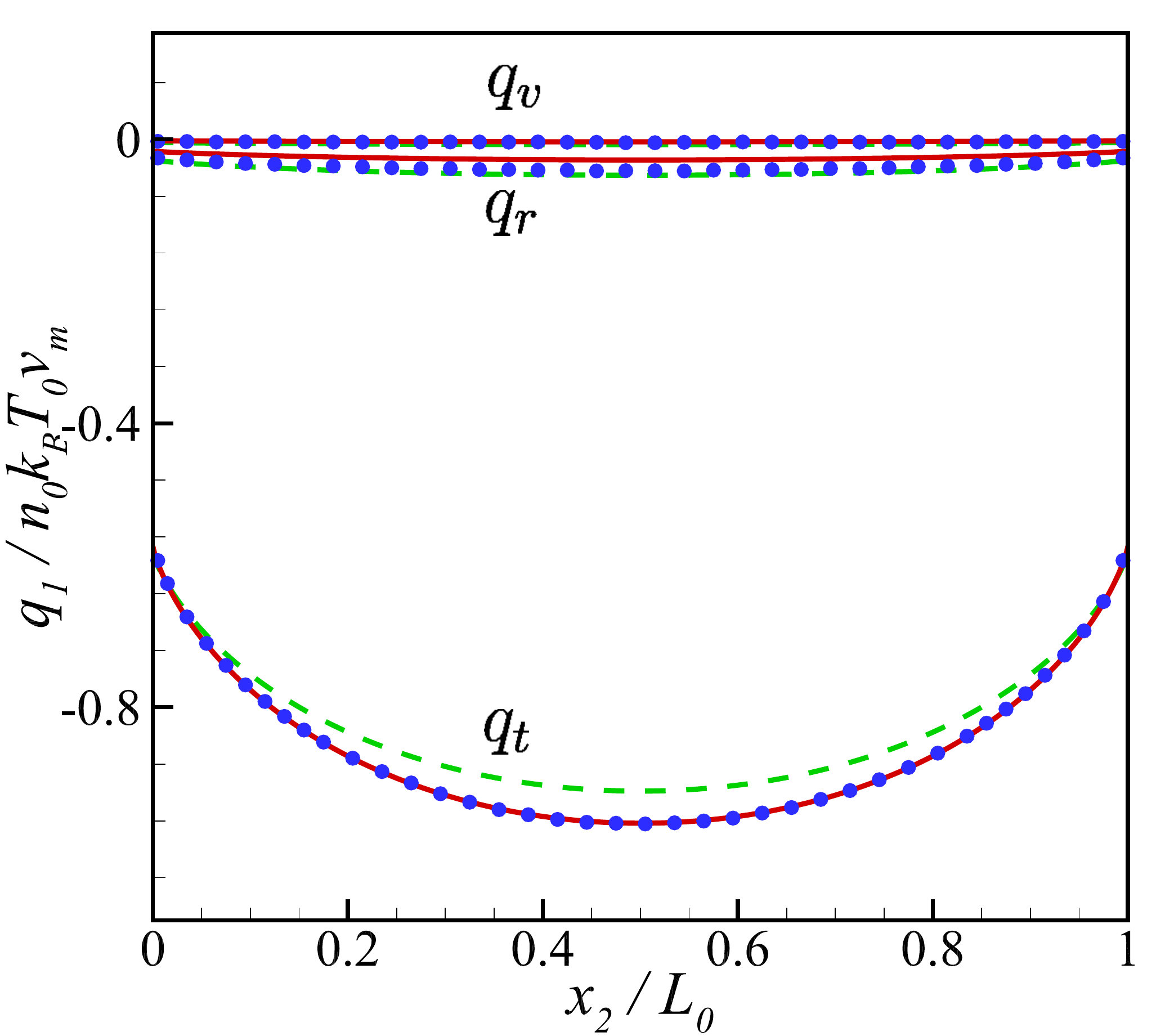}\label{fig:1DCreepFlow:b}} 
	\caption{Comparisons of the (a) velocity and (b) heat flux in flow direction of nitrogen between kinetic model I (green lines), kinetic model II (red lines) and DSMC (blue circles) for one-dimensional creep flow driven by the Maxwell demon at $\text{Kn}_{\text{gas}}=1$. Both the flow velocity and the heat flux have been further normalized by ${2a_0L_0}/{v_m^2}$.}
	\label{fig:1DCreepFlow}
\end{figure}

The creep flow driven by the Maxwell demon is a thought test~\citep{Li2021JFM}, where each gas molecule is subjected to an external acceleration based on its kinetic energy:
\begin{equation}\label{eq:CreepFlow_acceleration}
	a_1 = a_0\left(\frac{v_1^2}{v_m^2}-\frac{3}{2}\right).
\end{equation}
That is, fast molecules are forced towards the positive direction, while the slow molecules move in opposite direction. 

Consider the nitrogen flow driven by the Maxwell demon confined between two parallel plates with distance $L_0$ apart. To solve the force-driven flow, we choose small values of $a_0$ so that the gas flow deviates only slightly from the global equilibrium; the acceleration acting on the molecules is linearised, which results in,
\begin{equation}\label{eq:CreepFlow_equation}
	\begin{aligned}[b]
		\frac{\partial{(\bm{a}f_0)}}{\partial{\bm{v}}}=&~\frac{2a_0L_0}{v_m^2}v_1E_t(T_0)\left(\frac{v_1^2}{v_m^2}-\frac{5}{2}\right), \\
		\frac{\partial{(\bm{a}f_1)}}{\partial{\bm{v}}}=&~\frac{d_ra_0L_0}{v_m^2}v_1k_BT_0E_t(T_0)\left(\frac{v_1^2}{v_m^2}-\frac{5}{2}\right), \\
		\frac{\partial{(\bm{a}f_2)}}{\partial{\bm{v}}}=&~\frac{d_va_0L_0}{v_m^2}v_1k_BT_0E_t(T_0)\left(\frac{v_1^2}{v_m^2}-\frac{5}{2}\right).
	\end{aligned}
\end{equation}
The plates at rest are fully diffuse, then the boundary conditions are simply given by \eqref{eq:BC_FourierFlow} and \eqref{eq:BC_FourierFlow_n_in}, but with the wall temperature replaced by $T_0$. 

\begin{figure}[t]
	\centering
	\subfloat[]{\includegraphics[scale=0.24,clip=true]{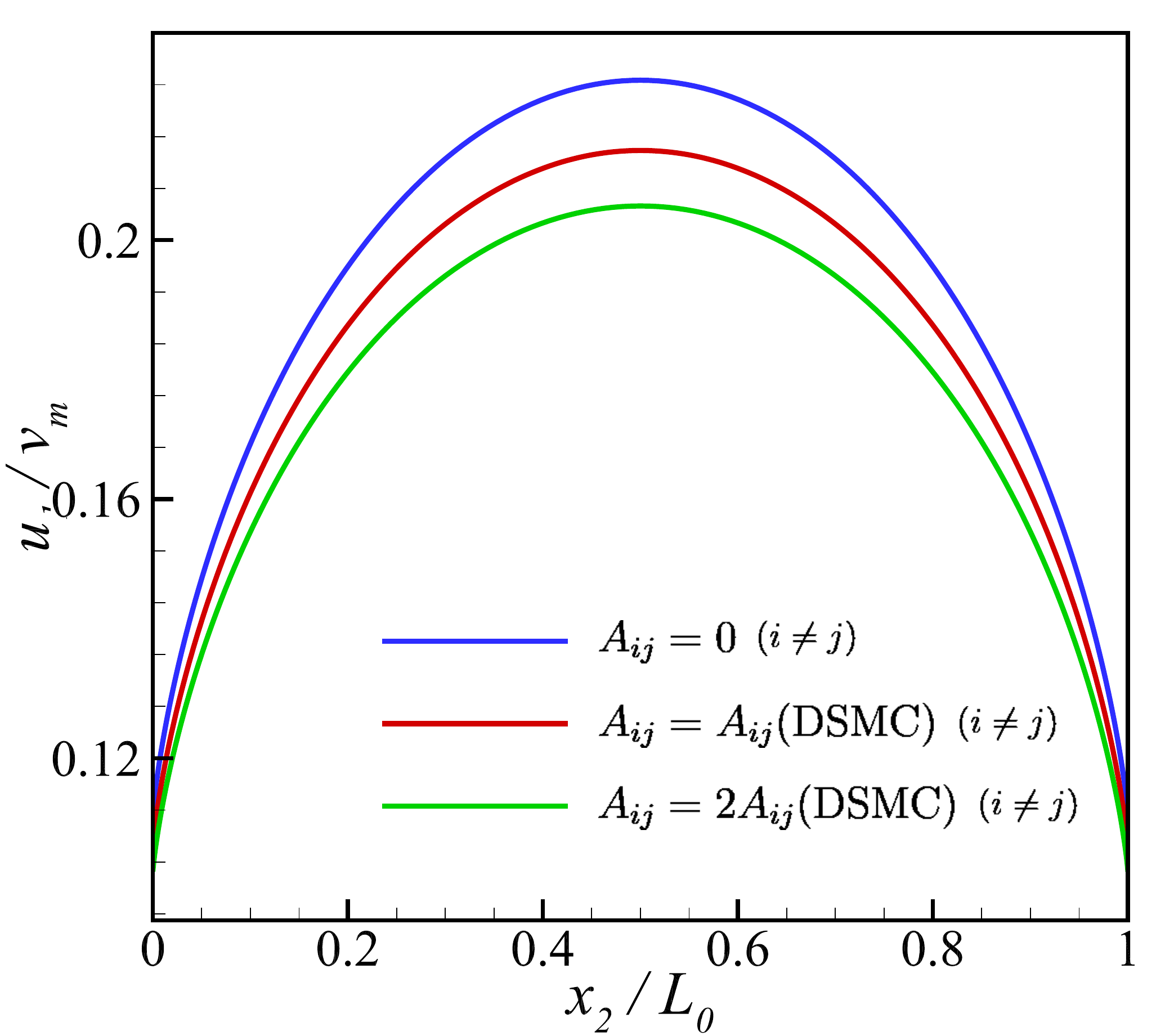}\label{fig:1DCreepFlow_Aij:a}} \quad
	\subfloat[]{\includegraphics[scale=0.24,clip=true]{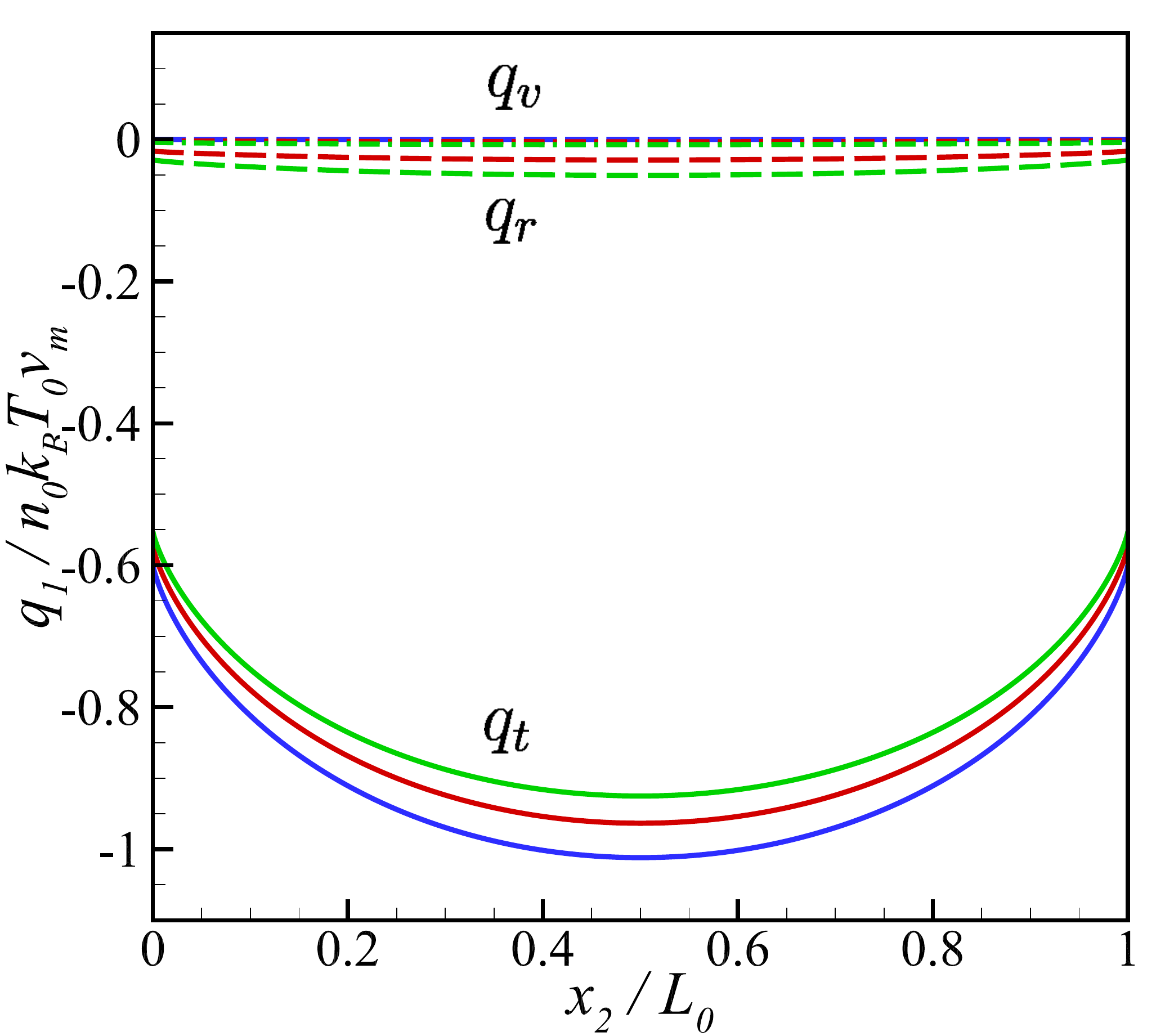}\label{fig:1DCreepFlow_Aij:b}} 
	\caption{Same as figure~\ref{fig:1DCreepFlow}, except that the off-diagonal elements in $A$ are set to be zero (blue), the values from DSMC (red) and double of those from DSMC (green), respectively.}
	\label{fig:1DCreepFlow_Aij}
\end{figure}

Figure \ref{fig:1DCreepFlow} shows the good agreement between the solutions of kinetic model II and DSMC at $\text{Kn}_{\text{gas}}=1$. The rotational/vibrational heat flux is one/two order of magnitude smaller than the translational heat flux, and hence makes negligible contribution to the total heat transfer in this problem. However, significant discrepancy are observed in both velocity and heat flux profiles predicted by kinetic model I. It demonstrates the importance to use correct velocity-dependent collision rates based on the intermolecular potential, as it is realized in kinetic model II.

To assess the influence of the thermal relaxation rates on the creep flow, two more cases are conducted using kinetic model II by varying the values of the matrix $A$ but keeping the Eucken factors fixed. More specifically, the off-diagonal elements in $A$ in the two cases are set to be zero and double of those given by DSMC, respectively. The values of diagonal elements are calculated based on \eqref{eq:EuckenFactor_A} using the fixed Eucken factors. Figure \ref{fig:1DCreepFlow_Aij} shows that these relaxation rates affect the  flow velocity and heat fluxes, despite that the thermal conductivities are fixed. In particular, when the off-diagonal elements in $A$ are zero, the heat fluxes of different types of DoF are decoupled, so that the internal heat fluxes are exactly zero. This situations occur in  many traditional kinetic models, such as the Rykov model and the ellipsoidal-statistical BGK model. This example demonstrates the importance of recovering the fundamental thermal relaxation process rather than the apparent thermal conductivities in rarefied gas flow simulations. 

\subsection{Normal shock wave}\label{subsec:validation_ShockWave}

\begin{figure}[t]
	\centering
	\subfloat[]{\includegraphics[scale=0.24,clip=true]{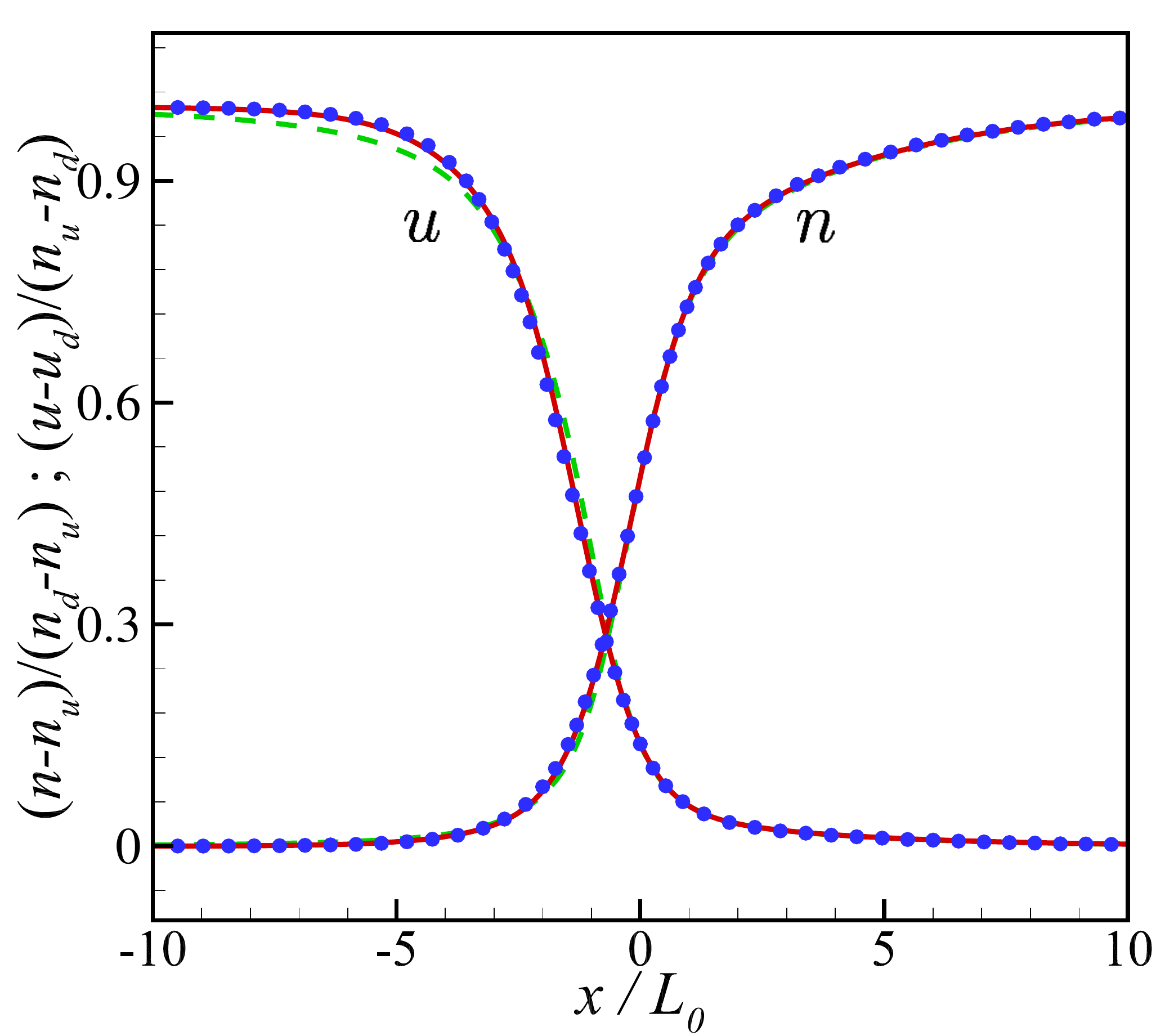}\label{fig:1DShockWave:a}} \quad
	\subfloat[]{\includegraphics[scale=0.24,clip=true]{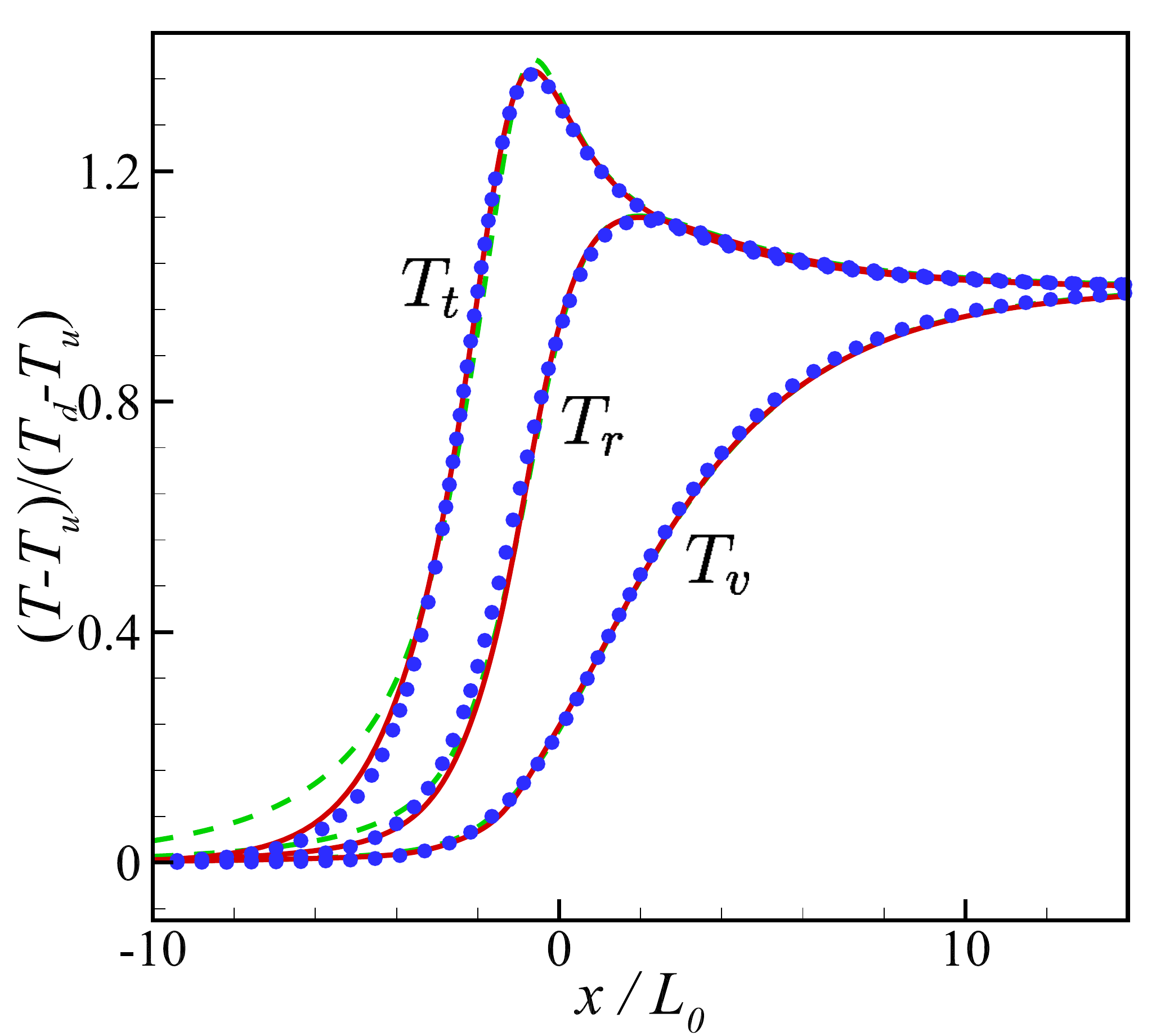}\label{fig:1DShockWave:b}} \\
	\subfloat[]{\includegraphics[scale=0.24,clip=true]{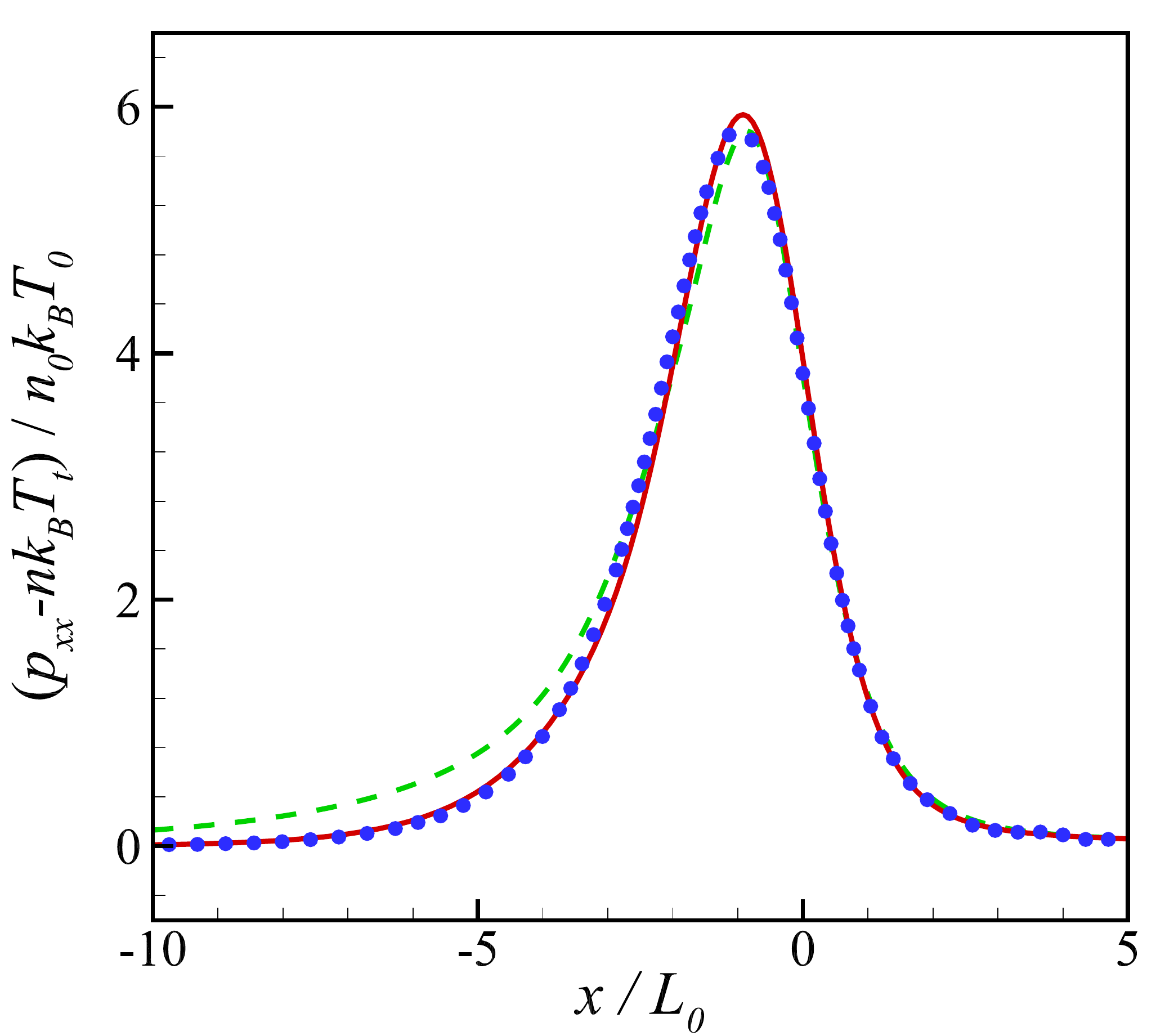}\label{fig:1DShockWave:c}} \quad
	\subfloat[]{\includegraphics[scale=0.24,clip=true]{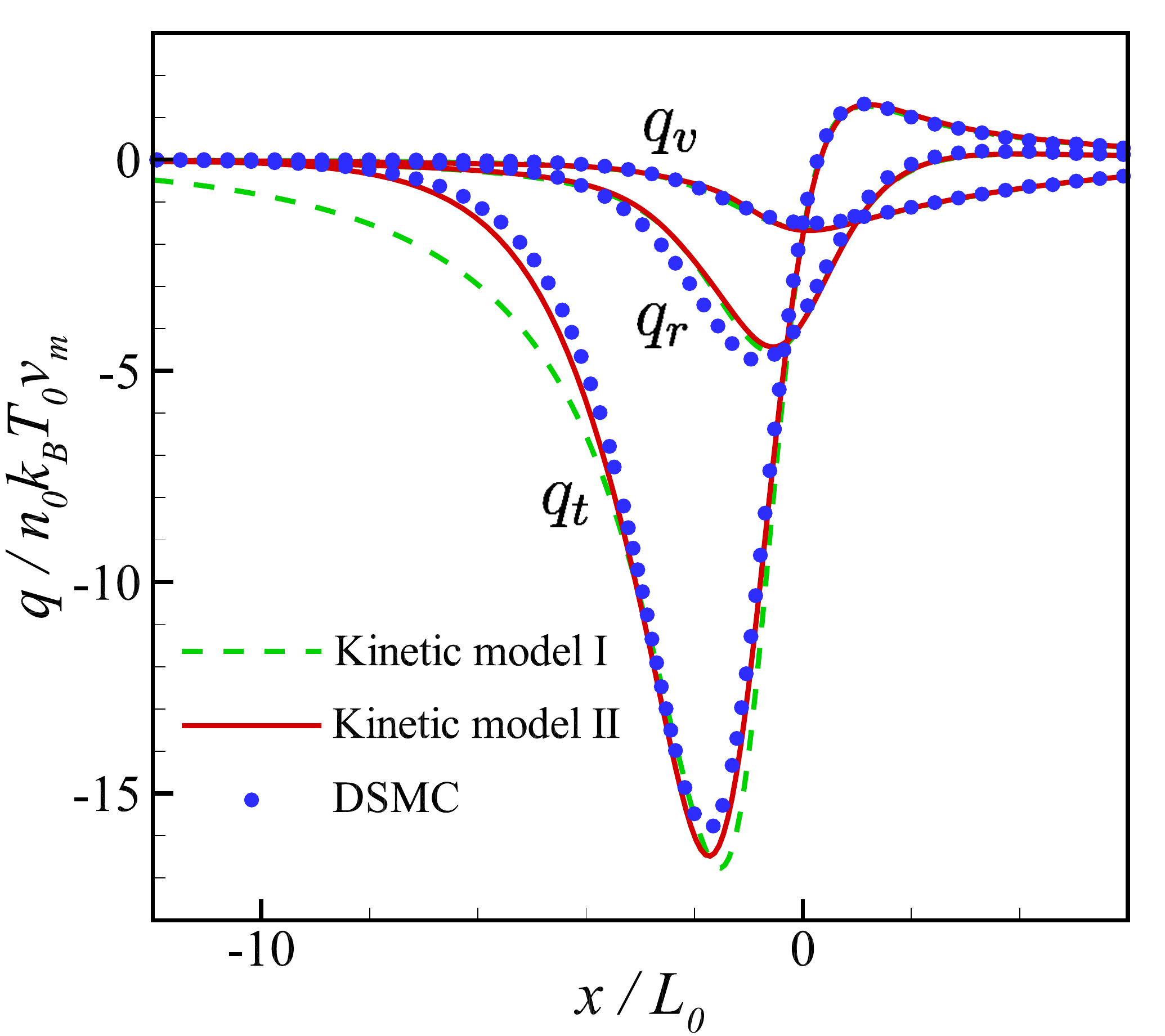}\label{fig:1DShockWave:d}}
	\caption{Comparisons of the (a) density and velocity, (b) temperature, (c) deviated pressure and (d) heat flux of nitrogen between kinetic model I (green lines), kinetic model II (red lines) and DSMC (blue circles) for normal shock wave at $\text{Ma}=5$.}
	\label{fig:1DShockWave}
\end{figure}

In the simulations of normal shock wave of nitrogen, the upstream number density $n_u=n_0$ and temperature $T_u=T_0$ are chosen to be the reference values, which also determine the characteristic length to be $L_0={16\mu(T_0)}/({5n_0\sqrt{2\pi mk_BT_0}})$ and hence $\text{Kn}_{\text{gas}}={5\pi}/{16}$ in this problem. The total length of the simulation domain is $90L_0$, so that the boundary conditions at both ends can be approximated by equilibrium states (the wave front is initially located at $x=0$):
\begin{equation}\label{eq:BC_ShockWave}
	\begin{aligned}[b]
		&x=-30L_0,~v\ge{}0: \quad f_0=\frac{n_{u}}{n_0}E_t(T_u), ~f_1=\frac{d_r}{2}k_BT_uf_0, ~f_2=\frac{d_v}{2}k_BT_uf_0, \\
		&x=60L_0,~v\le{}0: \quad ~~f_0=\frac{n_{d}}{n_0}E_t(T_d), ~f_1=\frac{d_r}{2}k_BT_df_0, ~f_2=\frac{d_v}{2}k_BT_df_0,
	\end{aligned}
\end{equation}
where the subscripts $u, d$ represent the upstream and downstream end, respectively. Given the Mach number, macroscopic quantities at the downstream end are determined by the Rankine-Hugoniot relation.

Numerical results of both the kinetic model I, II and DSMC are compared in figure \ref{fig:1DShockWave}, when the Mach number is $\text{Ma}=5$. As expected, the kinetic model equations II reproduce the structure of normal shock wave with high accuracy. While kinetic model II significantly overestimates the temperature, heat flux and deviated pressure before the wave front, because the velocity dependence of the collision frequency are not involved in the RTA. The rotational and vibrational collision numbers, $Z_r$ and $Z_v$, which affect energy exchange rate between internal and translational modes, play roles in the  difference of rotational and vibrational temperatures. That is, the distance for vibrational temperature to reach equilibrium are much longer than that for rotational modes. This is consistent with the fact that we set $Z_v=10Z_r$.

\section{Rarefied molecular gas flow with radiation}\label{sec:radiation}

The influence of thermal radiation to the total heat transfer is investigated by solving kinetic model equation with BCO. Three typical problems are considered: one-dimensional Fourier flow, Couette flow and normal shock wave in nitrogen flow. The vibrational DoF change with temperature as equation \eqref{eq:harmonic_oscillator_dv}, and the effective absorptivity in photon gray model $k_{\text{gray}}$ is assumed to be constant for simplicity. The transport coefficients and the relaxation rates of gas are fixed as those used in \S\ref{sec:validation}, while the parameters $\text{Kn}_{\text{photon}}$ and $\tilde{\sigma}_R$ are varied to investigate the importance of radiation in rarefied molecular gas flows. 

\subsection{Fourier flow}

Consider the heat transfer between two plates maintained at different temperature $T_1=0.5T_0$ and $T_2=1.5T_0$, which emit photons with Planck distribution and reflect gas molecules diffusely. The Knudsen numbers of gas is $\text{Kn}_{\text{gas}}=0.1$, $\text{Kn}_{\text{photon}}$ varies from 0.1 to 100, $\tilde{\sigma}_R$ changes from 0.001 to 0.1, and $T_0/T_{\text{ref}}=2$. 

\begin{figure}[t]
	\centering
	\subfloat[]{\includegraphics[scale=0.21,clip=true]{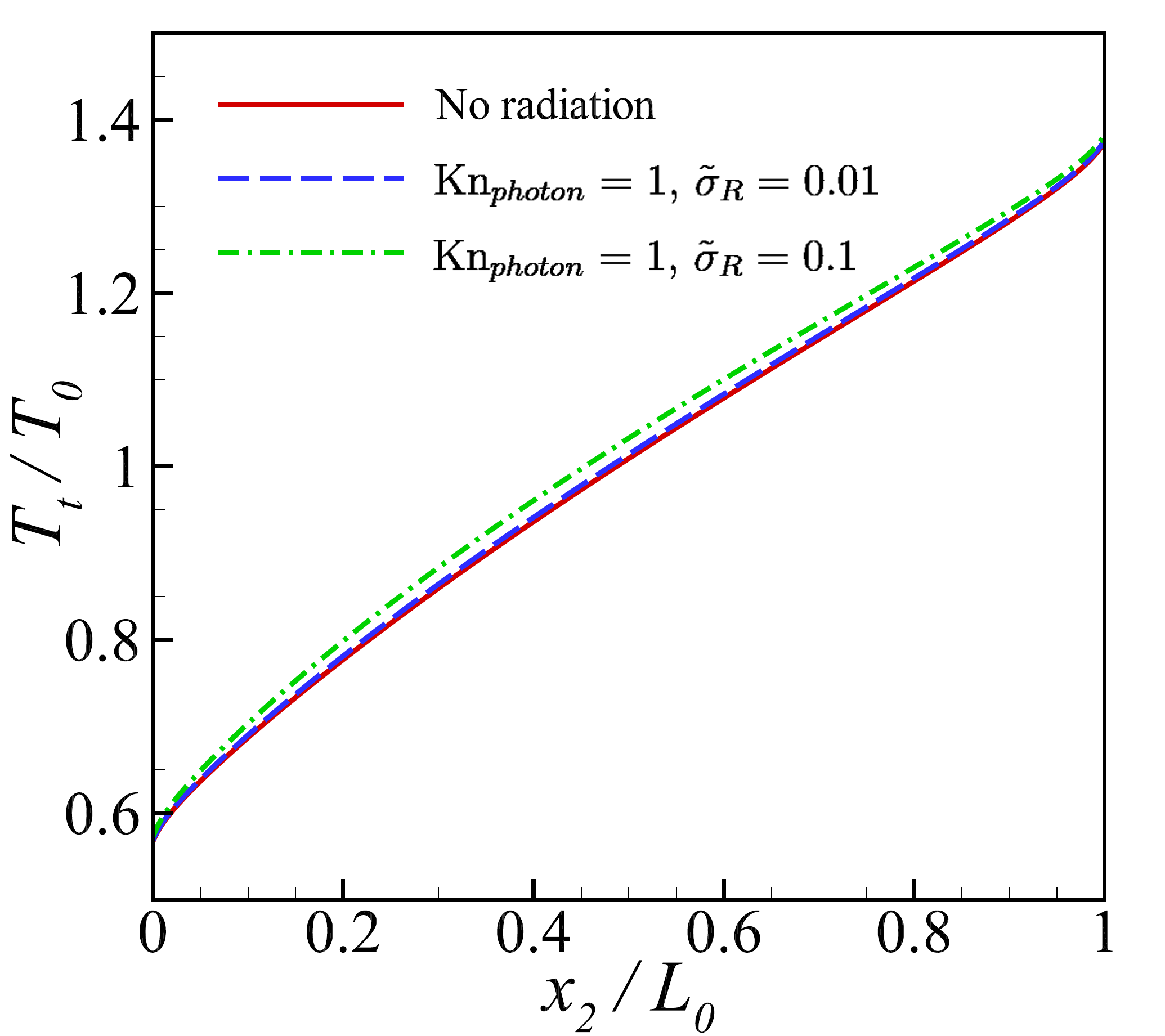}\label{fig:RadiativeFourierFlow:a}} 
	\subfloat[]{\includegraphics[scale=0.21,clip=true]{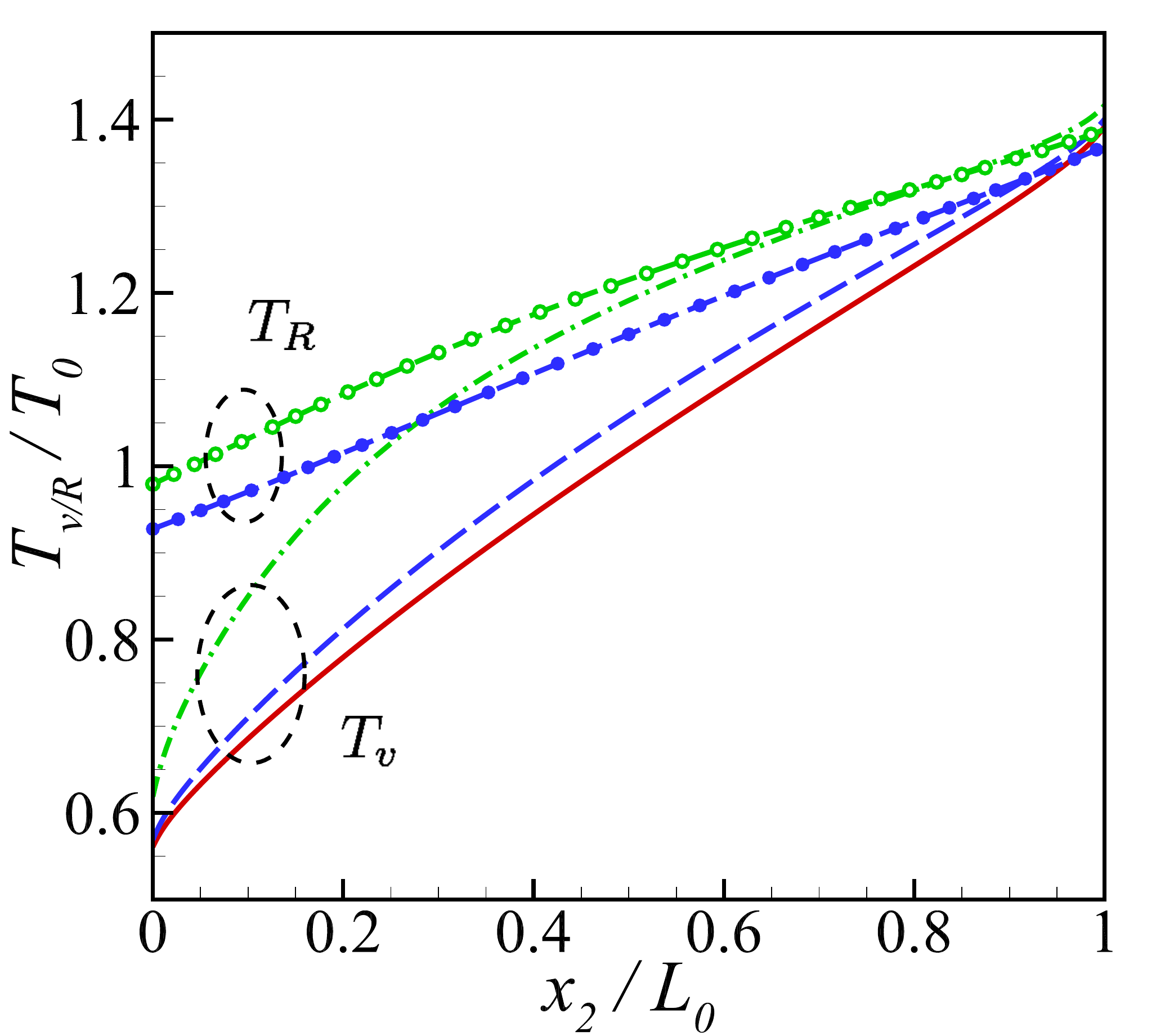}\label{fig:RadiativeFourierFlow:b}} 
	\subfloat[]{\includegraphics[scale=0.21,clip=true]{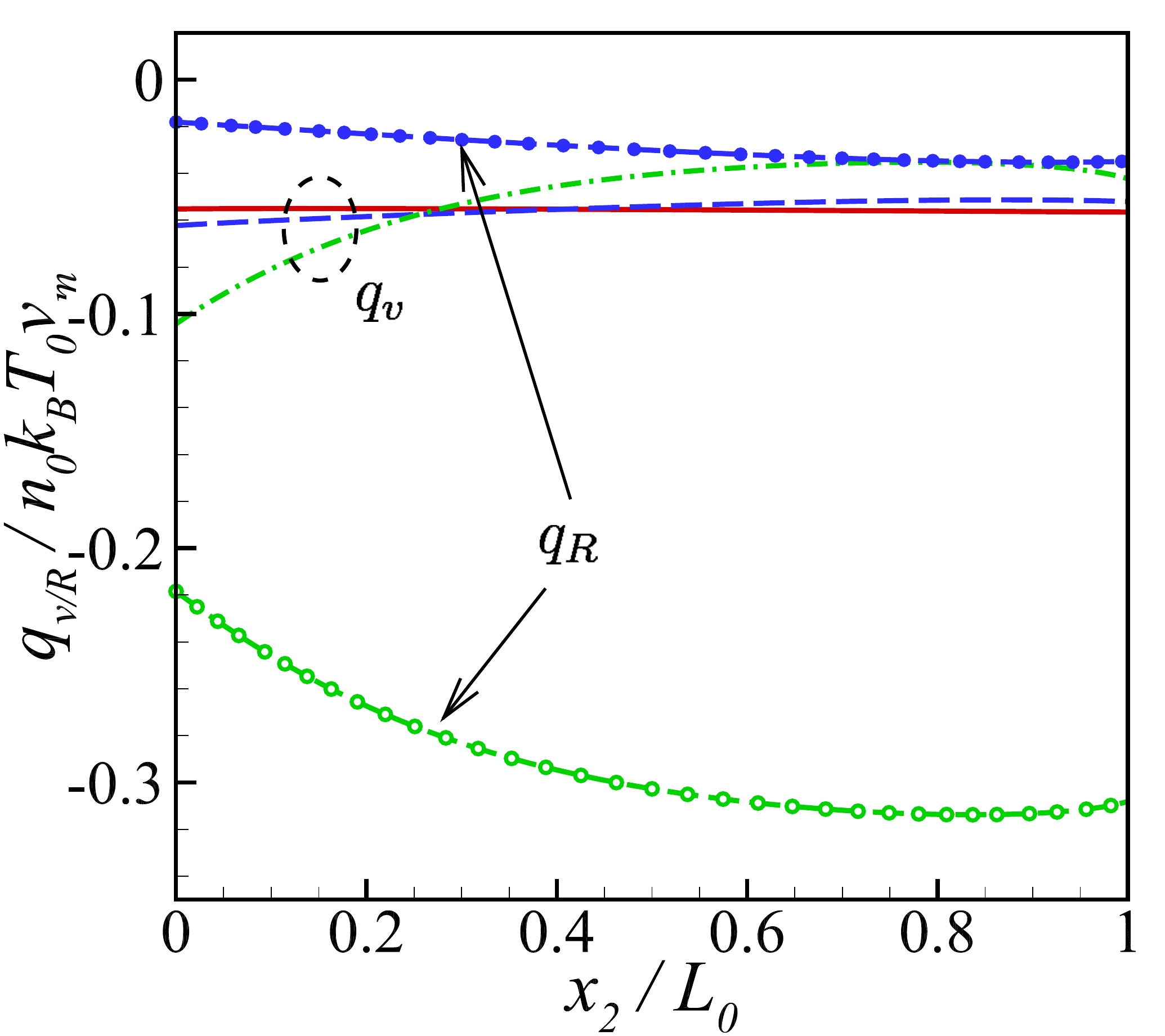}\label{fig:RadiativeFourierFlow:d}}
	\caption{Comparisons of the (a) translational temperature, (b) vibrational and radiative temperature, (c) vibrational and radiative heat flux of nitrogen in Fourier flow between different radiation strength, when $\text{Kn}_{\text{gas}}=0.1$ and $T_0/T_{\text{ref}}=2$. The results are obtained from kinetic model II.}
	\label{fig:RadiativeFourierFlow}
\end{figure}

\begin{figure}[h]
	\centering
	\subfloat[]{\includegraphics[scale=0.33,clip=true]{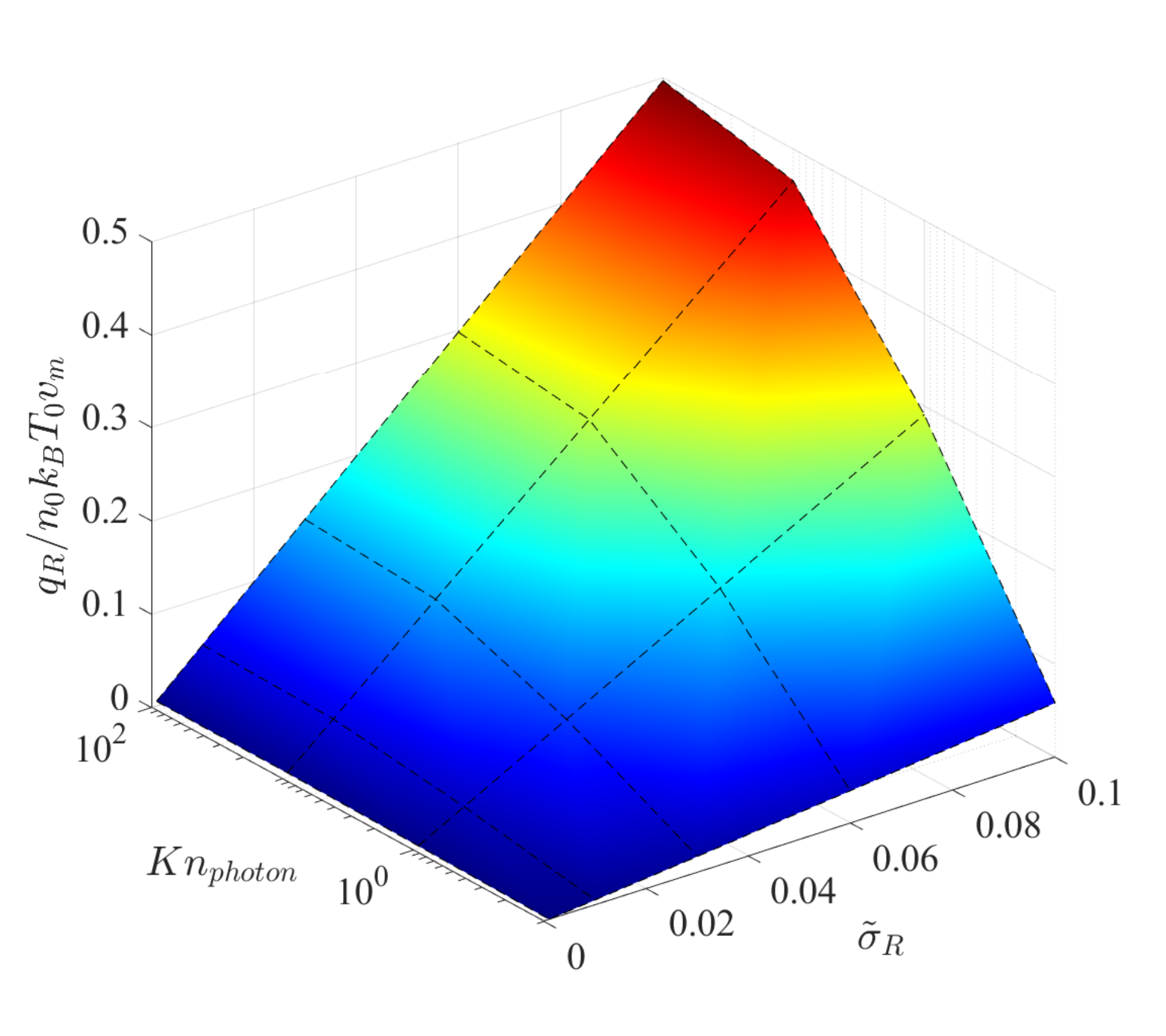}\label{fig:RadiativeFourierFlow_q:a}} 
	\subfloat[]{\includegraphics[scale=0.33,clip=true]{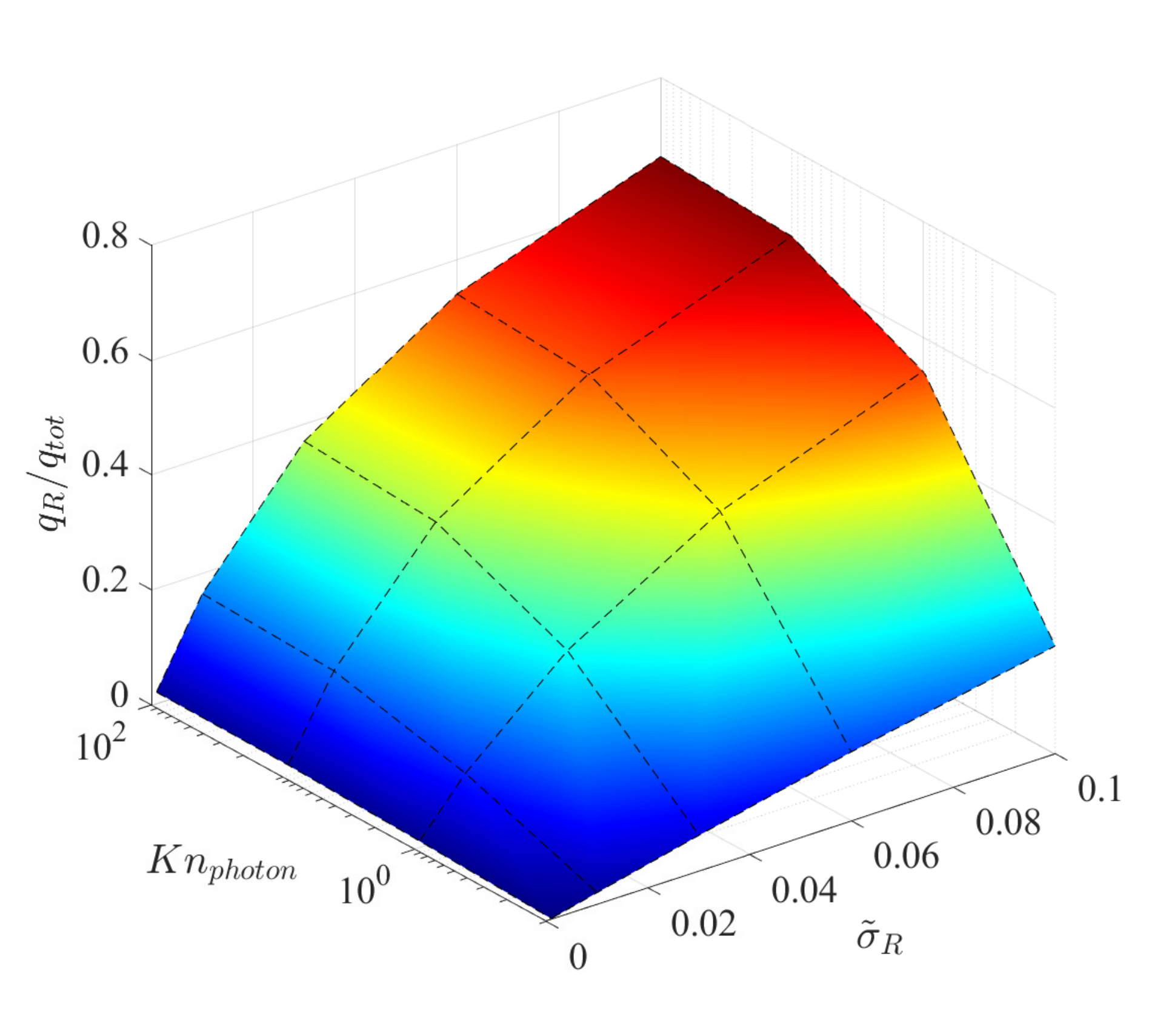}\label{fig:RadiativeFourierFlow_q:b}}
	\caption{The radiative heat flux change with $\text{Kn}_{\text{photon}}$ and $\tilde{\sigma}_R$ in the Fourier flow, when $\text{Kn}_{\text{gas}}=0.1$ and $T_0/T_{\text{ref}}=2$, (a) radiative heat flux, (b) ratio of radiative heat flux to the total one.}
	\label{fig:RadiativeFourierFlow_q}
\end{figure}

It can be seen from figure \ref{fig:RadiativeFourierFlow}, the radiation significantly changes the profiles of vibrational temperature, while its influence on translational temperature is relatively small. The radiative heat flux is calculated and compared with the total heat flux, as shown in figure \ref{fig:RadiativeFourierFlow_q}. Firstly, since $\tilde{\sigma}_R$ indicates the relative strength of radiative intensity in terms of characteristic heat flux of gas flow, the radiative heat flux $q_R$ and its proportion to the total one increase with $\tilde{\sigma}_R$. Secondly, When $\text{Kn}_{\text{photon}}\rightarrow\infty$, no interactions between gas molecules and photons, the gas flow is transparent to radiation, and radiative heat flux is linearly proportional to $\tilde{\sigma}_R$. Thirdly, when $\text{Kn}_{\text{photon}}\rightarrow 0$, the transport distance of radiative energy is extremely short, which means that the photons emitted by vibrational transition are absorbed immediately. Consequently, the energy transported by radiation field becomes negligible. Therefore, in the Fourier flow, the radiative heat transport is important and even dominated when both $\tilde{\sigma}_R$ and $\text{Kn}_{\text{photon}}$ are large. The relative difference between the radiative heat flux from kinetic model I and II are less than $0.1\%$ in these cases.

\subsection{Couette flow}

For the Couette flow between two plates with the same temperature $T_0$, the velocity of lower and upper plates are $v_1=-v_m$ and $v_1=v_m$, respectively. Two vibrational collision numbers are considered $Z_v=2Z_r$ and $Z_v=10Z_r$ here. The Knudsen numbers of gas is $\text{Kn}_{\text{gas}}=0.1$, $\text{Kn}_{\text{photon}}$ varies from 0.1 to 100, $\tilde{\sigma}_R$ changes from 0.01 to 0.1, and $T_0/T_{\text{ref}}=2$.

\begin{figure}[t]
	\centering
	\subfloat[]{\includegraphics[scale=0.24,clip=true]{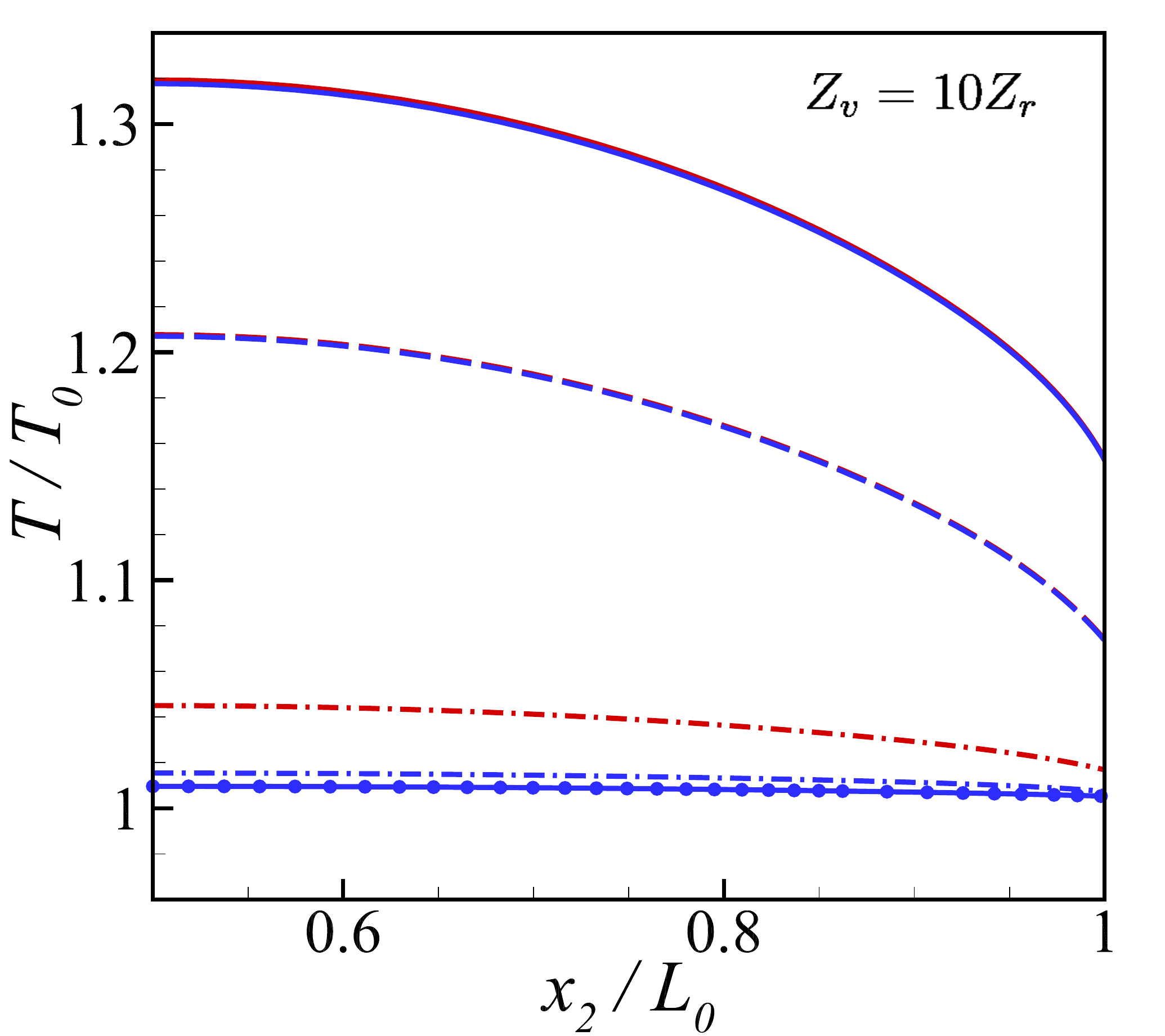}\label{fig:RadiativeCouetteFlow:a}} \quad
	\subfloat[]{\includegraphics[scale=0.24,clip=true]{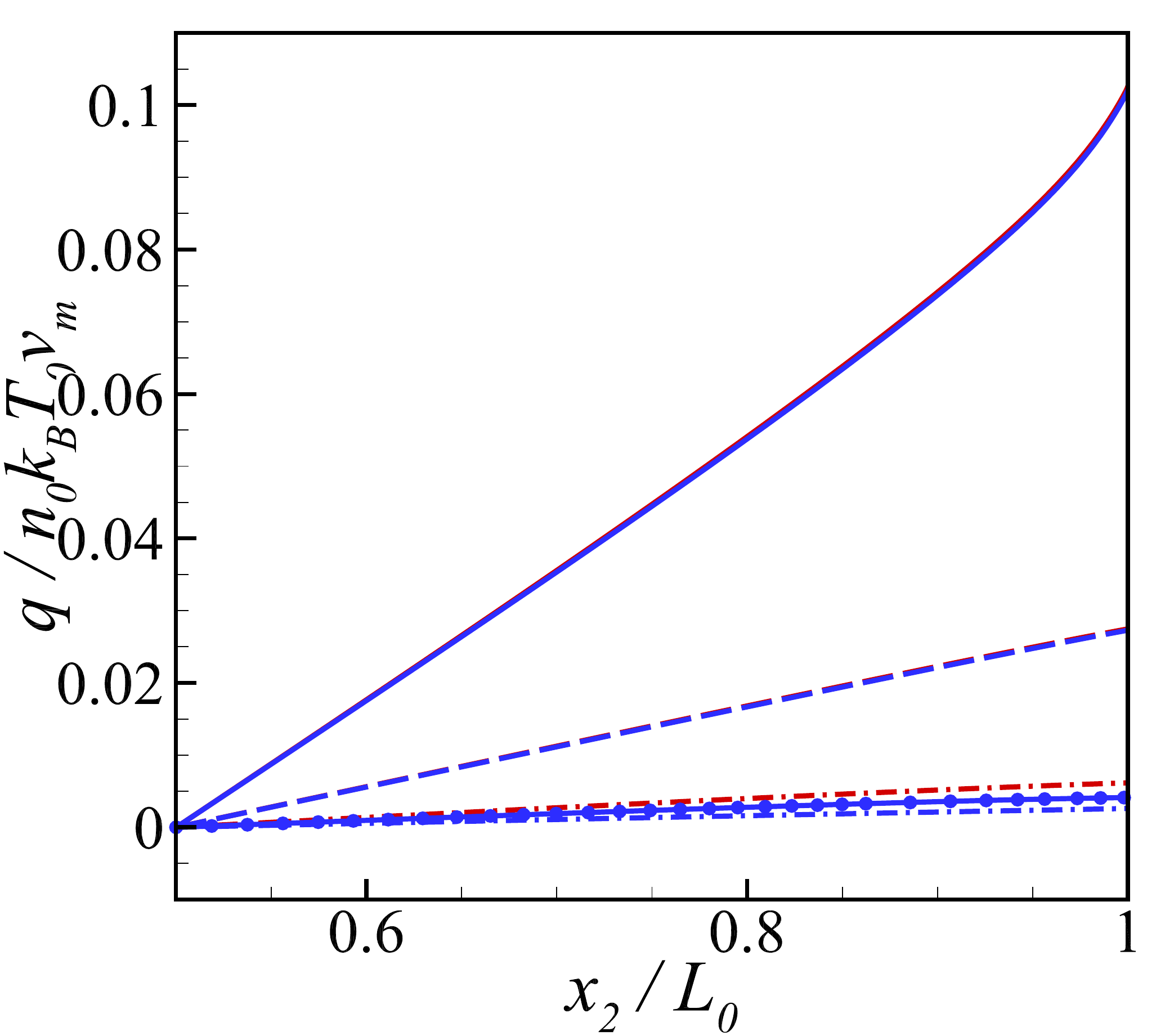}\label{fig:RadiativeCouetteFlow:b}} \\
	\subfloat[]{\includegraphics[scale=0.24,clip=true]{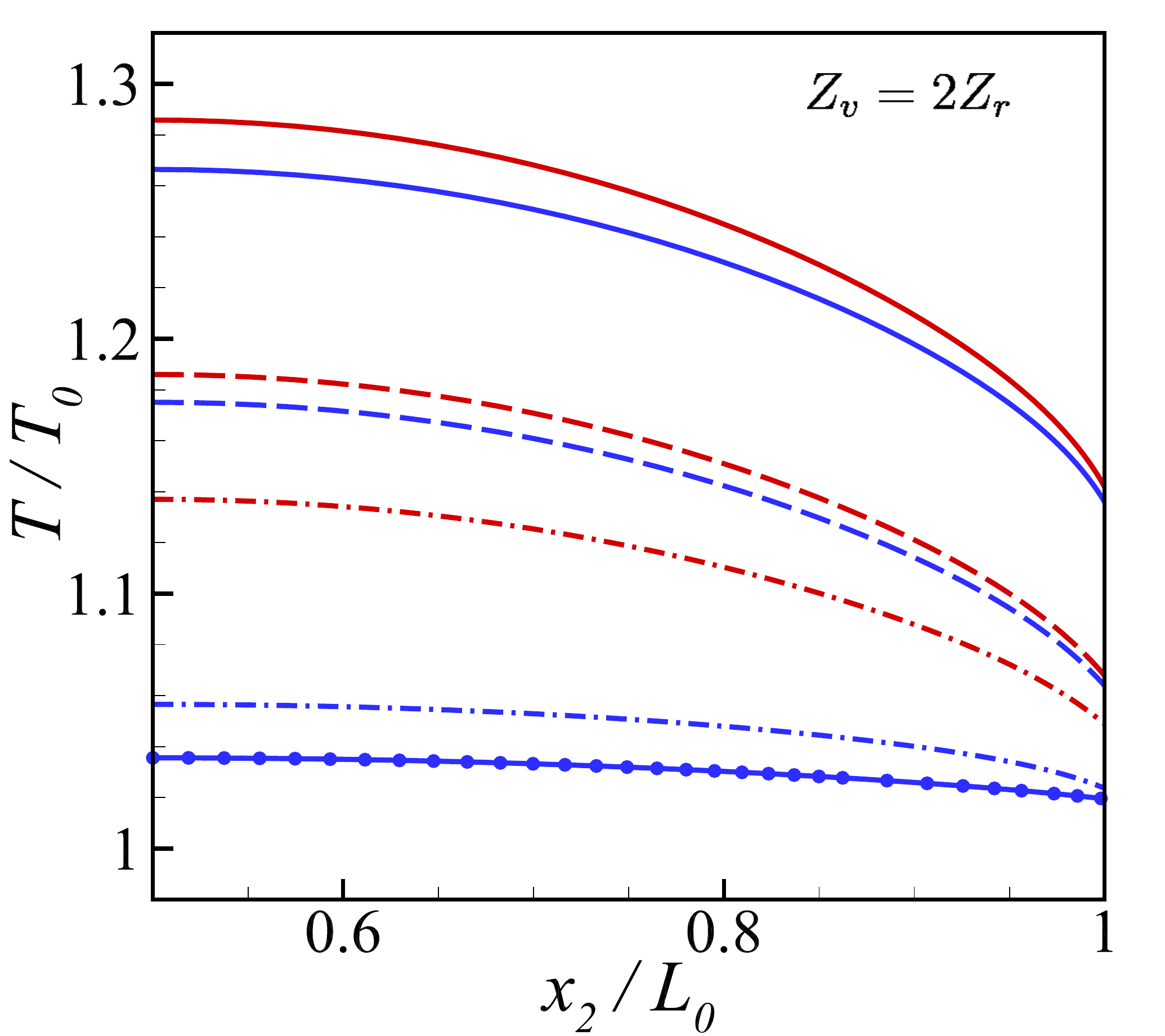}\label{fig:RadiativeCouetteFlow:c}} \quad
	\subfloat[]{\includegraphics[scale=0.24,clip=true]{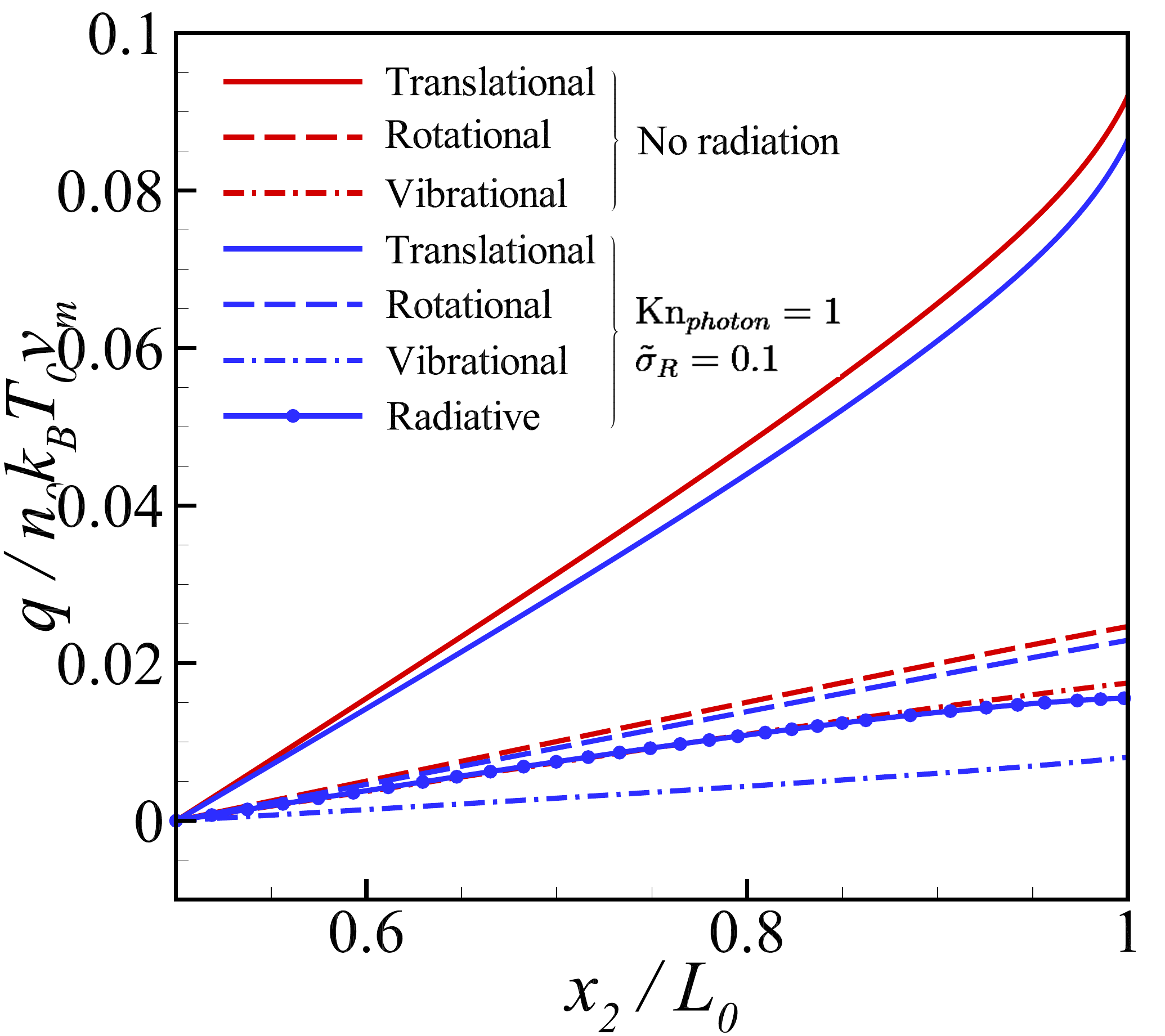}\label{fig:RadiativeCouetteFlow:d}}
	\caption{Comparisons of the (a,c) temperature, (b,d) heat flux of nitrogen Couette flow, when $\text{Kn}_{\text{photon}}=1$, $\tilde{\sigma}_R=0.1$, $\text{Kn}_{\text{gas}}=0.1$ and $T_0/T_{\text{ref}}=2$; (a,b) $Z_v=10Z_r$, (c,d) $Z_v=2Z_r$.}
	\label{fig:RadiativeCouetteFlow}
\end{figure}

\begin{figure}[t]
	\centering
	\subfloat[]{\includegraphics[scale=0.33,clip=true]{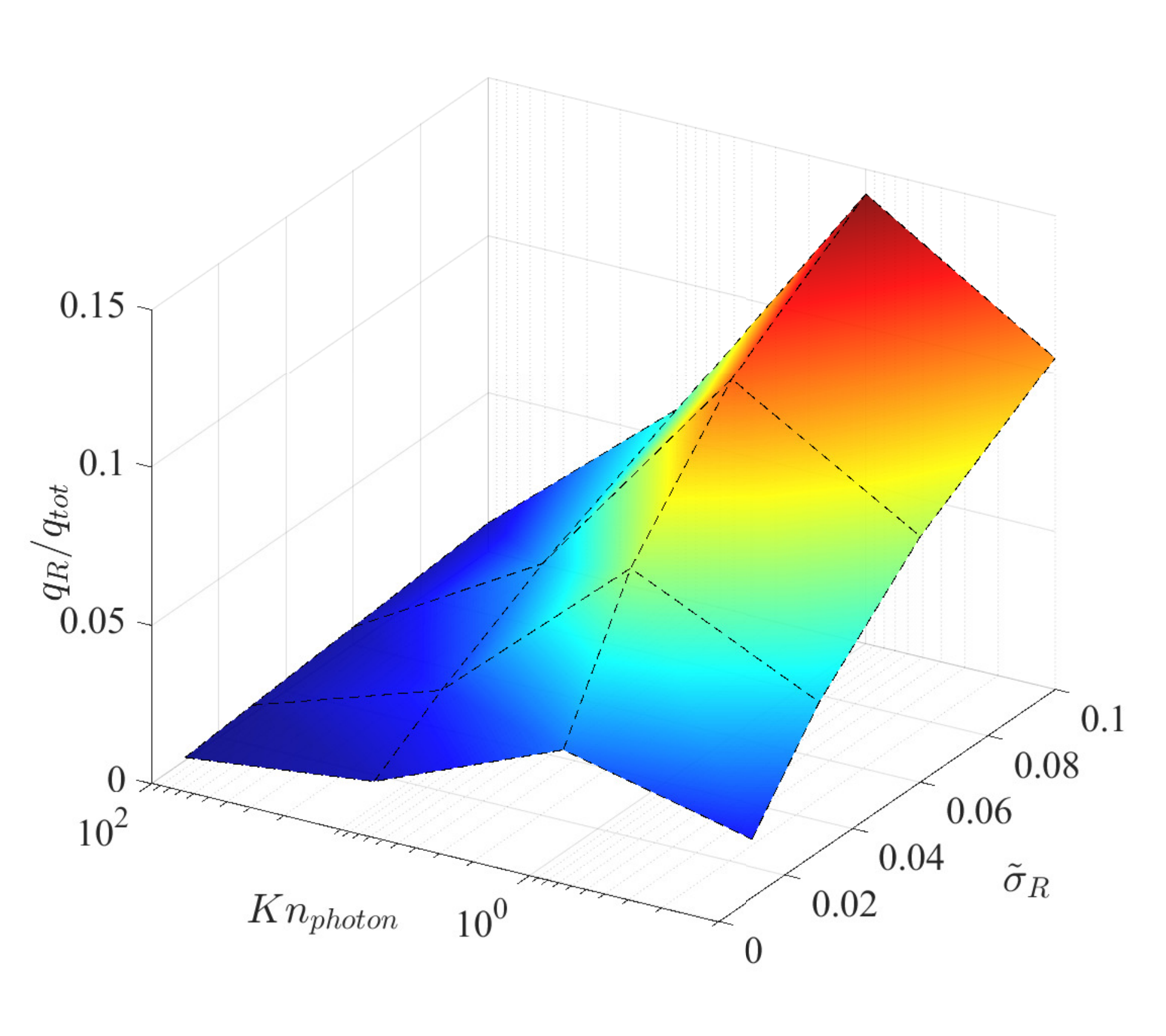}\label{fig:RadiativeCouetterFlow_q:a}} 
	\subfloat[]{\includegraphics[scale=0.33,clip=true]{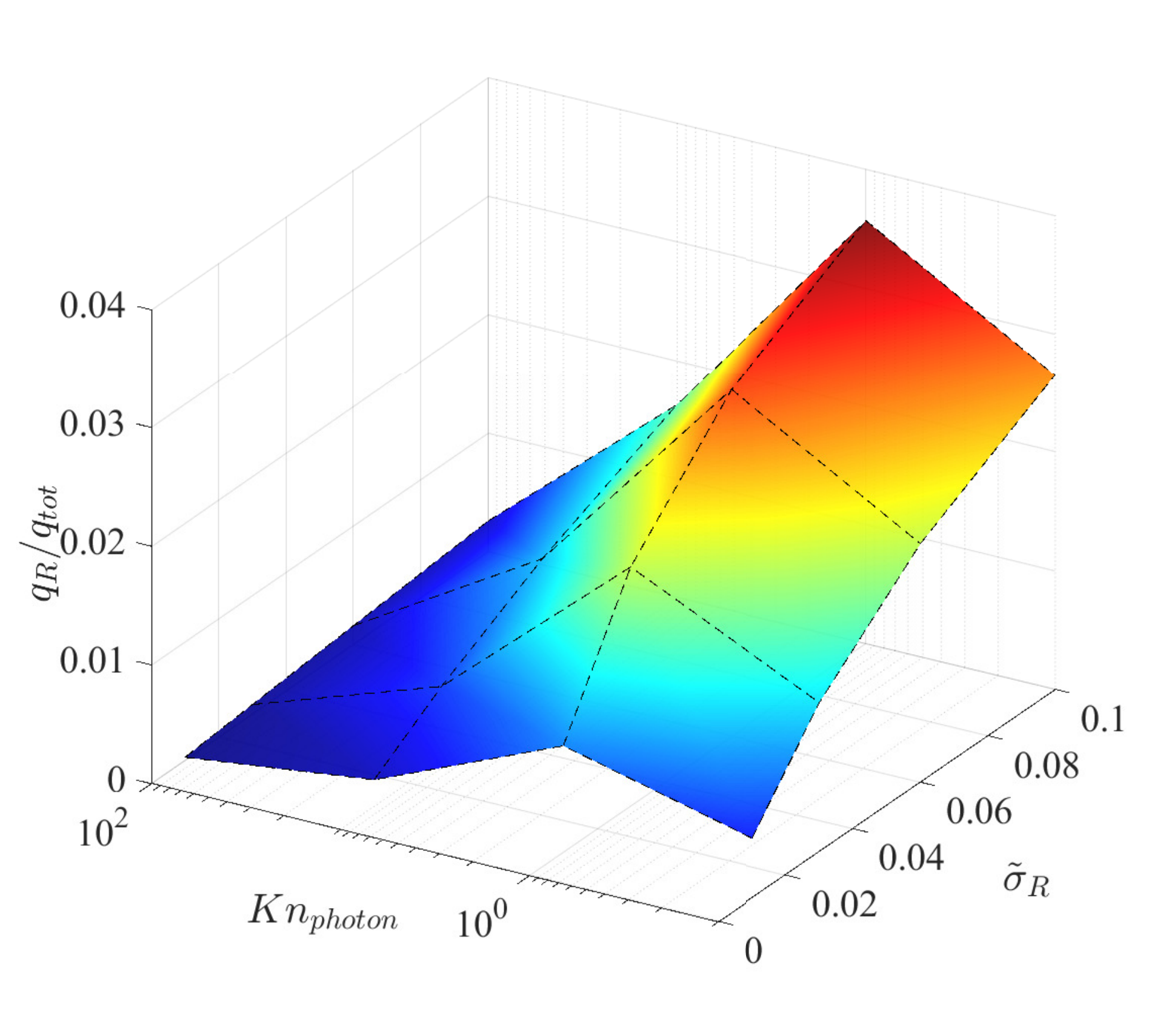}\label{fig:RadiativeCouetterFlow_q:b}}
	\caption{The ratio of radiative heat flux to the total heat flux change with $\text{Kn}_{\text{photon}}$ and $\tilde{\sigma}_R$ in the Couette flow, when $\text{Kn}_{\text{gas}}=0.1$ and $T_0/T_{\text{ref}}=2$; (a) $Z_v=2Z_r$, (b) $Z_v=10Z_r$.}
	\label{fig:RadiativeCouetterFlow_q}
\end{figure}

Due to the symmetry, only half of the  domain $(L_0/2\le{}x_2\le{}L_0)$ is shown in figure \ref{fig:RadiativeCouetteFlow}. Large vibrational collision number ($Z_v=10Z_r$) suppresses the rise of the vibrational temperature, and hence limits the radiation effect. When $Z_v=2Z_r$, the radiative effect lower the vibrational temperature and heat flux significantly compared with those of the non-radiative gas flow. However, the radiation itself may contribute a nonnegligible part to the total heat flux.

Contradict to the results in Fourier flow problems, the proportion of radiative heat flux is not monotonically change with $\text{Kn}_{\text{photon}}$, as shown in figure \ref{fig:RadiativeCouetterFlow_q}. With the increase of $\text{Kn}_{\text{photon}}$, the radiative heat flux firstly becomes significant due to the longer distance the photons can propagate before absorbed by gas molecules. However, it decreases gradually when $\text{Kn}_{\text{photon}}$ further increases, since the rise of radiative temperature is generated by the gas-photon interaction in Couette flow, and the lower emissivity of photon (larger $\text{Kn}_{\text{photon}}$) reduces the radiative temperature difference between the gas and the wall. As the results of the two opposite mechanisms, the radiative heat flux reaches the maximum value at an intermediate value of $\text{Kn}_{\text{photon}}$, which is around 1 in these cases.

\subsection{Normal shock wave}

When the temperature dependent vibrational DoF is considered, the specific heat ratio changes with the temperature across the shock wave structure, thus the Rankine-Hugoniot relation for the upstream and downstream macroscopic quantities is no long applicable. In this situation, consider the conservation of mass, momentum and energy of the normal shock wave,
\begin{equation}\label{eq:conservation_shockwave}
	\begin{aligned}
		n_1u_1&=n_2u_2, \\
		mn_1u_1^2+p_1&=mn_2u_2^2+p_2, \\
		c_{p,1}T_1+\frac{1}{2}u_1^2+2\pi\int{I^R_1\cos{\theta}\mathrm{d}\theta}&=c_{p,2}T_2+\frac{1}{2}u_2^2+2\pi\int{I^R_2\cos{\theta}\mathrm{d}\theta},
	\end{aligned}
\end{equation}
where $c_p$ is the specific heat capacity with constant pressure, which relates to the specific heat ratio $\gamma(T)=(5+d_r+d_v(T))/(3+d_r+d_v(T))$ as $c_p=[\gamma/(\gamma-1)]k_B/m$. The heat flux due to radiation vanish far away from the shock wave layer, then the equations \eqref{eq:conservation_shockwave} are solved,
\begin{equation}\label{eq:shockwave_n_u_T}
	\begin{aligned}
		\frac{n_2}{n_1}&=1-\frac{1}{\gamma(T_1)\text{Ma}^2}\left(\frac{p_2}{p_1}-1\right), \quad
		\frac{u_2}{u_1}&=\frac{n_1}{n_2}, \quad
		\frac{T_2}{T_1}&=\frac{p_2}{p_1}\frac{n_1}{n_2}.
	\end{aligned}
\end{equation}
with
\begin{equation}\label{eq:shockwave_p}
	\begin{aligned}
		\frac{p_2}{p_1}=~&\frac{\left((\gamma(T_1)^2\text{Ma}^4+\gamma(T_2)^2)(\gamma(T_1)-1)-2\text{Ma}^2\gamma(T_1)(\gamma(T_2)^2-\gamma(T_1))\right)^{1/2}}{(\gamma(T_1)-1)^{1/2}(\gamma(T_2)+1)} \\
		&+\frac{\gamma(T_1)\text{Ma}^2+1}{\gamma(T_2)+1},
	\end{aligned}
\end{equation}
where Mach number is defined by the velocity of upstream flow $\text{Ma}=u_1/\sqrt{\gamma(T_1)k_BT_1/m}$.

\begin{figure}[t]
	\centering
	\subfloat[]{\includegraphics[scale=0.24,clip=true]{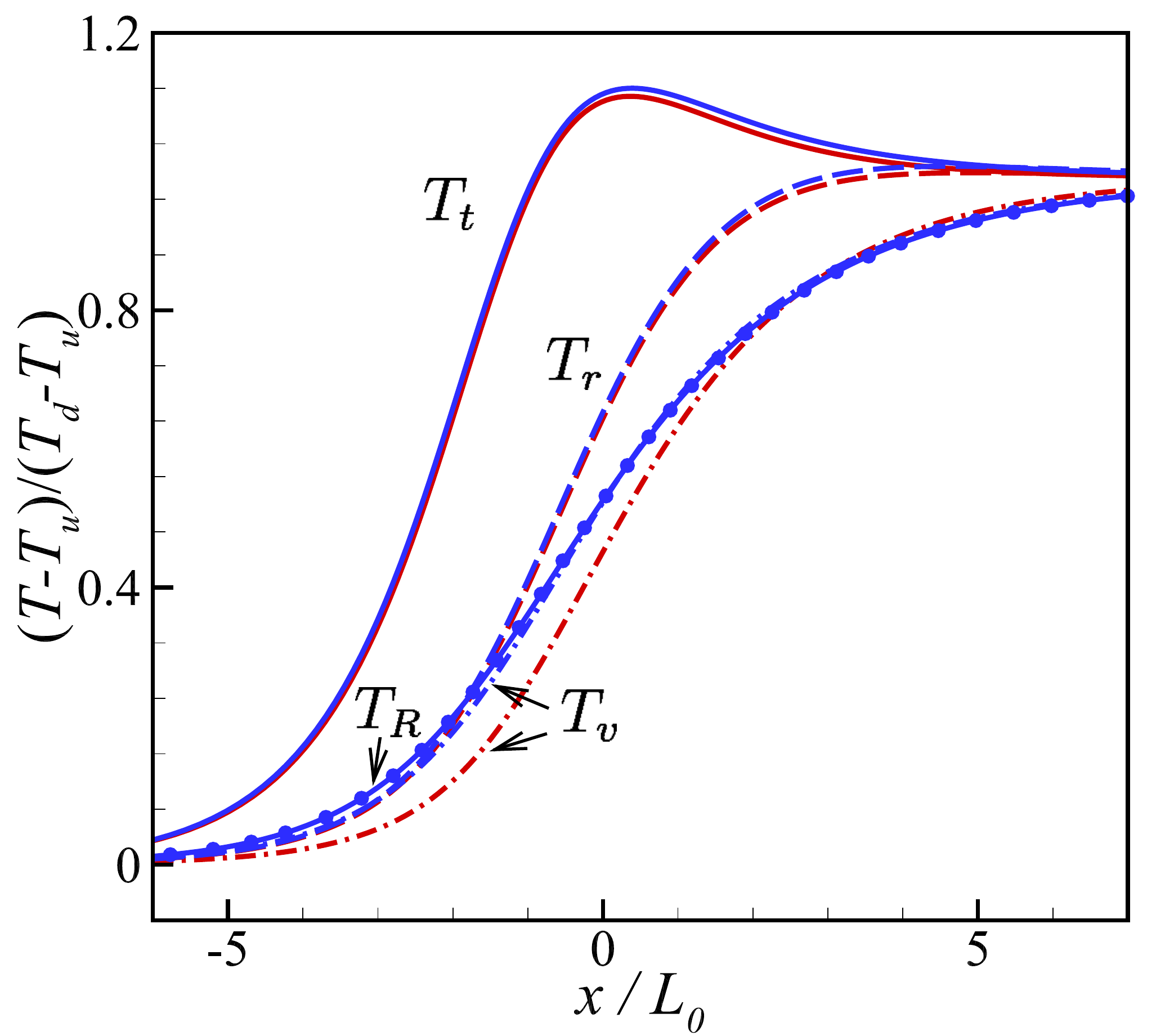}\label{fig:RadiativeNormalShockwave_T1}} \quad
	\subfloat[]{\includegraphics[scale=0.24,clip=true]{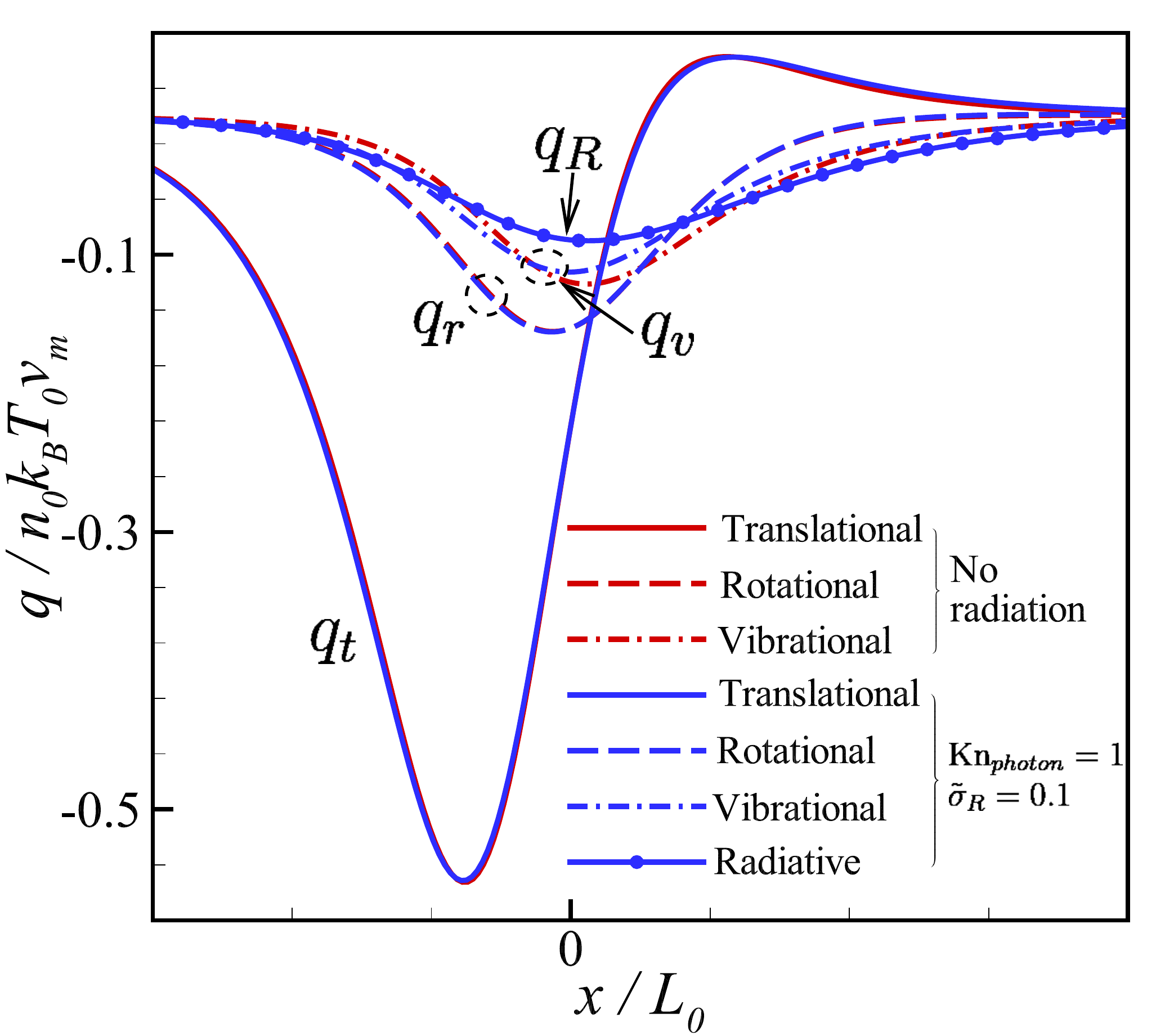}\label{fig:1RadiativeNormalShockwave_Q1}} \\
	\subfloat[]{\includegraphics[scale=0.24,clip=true]{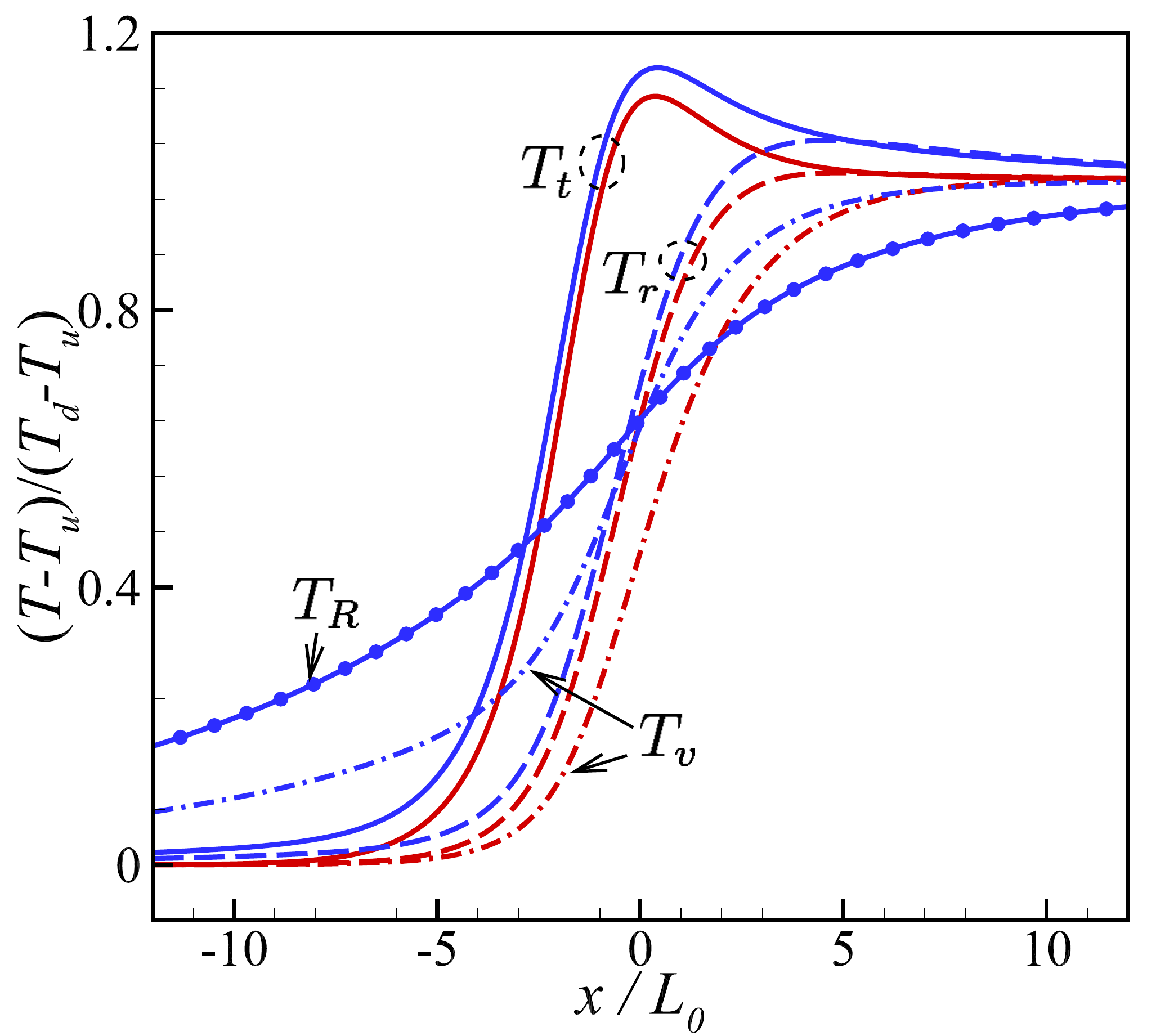}\label{fig:RadiativeNormalShockwave_T2}} \quad
	\subfloat[]{\includegraphics[scale=0.24,clip=true]{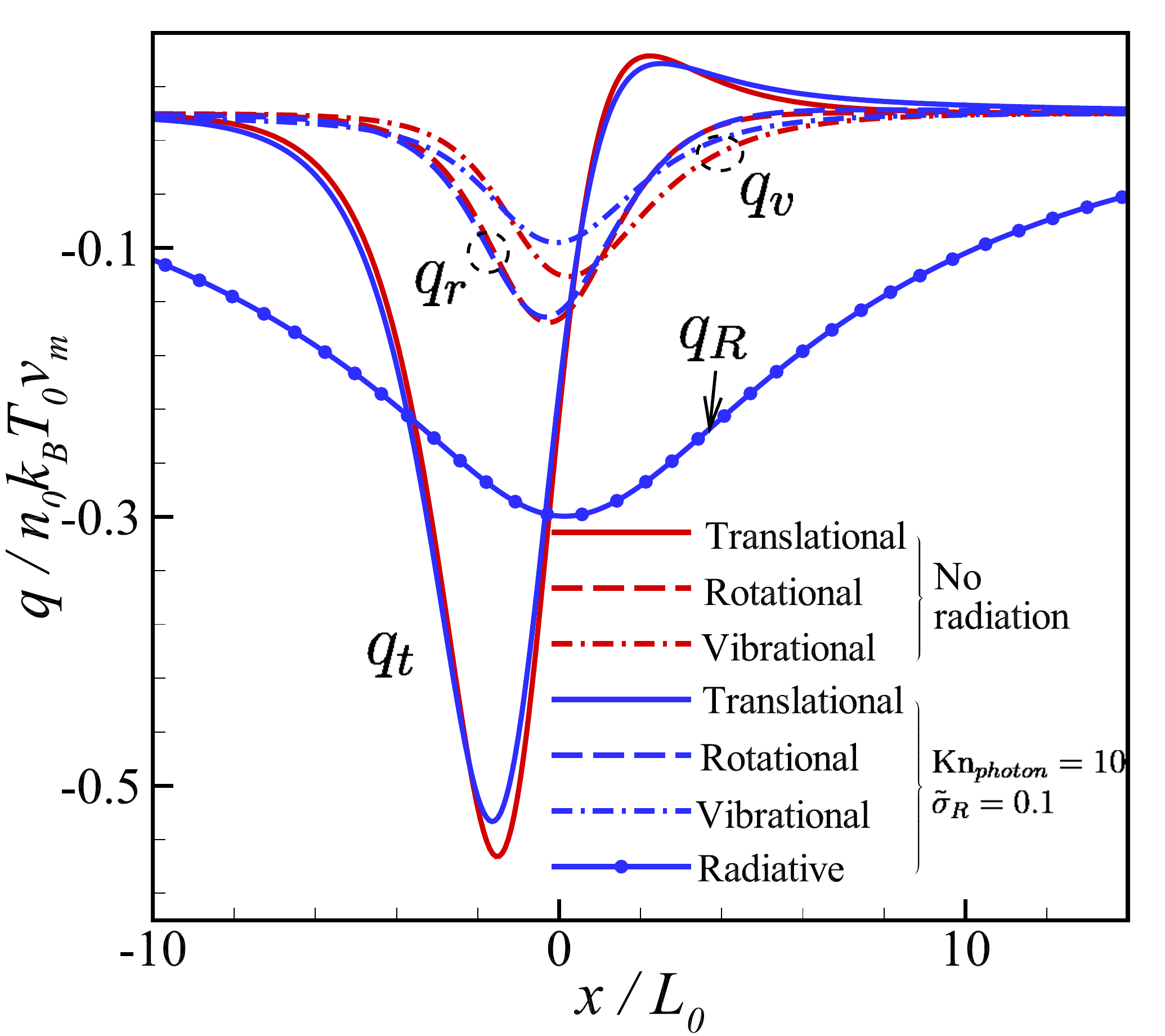}\label{fig:RadiativeNormalShockwave_Q2}} \\
	\caption{Temperature and heat flux distribution across the shock wave structure, when $\text{Ma}=2$, $\tilde{\sigma}_R=0.1$ and $T_0/T_{\text{ref}}=1$. (a,b) $\text{Kn}_{\text{photon}}=1$, (c,d) $\text{Kn}_{\text{photon}}=10$.}
	\label{fig:RadiativeNormalShockwave}
\end{figure}

Figure \ref{fig:RadiativeNormalShockwave} shows the temperature and heat flux across the shock wave structure with different $\text{Kn}_{\text{photon}}$. The smaller $\text{Kn}_{\text{photon}}$ means a more frequent interaction between vibrational mode and radiation field, and thus makes the radiative temperature more close to the vibrational one. While the larger $\text{Kn}_{\text{photon}}$ leads to a much thicker layer of the normal shock wave, and profound radiative heat flux.

\section{Applications to two-dimensional hypersonic radiative flows}\label{sec:2D}

One of the most important situations in which the radiative energy interacts strongly with the rarefied gas flow is radiative shock wave passing obstacles. The gas is heated sufficiently for the radiative flux to be comparable to the flux of kinetic energy. With the ability to perform coupled simulations of rarefied gas dynamics and radiative field, the effect of radiation on the flow structure of hypersonic non-equilibrium can be understood. Since our objective here is to examine the significance of radiative transfer, for the sake of computational simplicity, the kinetic model I with RTA is adopted. The photon absorptivity is stilled approximated by gray model, but its value is proportional to the product of gas number density and vibrational DoF, as indicated in \eqref{eq:absorptivity_Einstein_coefficient}. When the spectrum absorptivities are available, the extension to non-gray model of radiation field is described by the equations \eqref{eq:ke_kne}.

We consider the hypersonic gas flow with density $n_0$ at $\text{Ma}=15$ passing a cylinder with side length $L_0$, the temperatures of both the incoming flow and surfaces of the cylinder are maintained at $T_0=T_{\text{ref}}/2$. The Knudsen numbers of gas of the incoming flow are $\text{Kn}_{\text{gas}}=0.05$, the Knudsen numbers of photon at reference state $\text{Kn}_{\text{photon,ref}}$, when the gas density is $n_0$ and the vibrational DoF is 1, varies from 0.1 to 1000, and the dimensionless radiative strength $\tilde{\sigma}_R$ changes from 0.1 to 10. 

The discretized velocity method is applied to solve the kinetic model equations. The simulation domain has the radius ten times larger than that of the cylinder ($r=L_0/2$). Only the upper-half domain is used in the simulation due to symmetry, which is divided into $40\times60$ structured quadrilateral meshes with refinement near the cylinder surface. The reduced 2D molecular velocity space is truncated by $[-25\sqrt{2}v_m/2,25\sqrt{2}v_m/2]\times[-7\sqrt{2}v_m,7\sqrt{2}v_m]$, and discretized uniformly by $120\times50$ velocity points. The properties of photon over the solid angle is obtained by Gauss-Legendre integral with $48\times48$ discretized points. In the numerical implementation, a finite volume method with second-order reconstruction scheme is adopted. The convention fluxes are evaluated implicitly by the point relaxation technique, while the collision terms are calculated with the Venkata limiter.

\subsection{Effect of radiation}

\begin{figure}[t]
	\centering
	\subfloat[]{\includegraphics[scale=0.28,clip=true]{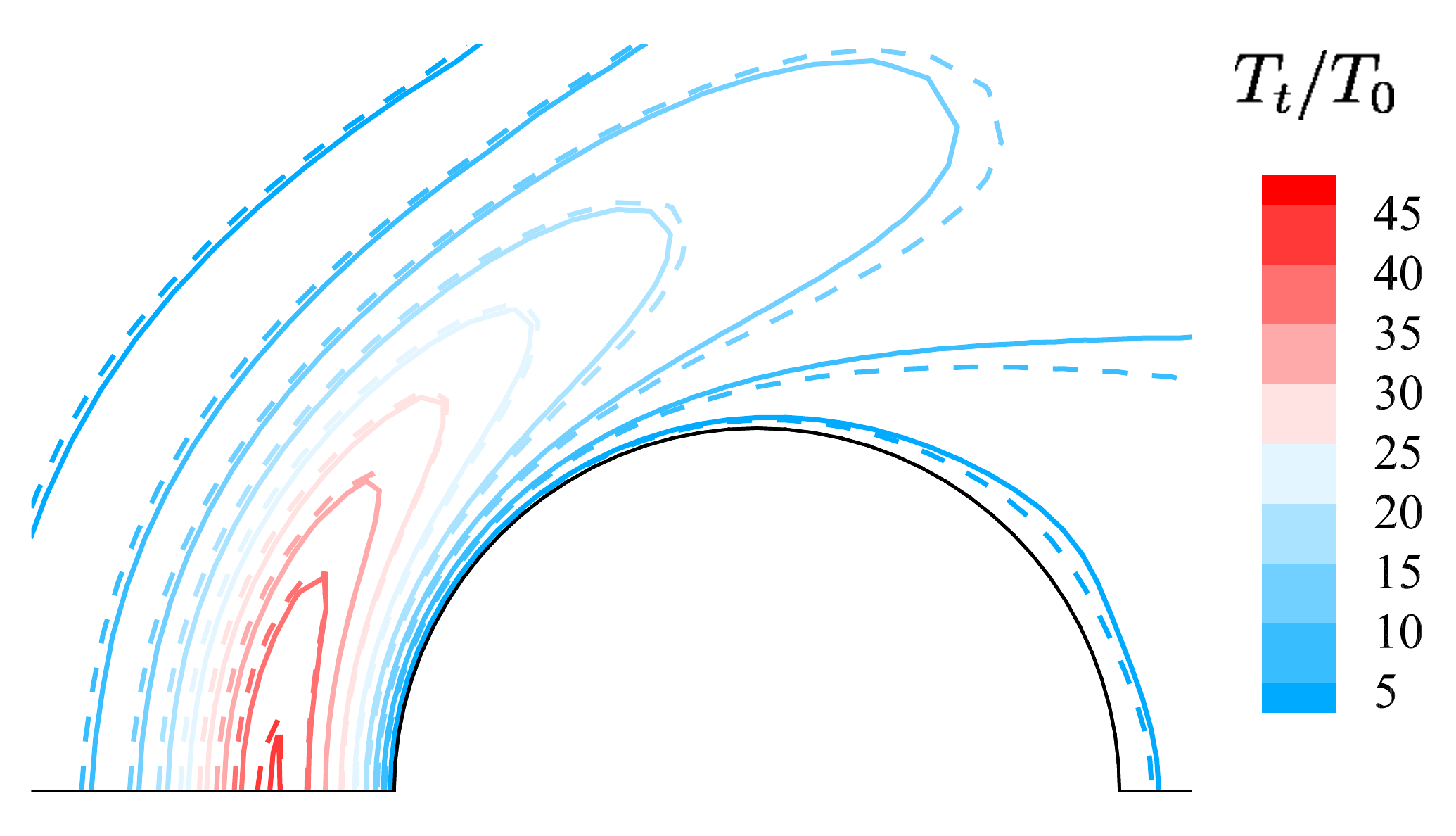}\label{fig:2DShockWave_compare:a}} \quad
	\subfloat[]{\includegraphics[scale=0.28,clip=true]{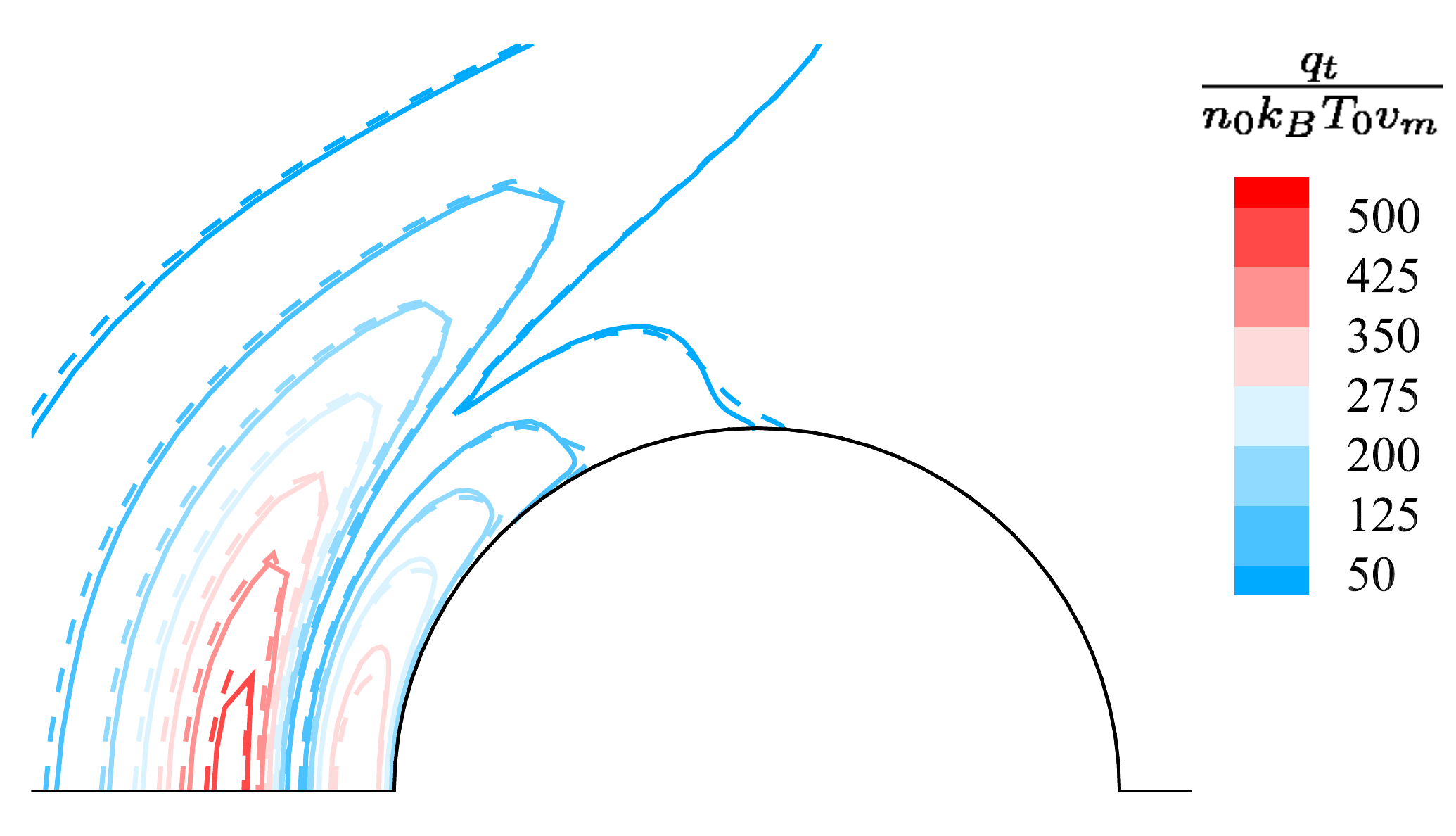}\label{fig:2DShockWave_compare:b}} \\
	\subfloat[]{\includegraphics[scale=0.28,clip=true]{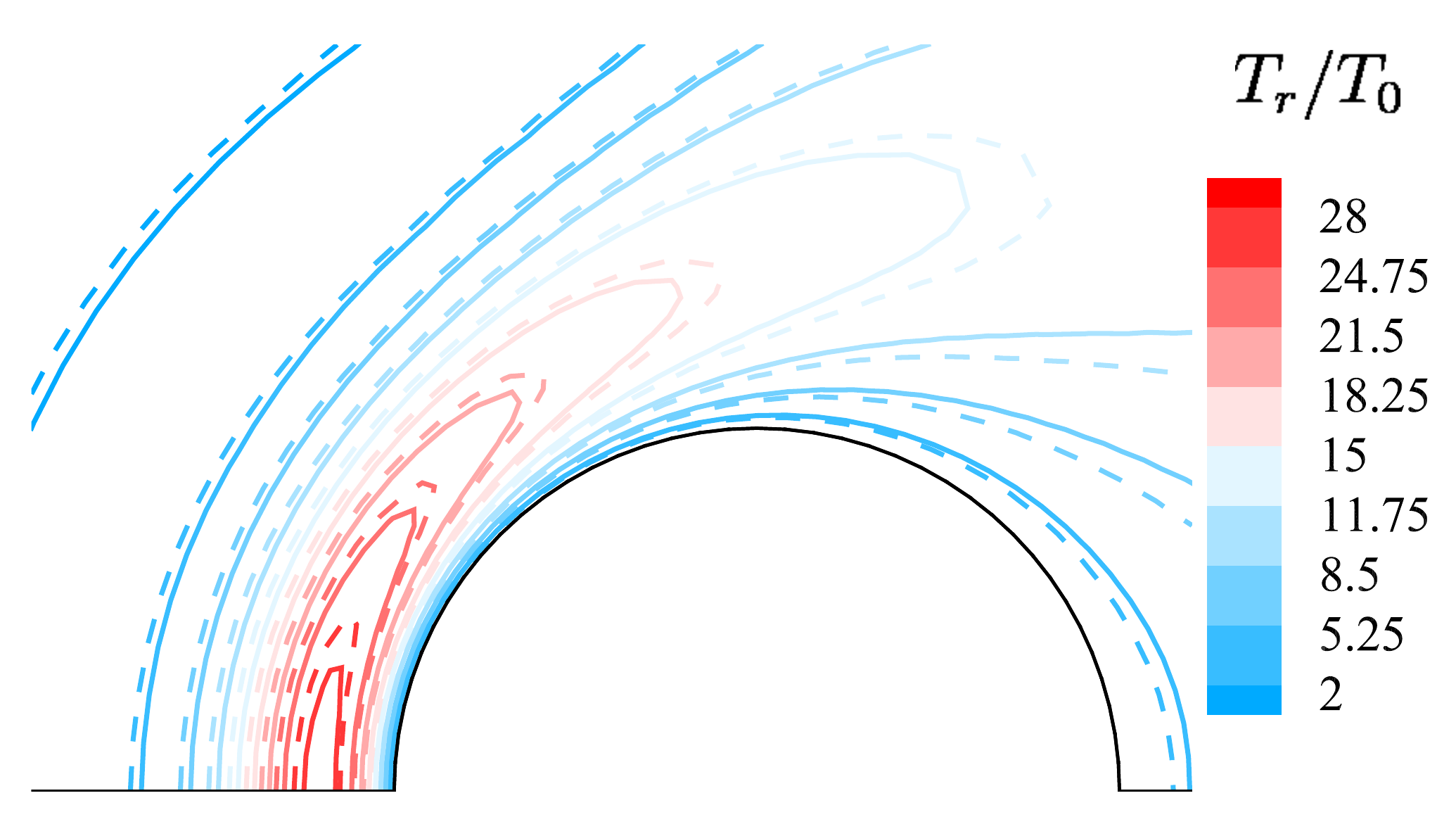}\label{fig:2DShockWave_compare:c}} \quad
	\subfloat[]{\includegraphics[scale=0.28,clip=true]{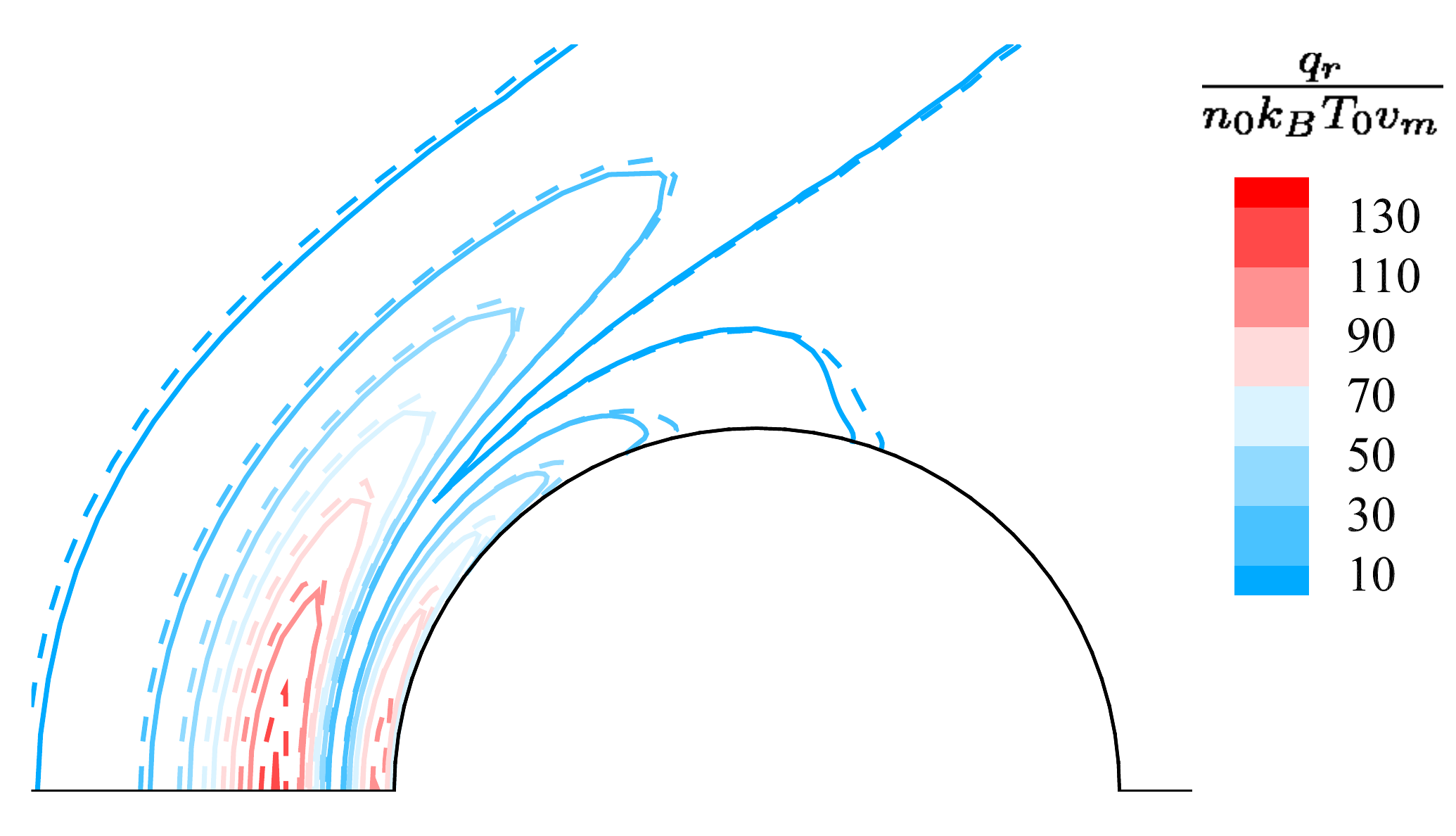}\label{fig:2DShockWave_compare:d}} \\
	\subfloat[]{\includegraphics[scale=0.28,clip=true]{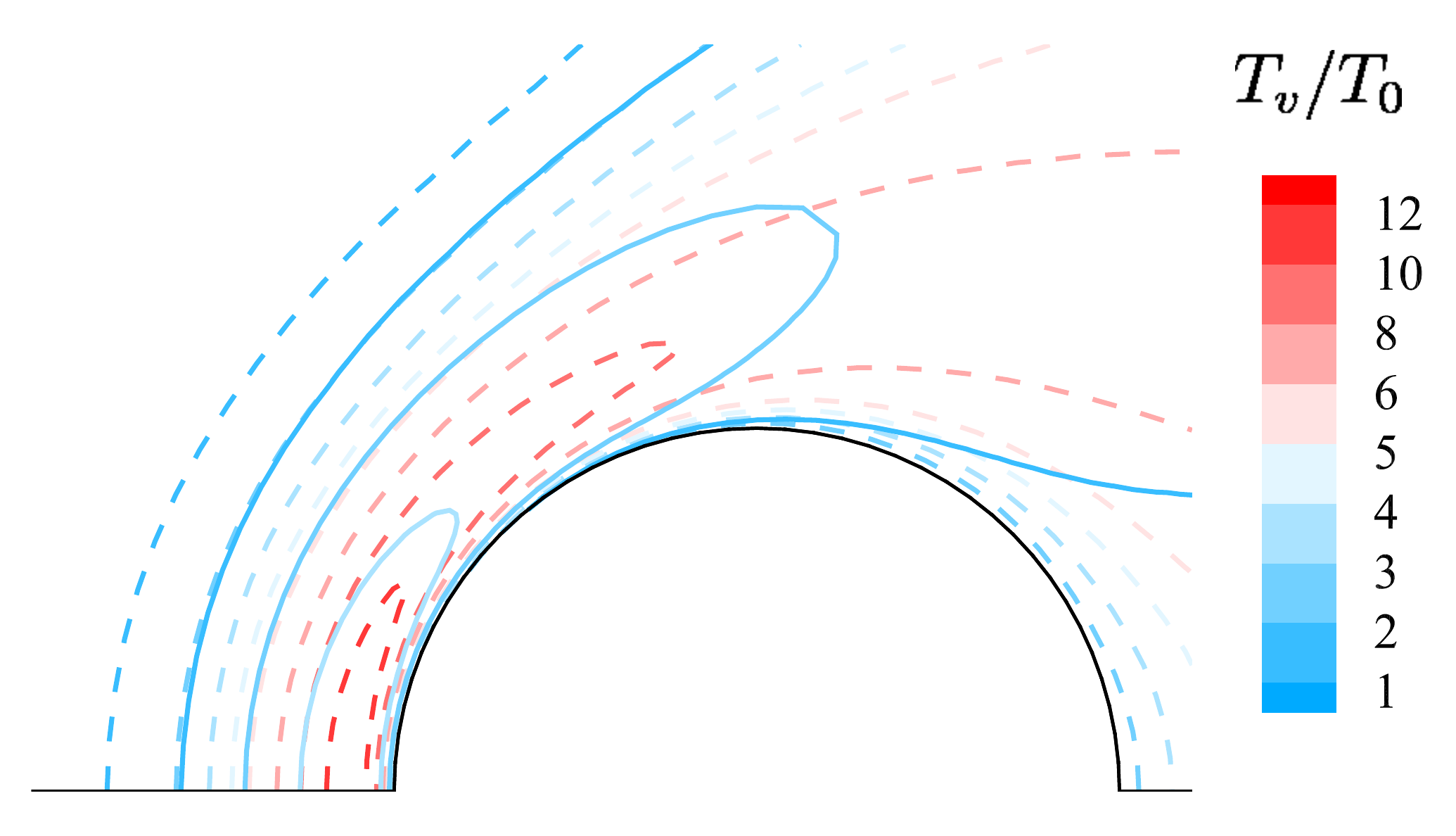}\label{fig:2DShockWave_compare:e}} \quad
	\subfloat[]{\includegraphics[scale=0.28,clip=true]{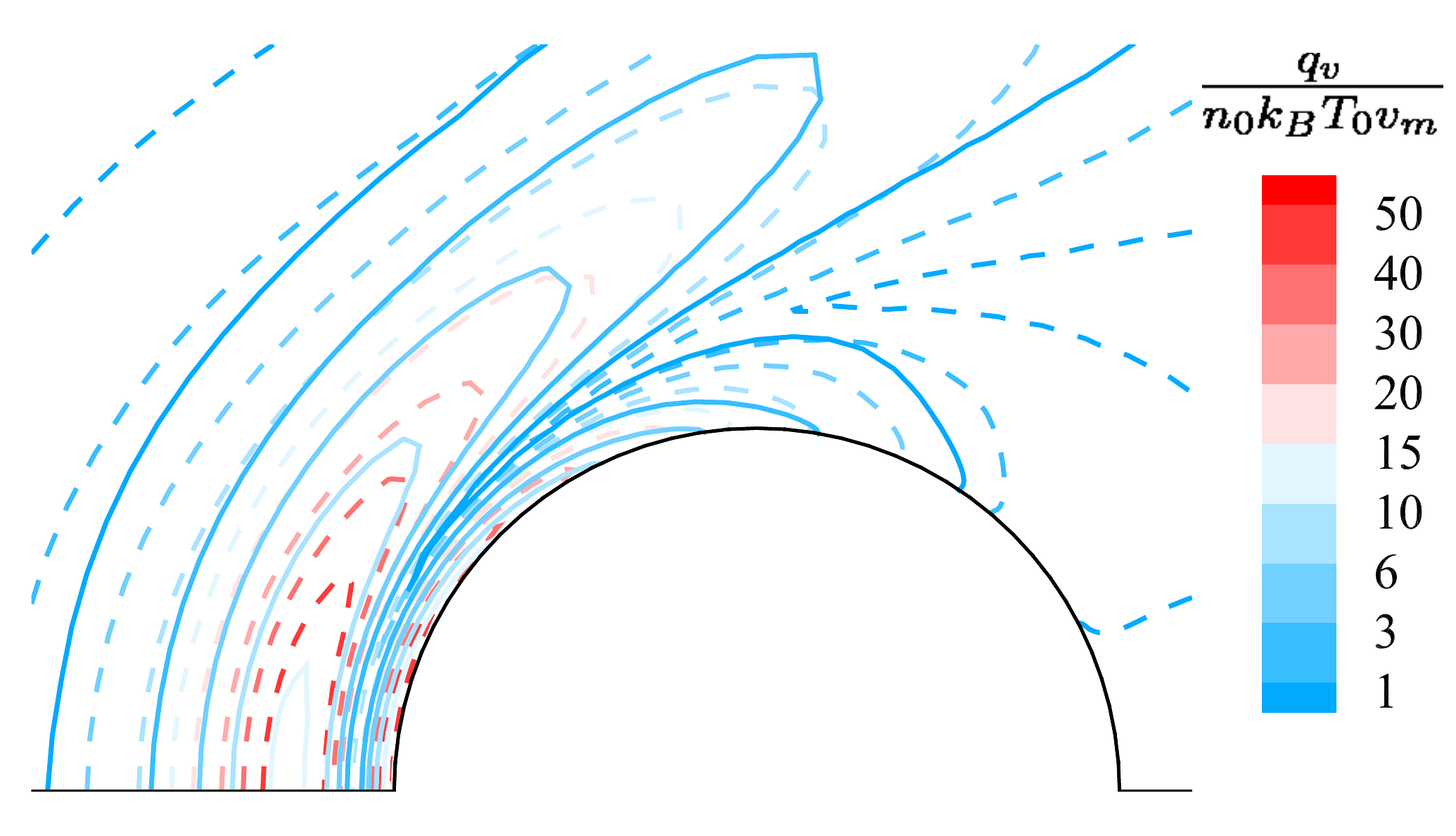}\label{fig:2DShockWave_compare:f}}\\
	\caption{A $\text{Ma}=15$ shock wave passing through the cylinder. The distribution of the (a) translational, (c) rotational (e) vibrational temperatures (normalized by $T_0$), and the (b) translational, (d) rotational (f) vibrational heat fluxes (normalized by $n_0k_BT_0v_m$) around the cylinder solved by kinetic model equations when the incoming flow has $\text{Kn}_{\text{gas}}=0.05$ and $T_0=T_{\text{ref}}/2$. The dashed lines represent the results of the case without radiation, while the solid lines are the results of the case with $\text{Kn}_{\text{photon,ref}}=100$ and $\tilde{\sigma}_R=10$. Note that the colorbars for vibrational temperature and heat flux are not in the linear scale.}
	\label{fig:2DShockWave_compare}
\end{figure}

When $\text{Kn}_{\text{photon,ref}}=100$ and $\tilde{\sigma}_R=10$, the temperature and heat flux of each mode of the surrounding gas are shown in figure \ref{fig:2DShockWave_compare}, for both non-radiative and radiative gas flow. Due to the direct energy exchange between the vibrational mode of gas molecules and photons, the vibrational temperature is significantly reduced by radiation, while the influence on both the translational and rotational modes are relatively small. Therefore, the difference between the translational and vibrational temperatures enlarges in the radiative hypersonic gas flow, since both $\mu_b$ and $\mu_b^R$ contribute to the thermal non-equilibrium. To be specific, the bulk viscosity $\mu_b$ resisting the volume change makes the maximum vibrational temperature about 1/4 of the translational one. Meanwhile, the bulk viscosity $\mu_b^R$ resisting the radiation transition further lower the maximum vibrational temperature to be 1/12 of the translational one. Therefore, the heat flux from the vibrational mode becomes negligible in this case.

\begin{figure}[h]
	\centering
	\subfloat[]{\includegraphics[scale=0.2,clip=true]{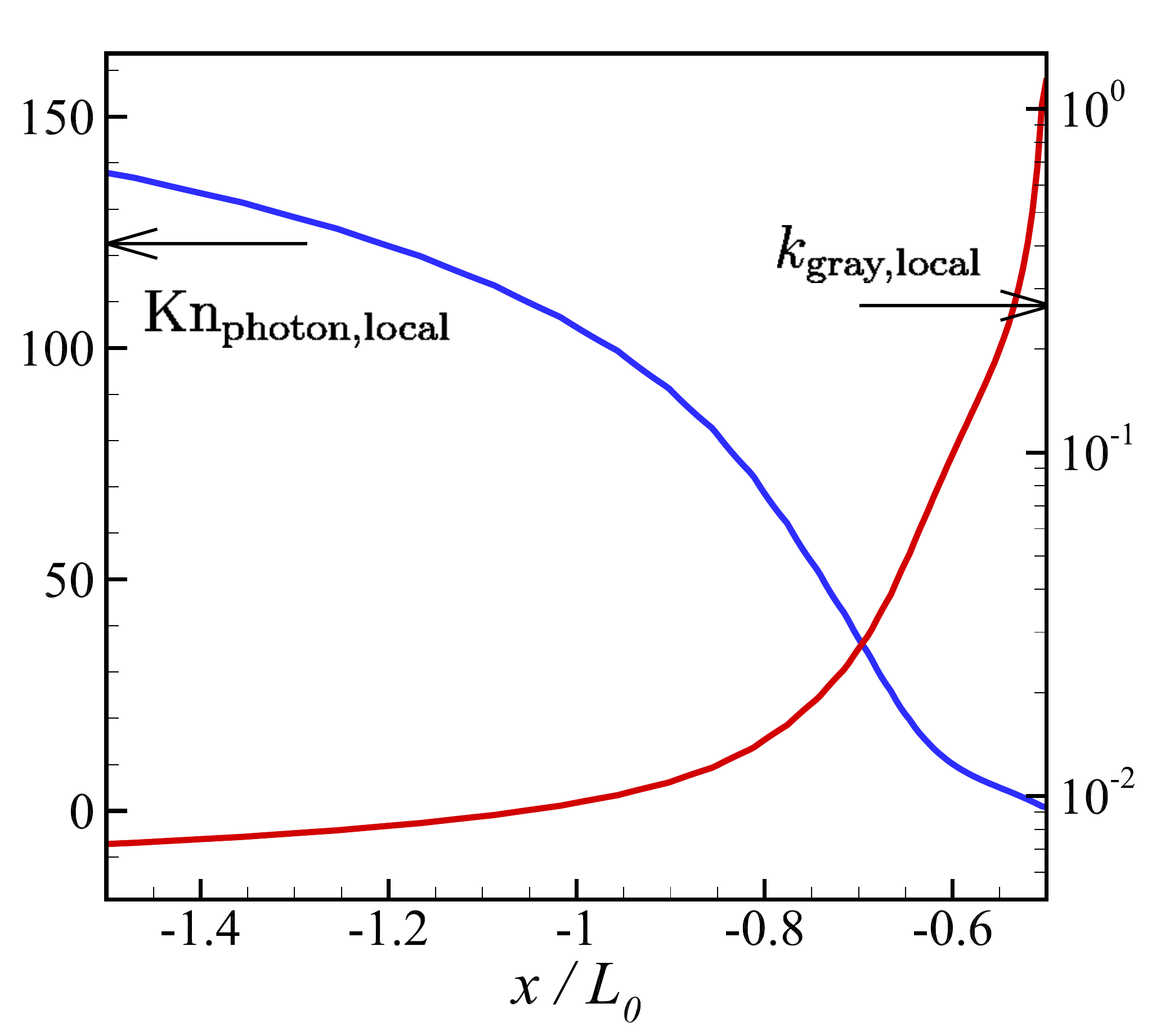}\label{fig:2DShockWave_vR:a}}
	\subfloat[]{\includegraphics[scale=0.2,clip=true]{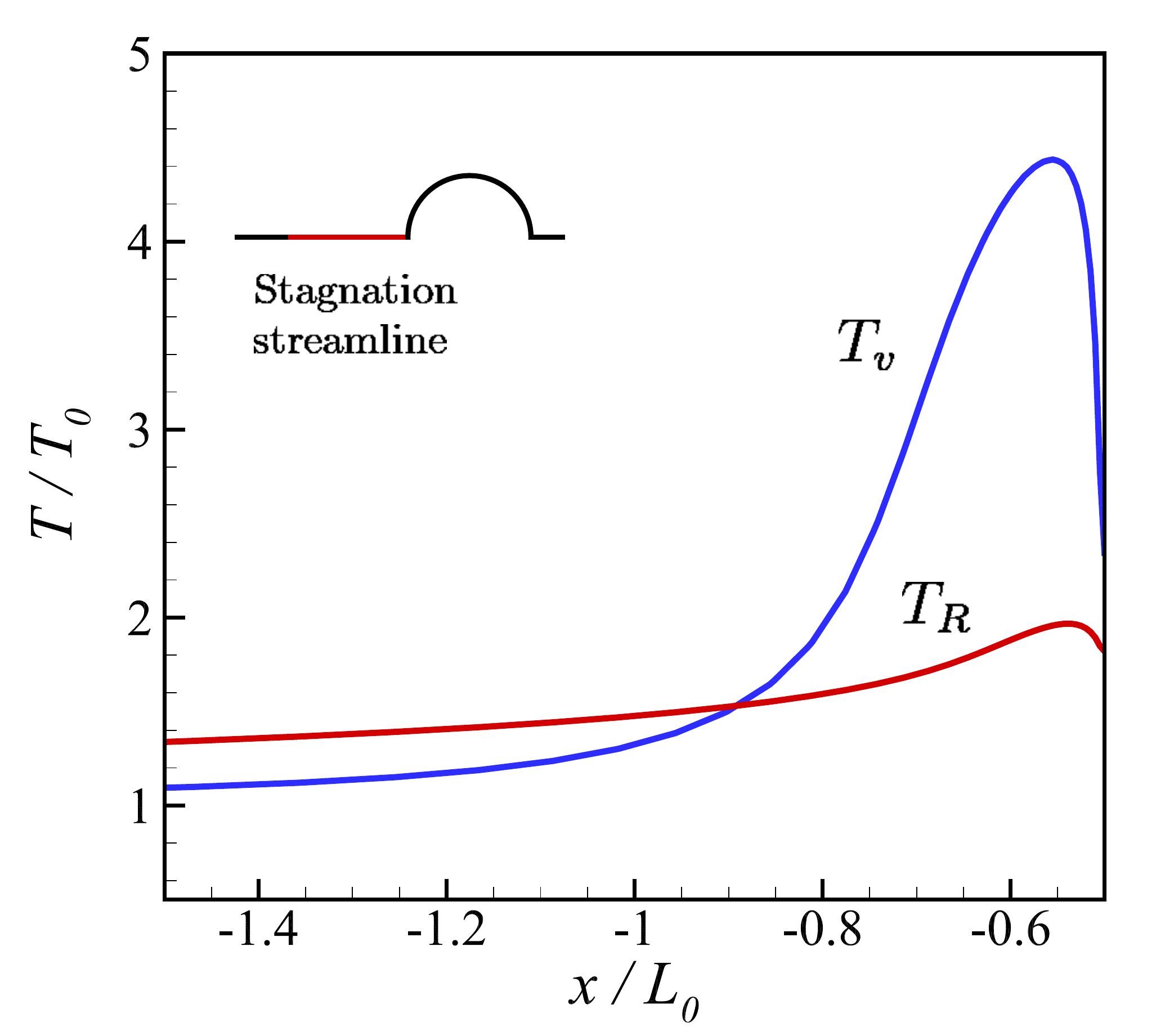}\label{fig:2DShockWave_vR:b}} 
	\subfloat[]{\includegraphics[scale=0.2,clip=true]{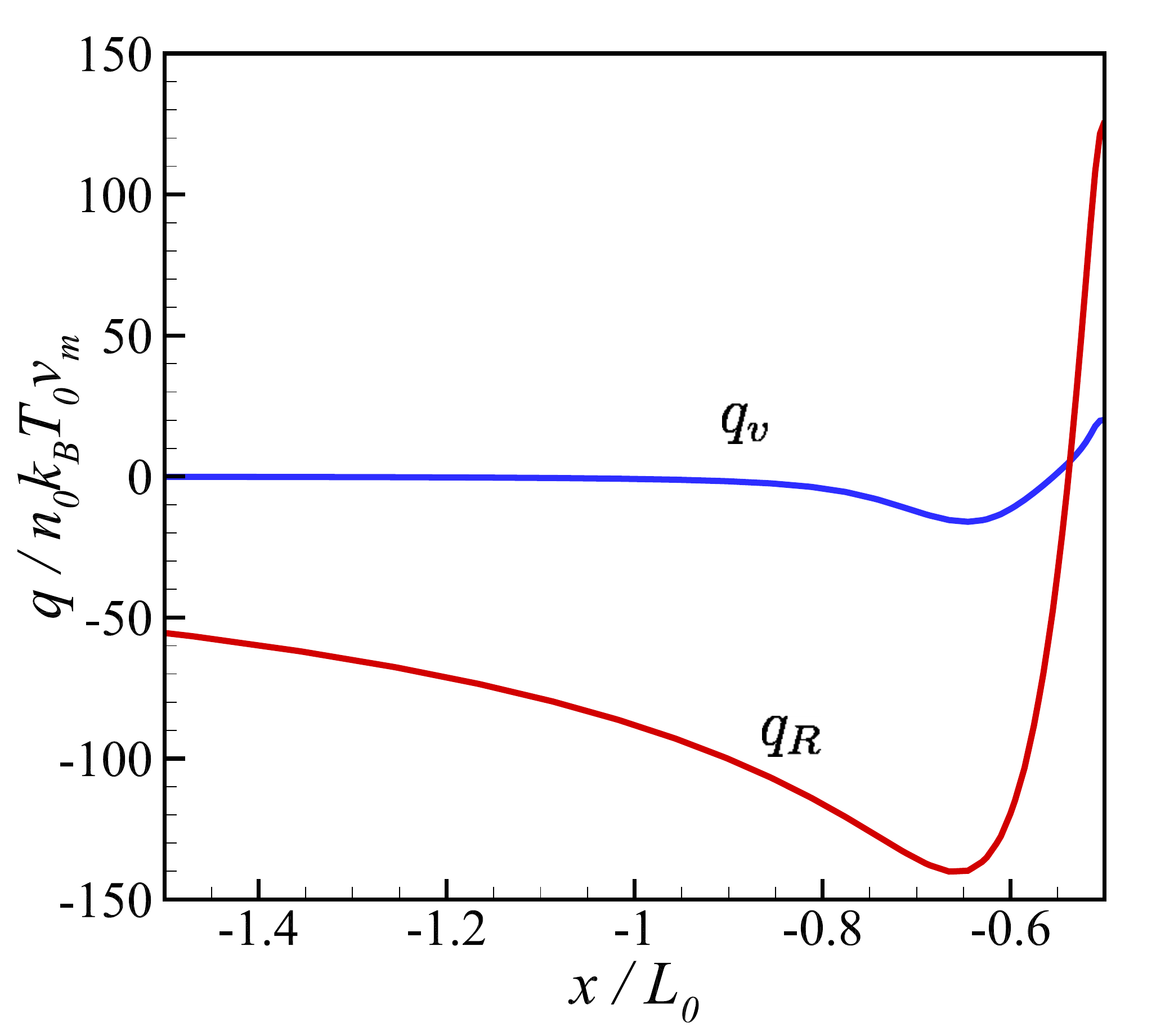}\label{fig:2DShockWave_vR:c}} 
	\caption{The distribution of (a) the local photon Knudsen number and absorptivity (normalized by $1/L_0$), vibrational and radiative (b) temperature (normalized by $T_0$), and (c) heat flux (normalized by $n_0k_BT_0v_m$) along the stagnation streamline before the cylinder, when the incoming flow has $\text{Kn}_{\text{gas}}=0.05$, $\text{Ma}=15$, $T_0=T_{\text{ref}}/2$, $\text{Kn}_{\text{photon,ref}}=100$ and $\tilde{\sigma}_R=10$.}
	\label{fig:2DShockWave_vR}
\end{figure}

The photon absorptivity is proportional to the population of vibrational energy state (equation \eqref{eq:absorptivity_Einstein_coefficient}), thus it increases with the gas density and effective vibrational DoF. As shown in figure \ref{fig:2DShockWave_vR:c}, the order of magnitudes of the photon absorptivity increases significantly in the stagnation region, and hence lower the local photon Knudsen number there. The vibrational/radiative temperatures and heat fluxes along the stagnation streamline are shown in figure \ref{fig:2DShockWave_vR}(a,b). The variation of radiative temperature along the streamline is very small, since the photon Knudsen number is much larger than that of the gas over the entire domain, while the vibrational and radiative temperatures are getting close near the stagnation point, as a result of the high photon absorptivity. However, owing to the relatively high radiative strength $\tilde{\sigma}_R$, the radiative heat flux is one order of magnitude larger than that of the vibrational mode, and contribute around 32\% to the total heat flux over the cylinder surface.

\subsection{Influence of $\text{Kn}_{\text{photon}}$ and $\tilde{\sigma}_R$}

\begin{figure}[h]
	\centering
	\subfloat[]{\includegraphics[scale=0.28,clip=true]{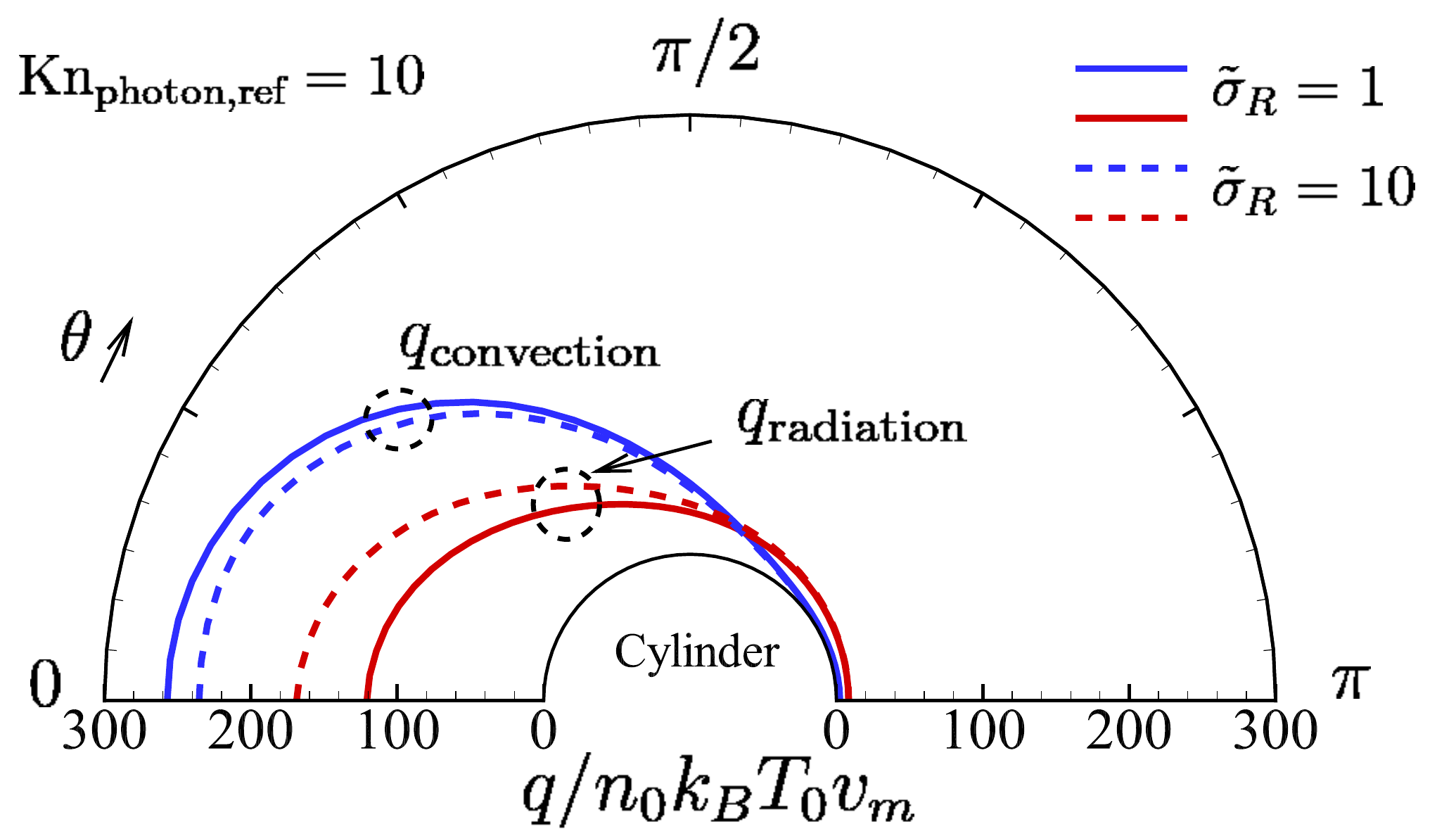}\label{fig:2DShockWave_Qwall:a}} \quad
	\subfloat[]{\includegraphics[scale=0.28,clip=true]{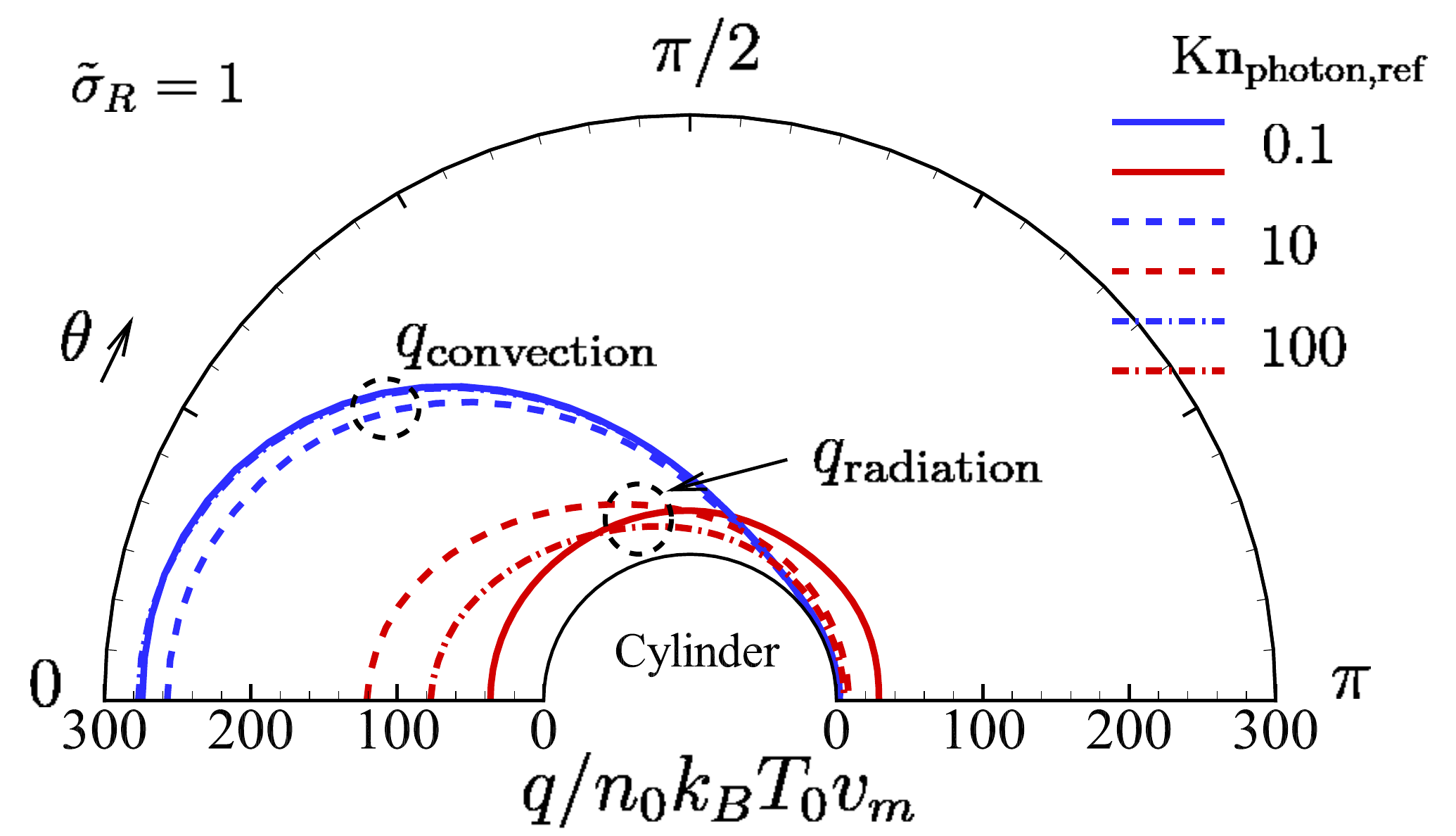}\label{fig:2DShockWave_Qwall:b}} \\
	\caption{
		The heat flux (normalized by $n_0k_BT_0v_m$) from convection (blue lines) and radiation (red lines) along the cylinder surface.  (a) $\text{Kn}_{\text{photon,ref}}=10$, $\tilde{\sigma}_R=$1 (solid lines), 10 (dashed lines). (b) $\tilde{\sigma}_R=1$, $\text{Kn}_{\text{photon,ref}}=$0.1 (solid lines), 10 (dashed lines), 100 (dashed dot lines). Note that $\theta$ is the clockwise angle measured from the stagnation streamline. Also note that the two lines of convective heat fluxes when $\text{Kn}_{\text{photon,ref}}=0.1$ and 100 are overlapped in (b).}
	\label{fig:2DShockWave_Qwall}
\end{figure}

By varying the reference photon Knudsen numbers $\text{Kn}_{\text{photon,ref}}$ from $10^{-1}$ to $10^3$, and the dimensionless radiative strength $\tilde{\sigma}_R$ from $10^{-1.5}$ to $10^1$, their influence on the radiative heat flux on the surface of cylinder are systematically revealed. Figure \ref{fig:2DShockWave_Qwall} shows the magnitude of normal heat flux from both convection and radiation to the cylinder. Note that $\text{Kn}_{\text{photon,ref}}$ indicates the frequency of gas-photon interaction at reference state, which is determined by the molecular structure and number density. 

When $\text{Kn}_{\text{photon,ref}}$ is fixed (figure \ref{fig:2DShockWave_Qwall:a}), higher value of $\tilde{\sigma}_R$ not only increases the relative radiative intensity and heat flux, but also leads to stronger coupling between the gas flow and the radiation field, and hence decreases the convection heat flux. Therefore, the relative importance of the radiative heat transfer becomes significant. 

When $\tilde{\sigma}_R$ is fixed (figure \ref{fig:2DShockWave_Qwall:b}), the influence of $\text{Kn}_{\text{photon,ref}}$ is different at different regimes of photon Knudsen number. Two mechanisms are summarized below:
\begin{enumerate}
	\item the gas-photon interaction becomes weaker with the increase of $\text{Kn}_{\text{photon,ref}}$, and thus makes the radiative temperature much lower than the gas temperature. Consequently, the radiative heat flux decreases while the convective heat flux increases. This mechanism dominates the influence of $\text{Kn}_{\text{photon,ref}}$ when the photon Knudsen number is relatively large, e.g., when $\text{Kn}_{\text{photon,ref}}$ changes from 10 to 100. 
	\item the mean free path of photon (reciprocal of absorptivity) increases with $\text{Kn}_{\text{photon,ref}}$, and then the transportation distance of energy carried by the photon before it is absorbed by the gas becomes longer. 
\end{enumerate} 

Therefore, the radiative heat flux increases and surpasses the first mechanism when the photon Knudsen number is relatively small, where the radiative temperature is almost the same as the vibrational temperature ($\text{Kn}_{\text{photon,ref}}$ changing from 0.1 to 10 shown in figure \ref{fig:2DShockWave_Qwall:b}). It is noticed that in the leeward side of the cylinder, the radiative heat flux still decreases when $\text{Kn}_{\text{photon,ref}}$ changes from 0.1 to 10, which can be understood that the local photon Knudsen number is much higher (due to its low gas density in the leeward region) than that in the windward side.

\begin{figure}[t]
	\centering
	\includegraphics[scale=0.37,clip=true]{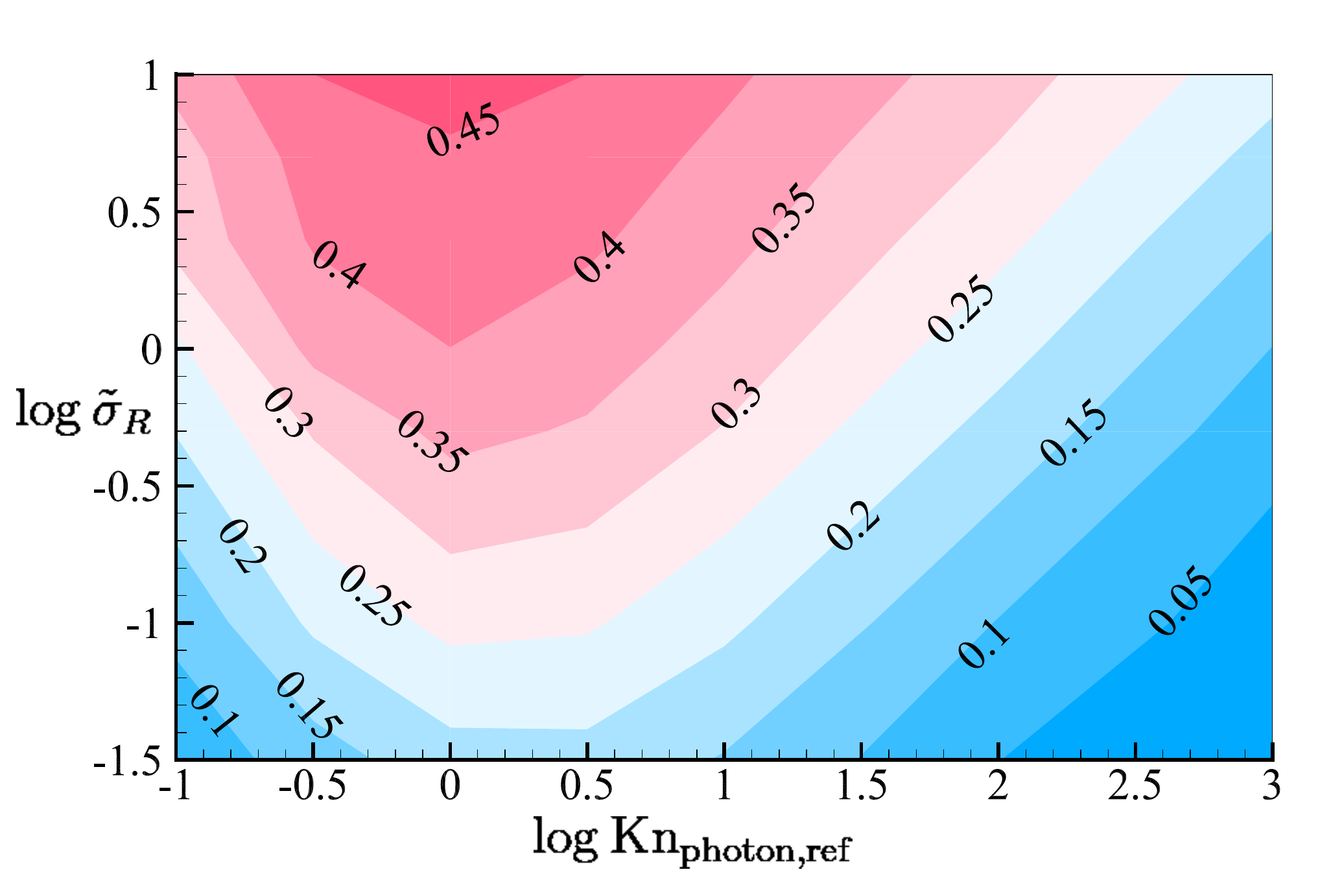}
	\caption{The ratio of radiative heat flux to the total heat flux on the surface of the cylinder in hypersonic gas flow with the variation of $\text{Kn}_{\text{photon,ref}}$ and $\tilde{\sigma}_R$.} 
	\label{fig:2DShockWave_Qwall_qRqtot}
\end{figure}

To quantitatively measure the importance of the radiation in hypersonic gas flow as $\text{Kn}_{\text{photon,ref}}$ and $\tilde{\sigma}_R$ change, the ratio of normal radiative heat flux to the total one on the surface of the cylinder is calculated and shown in figure \ref{fig:2DShockWave_Qwall_qRqtot}. The contour lines show two regimes of the influence from the two parameters: when $\text{Kn}_{\text{photon,ref}}\gg 1$, the proportion of the radiative heat flux is constant along $\tilde{\sigma}_R/\text{Kn}_{\text{photon,ref}}$; when $\text{Kn}_{\text{photon,ref}}\ll 1$, it is constant along $\tilde{\sigma}_R\cdot\text{Kn}_{\text{photon,ref}}$. Therefore, it gives the clear criterions to determine the importance of radiative heat transfer in these problems.

\section{Conclusions}\label{sec:conclusion}

In summary, we have proposed two tractable kinetic models to describe the high-temperature rarefied  gas flows with radiation, where the gas flow and radiation are coupled self-consistently from mesoscopic perspective, for the first time in literatures. Both the two kinetic models can recover not only the transport coefficients, but also the fundamental relaxation processes. While they differ in the manner dealing with the elastic intermolecular collisions: one use Boltzmann collision operator that has the ability to capture the influence of intermolecular potentials; the other adopts relaxation time approximation that has higher computational efficiency.

The accuracy of the proposed models have been assessed by comparing with DSMC simulations for one-dimensional non-radiative Fourier flow, Couette flow, creep flow driven by the Maxwell demon and normal shock wave. The kinetic model with Boltzmann collision operator demonstrates its excellent accuracy in all these benchmarks, while the one with relaxation time approximation shows slight discrepancy in normal shock wave and creep flows, due to its ignoring of the velocity-dependence of intermolecular collision rates. 

Then, the study on the effect of radiation has been conducted by solving the kinetic models in both one-dimensional typical rarefied gas flows and two-dimensional radiative hypersonic flow passing cylinder. It is found that the presence of radiation can double the heat load on obstacle surface compared to that with convective heat transfer only. Additionally, the influences of photon Knudsen number and relative radiation strength are systemically investigated, and the parameter regions determining the importance of radiative heat transfer are displayed at different regimes of photon Knudsen number.

Although only the radiation transitions of vibrational mode are involved in the coupling of gas and radiation field in this work, a full consideration of all possible radiation transitions can be achieved straightforwardly in the framework of our general kinetic models. Meanwhile, accurate prediction of the radiative environment requires detailed understanding and reliable data of the radiation transitions at the molecular level, which can be acquired from the mathematical models \citep{Zalogin2001TP, Babou2009JQSRT}, as well as experimental measurements in relevant conditions \citep{Reynier2021Galaxies}. In the future work, these realistic parameters will be incorporated into our general kinetic models to study the high-temperature rarefied (non-equilibrium) gas flow with strong radiation.

\section*{Acknowledgements} 
This work is supported by the National Natural Science Foundation of China under the grant No. 12172162.

\section*{Declaration of interests} 
The authors report no conflict of interest.

\appendix{

\section{Transport coefficients from kinetic model equations}\label{app:A}

The transport coefficients given by the kinetic model equations can be obtained by the Chapman-Enskog expansion~\citep{CE}. To the second approximation of the distribution function, it is assumed the velocity distribution function can be written as $f_i=f_i^{(0)}+f_i^{(1)}$, where $f_i^{(0)}=E_t(T)E_r(T)E_{v,i}(T)$ is the equilibrium distribution of the vibrational state $i$ at temperature $T$. Let $\mathcal{D}f_i\equiv{\partial{f_i}}/{\partial{t}}+\bm{v} \cdot {\partial{f_i}}/{\partial{\bm{x}}}+\bm{a} \cdot {\partial{f_i}}/{\partial{\bm{v}}}$, and consider $\mathcal{D}^{(0)}f_i=0$, according to Chapman-Enskog expansion and the kinetic model equation, we have
\begin{equation}\label{eq:Df(1)_J}
	\begin{aligned}[b]	
		f_i^{(1)} = g_{t,i} - f_i^{(0)} +\frac{1}{Z_r}\left(g_{r,i}-g_{t,i}\right) +\frac{1}{Z_v}\left(g_{v,i}-g_{t,i}\right)-\tau\mathcal{D}^{(1)}f_i+\tau J_{photon,i},
	\end{aligned}
\end{equation}
where
\begin{equation}\label{eq:Df(1)}
	\begin{aligned}[b]
		\mathcal{D}^{(1)}f_i=~&\frac{\partial{f_i^{(0)}}}{\partial{t}}+\bm{v} \cdot \frac{\partial{f_i^{(0)}}}{\partial{\bm{x}}}+\bm{a} \cdot \frac{\partial{f_i^{(0)}}}{\partial{\bm{v}}} \\
		=~&f_i^{(0)}\left[\left( \left(\frac{mc^{2}}{2k_BT}-\frac{5}{2}\right) +\left(\frac{I_r}{k_BT}-\frac{d_r}{2}\right) +\left(\frac{\varepsilon_i}{k_BT} -\frac{d_v}{2} \right)\right)\bm{c}\cdot\nabla\ln{T} \right. \\
		&\left. +\frac{2}{(3+d_r+d_v)}\left(\frac{d_r+d_v}{3}\left(\frac{mc^{2}}{2k_BT}-\frac{3}{2}\right) -\left(\frac{I_r}{k_BT}-\frac{d_r}{2}\right) -\left(\frac{\varepsilon_i}{k_BT}-\frac{d_v}{2}\right)\right)\nabla\cdot\bm{u} \right. \\
		&\left. +\frac{m}{k_BT}\left(\bm{c}\bm{c}-c^2\mathrm{I}\right):\nabla\bm{u} \right],
	\end{aligned}
\end{equation}
and $\mathrm{I}$ is a $3\times3$ identity matrix.

The pressure tensor $\bm{P}$ is then calculated according to \eqref{eq:macroscopic_variables_f},
\begin{equation}
	\begin{aligned}[b]
		\bm{P}=&\sum_{i}\int_{0}^{\infty}\int_{-\infty}^{\infty}{m\bm{c}\bm{c}(f_i^{(0)}+f_i^{(1)})}\mathrm{d}\bm{v}\mathrm{d}I_r \\
		=& \left(p_t+\frac{1}{Z_r}(p_{tr}-p_t)+\frac{1}{Z_v}(p_{tv}-p_t)\right)\mathrm{I} \\
		&-p\tau\left(\nabla\bm{u}+\nabla\bm{u}^{\mathrm{T}}-\frac{2}{3}\nabla\cdot\bm{u}\mathrm{I}\right)-p\tau\frac{2(d_r+d_v)}{3(3+d_r+d_v)}\nabla\cdot\bm{u}\mathrm{I} \\
		=& p\mathrm{I}-p\tau\left(\nabla\bm{u}+\nabla\bm{u}^{\mathrm{T}}-\frac{2}{3}\nabla\cdot\bm{u}\mathrm{I}\right)-2p\tau\frac{(3+d_r)d_rZ_r+(3+d_v)d_vZ_v}{3\left(3+d_r+d_v\right)^2}\nabla\cdot\bm{u}\mathrm{I},
	\end{aligned}
\end{equation}
where the integral term of $J_{photon,i}$ vanishes, since the gas-photon interaction is uncorrelated with the gas translational velocity. Then, the shear viscosity $\mu$ and bulk viscosity $\mu_b$ are obtained:
\begin{equation}
\begin{aligned}[b]
	\mu(T_t)&=p_t\tau, \\ \mu_b(T_t)&=2p_t\tau\frac{(3+d_r)d_rZ_r+(3+d_v)d_vZ_v}{3\left(3+d_r+d_v\right)^2}.
	\end{aligned}
\end{equation}

The translational, rotational and vibrational heat fluxes can be calculated according to \eqref{eq:macroscopic_variables_f},
\begin{equation}\label{eq:dq_dT}
	\begin{aligned}[b]
	\left[ 
      \begin{array}{ccc} 
        \bm{q}_t \\ \bm{q}_r \\ \bm{q}_v
      \end{array}
    \right]
	=& \sum_{i}\int_{0}^{\infty}\int_{-\infty}^{\infty}{\bm{c}\left[ 
      \begin{array}{ccc} 
        \frac{1}{2}mc^2 \\ I_r \\ \varepsilon_v
      \end{array}
    \right] \left(f_i^{(0)}+f_i^{(1)}\right)}\mathrm{d}\bm{v}\mathrm{d}I_r \\
	=& \sum_{i}\int_{0}^{\infty}\int_{-\infty}^{\infty}\bm{c}\left[ 
		\begin{array}{ccc} 
		  \frac{1}{2}mc^2 \\ I_r \\ \varepsilon_v
		\end{array}
	  \right] \times \\
	  & \left[\tau\left(\frac{g_{t,i} - f_i}{\tau} +\frac{g_{r,i}-g_{t,i}}{Z_r\tau} +\frac{g_{v,i}-g_{t,i}}{Z_v\tau}\right) + f_i +\tau J_{photon,i} -\tau\mathcal{D}^{(1)}f_i\right]\mathrm{d}\bm{v}\mathrm{d}I_r \\
	=& \tau\left[ 
      \begin{array}{ccc} 
        \partial{\bm{q}_{t}}/{\partial{t}} \\ \partial{\bm{q}_{r}}/{\partial{t}} \\ \partial{\bm{q}_{v}}/{\partial{t}}
      \end{array}
    \right] +\left[ 
		\begin{array}{ccc} 
		  \bm{q}_t \\ \bm{q}_r \\ \bm{q}_v
		\end{array}
	\right] - \frac{k_B\mu}{2m} \left[ 
      \begin{array}{ccc} 
        5 \\ d_r \\ d_v(T_v)
      \end{array}
    \right]
	\nabla{T}.
	\end{aligned}
\end{equation}
Consider $(\bm{q}_t, \bm{q}_r, \bm{q}_v)=-(\kappa_t, \kappa_r, \kappa_v)\nabla T$, according to \eqref{eq:heat_flux_relaxation}, the thermal conductivities are given by~\eqref{eq:kappa_A}.

}

\vskip 1cm


\bibliographystyle{jfm}
\bibliography{bibnew}

\end{document}